\documentclass[floats,floatfix,amssymb,prd,twocolumn,superscriptaddress,nofootinbib,preprintnumbers]{revtex4-1}

\usepackage{subcaption, ragged2e}
\DeclareCaptionJustification{justified}{\justifying}
\usepackage{amssymb,amsmath,verbatim,mathtools,needspace,enumitem,etoolbox,graphicx,microtype,afterpage,bm}
\usepackage[svgnames,dvipsnames,x11names,table]{xcolor}
\usepackage{tikz}
\usetikzlibrary{shapes.misc, positioning, arrows.meta}
\usepackage{framed}
\usepackage{setspace}
\usepackage{tcolorbox}
\usepackage[table]{xcolor}
\usepackage{pifont}

\usepackage{hyperref}
\hypersetup{
    colorlinks=true,
    linkcolor={Purple!80!black},
    citecolor={rosewood},
    urlcolor={Purple!50!black}
}

\definecolor{pyblue}{RGB}{31, 119, 180}
\definecolor{pyorange}{RGB}{255, 127, 14}
\definecolor{pygreen}{RGB}{44, 160, 44}
\definecolor{pyred}{RGB}{214, 39, 40}
\definecolor{lightgray}{gray}{0.9}
\definecolor{rosewood}{rgb}{0.4, 0.0, 0.04}

\usepackage{multirow,pifont,lmodern,float,tabularx,booktabs}
\usepackage[all]{hypcap}
\usepackage[T1]{fontenc}
\usepackage[utf8]{inputenc}

\allowdisplaybreaks

\usepackage{fontawesome}
\usepackage{ulem}

\usepackage{titlesec}
\titlespacing*{\section}{0pt}{*1.5}{*1}
\titlespacing*{\subsection}{0pt}{*1.5}{*0.8}
\titlespacing*{\subsubsection}{0pt}{*1.5}{*0.8}

\def\d{\mathrm{d}}
\def\P{\mathcal{P}}
\def\A{\mathcal{A}}
\def\Mpl{M_{\text{pl}}}
\def\Mpc{{\text{Mpc}}}
\def\kpc{{\text{kpc}}}

\newcommand{\xmark}{\ding{55}}

\newcommand{\montpellier}{
Laboratoire Univers \& Particules de Montpellier,
CNRS, Université de Montpellier, Montpellier, F-34095, France}

\newcommand{\strasbourg}{
Université de Strasbourg, CNRS, Observatoire Astronomique de Strasbourg, Strasbourg, F-67000, France}

\newcommand{\IAP}{Institut d'Astrophysique de Paris, Sorbonne Université, CNRS, Paris, F-75014, France}

\begin{document}

\title{Primordial Sharp Features through the Nonlinear Regime of Structure Formation}

\author{Cl\'ement Stahl}
\email{clement.stahl@unistra.fr}
\affiliation{\strasbourg}

\author{Denis Werth}
\email{denis.werth@iap.fr}
\affiliation{\IAP} 

\author{Vivian Poulin}
\email{vivian.poulin@umontpellier.fr}
\affiliation{\montpellier}
\date{\today}

\begin{abstract}
\noindent
Sudden violations of the slow-roll regime during inflation, a natural prediction of many UV-complete inflationary models, give rise to sharp features in the primordial power spectrum. At large scales, these features provide a unique window into the physics of inflation, with constraints primarily derived from Cosmic Microwave Background observations of linearly evolved primordial fluctuations. 
However, on smaller scales, it is less clear whether primordial features would survive the late-time nonlinear cosmological evolution, as they are expected to be washed out by mode coupling. 
In this paper, we run dedicated N-body simulations to tackle this question. We demonstrate that,  while oscillatory-like patterns are erased over time by nonlinearities, signatures of primordial sharp features can persist through the nonlinear regime of structure formation.
Those take the form of a localised power enhancement or decrease in the matter power spectrum---whose amplitude and position can in principle be used to recover the scale of the primordial feature---and an oscillatory pattern in the halo mass function.
While these findings highlight the power for constraining inflationary physics at small scales, they also show the challenges posed by potential degeneracies with other physical processes relevant in the nonlinear regime of structure formation such as non-cold dark matter candidates.
Our results open new avenues for probing inflationary physics in large scale structures and galactic physics and emphasise the need for refined theoretical tools to robustly constrain primordial features.
\end{abstract}

\maketitle

\begin{spacing}{1} 
{
\setcounter{tocdepth}{2}
  \hypersetup{linkcolor=rosewood}
  \tableofcontents
}
\end{spacing}

\section{Introduction}
\label{sec: Introduction}

The inflationary paradigm, in which primordial quantum fluctuations are stretched to cosmological scales and subsequently seed the large-scale structure of the Universe, stands as the leading framework for describing the early Universe~\cite{Starobinsky:1980te, Guth:1980zm, Linde:1981mu, Albrecht:1982wi}. Despite its success, the fundamental nature of inflation and the exact mechanism driving it remain elusive. Remarkably, current observations align well with the simplest inflationary model: a single scalar field slowly rolling down a sufficiently flat potential, generating an almost scale-invariant primordial power spectrum. This power spectrum is well constrained by Cosmic Microwave Background (CMB) data around $k_\star \sim 0.05 \,\Mpc^{-1}$~\cite{Planck:2018jri}. However, embedding such a minimalistic realisation of inflation within a fundamental UV-complete theory proves challenging. High-energy theories typically predict the existence of multiple fields, often accompanied by non-canonical kinetic terms, and the design of a sufficiently flat potential faces significant theoretical hurdles, including its sensitivity to Planck-scale corrections, see e.g.~\cite{Baumann:2014nda} for a review. As a result, concrete UV completions of inflation naturally lead to more complex scenarios, including the presence of \textit{sharp primordial features}: localised deviations from slow-roll dynamics that break scale invariance, see Refs.~\cite{Chen:2010xka, Chluba:2015bqa, Slosar:2019gvt, Achucarro:2022qrl} for reviews. These features are expected to show up in the primordial power spectrum and higher-order statistics, such as the bispectrum~\cite{Chen:2006xjb, Chen:2008wn, Adshead:2011jq, Achucarro:2012fd, Achucarro:2013cva, Bartolo:2013exa, Achucarro:2014msa, Torrado:2016sls, Chen:2022vzh, Werth:2023pfl, Pinol:2023oux}.

\vskip 4pt
The detection of primordial features would offer unprecedented insights into the physics of the early Universe and provide a powerful means to distinguish between competing models of inflation. Extensive efforts have been dedicated to searching for these features, mainly at the level of the power spectrum, across a range of cosmological observables. To date, the tightest constraints have been derived from measurements of CMB temperature anisotropies and E-mode polarisation~\cite{Wang:2000js, Adams:2001vc, Peiris:2003ff, Mukherjee:2003ag, Covi:2006ci, Hamann:2007pa, Meerburg:2011gd, Meerburg:2013dla, Benetti:2013cja, Miranda:2013wxa, Easther:2013kla, Chen:2014joa, Achucarro:2014msa, Hazra:2014goa, Hazra:2014jwa, Hu:2014hra, Ade:2015lrj, Gruppuso:2015zia, Gruppuso:2015xqa, Hazra:2016fkm, Torrado:2016sls, Ballardini:2018noo,Planck:2018jri, Zeng:2018ufm, Canas-Herrera:2020mme, Braglia:2021ckn, Braglia:2021sun, Naik:2022mxn, Hamann:2021eyw}. While no statistically significant detection of primordial features has yet been made from CMB data alone, deviations from a pure power-law primordial power spectrum have been constrained to within a few percent, primarily in the linear regime ($k < 0.1 \,h\,\Mpc^{-1}$). Notably, certain inflationary models have intriguingly reproduced anomalous oscillatory patterns observed in the temperature angular power spectrum residuals from {\it Planck} data (often dubbed `$A_{L}$ anomaly')~\cite{GallegoCadavid:2016wcz,Planck:2018jri,Domenech:2019cyh, Domenech:2020qay,Braglia:2021rej,Ballardini:2022vzh, Hazra:2022rdl}. Future observational missions are expected to improve constraints on the amplitude of primordial features by several percent~\cite{ACT:2020gnv, SPT-3G:2022hvq, SimonsObservatory:2018koc, LiteBIRD:2020khw, Abazajian:2019eic, Kogut:2024vbi, PRISM:2013fvg}, and prospects for detection remain promising~\cite{Braglia:2022ftm, Petretti:2024mjy}.

\vskip 4pt
Moreover, primordial features have also been invoked in the context of the recent ``cosmic tensions'' that have appeared over the last decade. Most notably, there exists a $5\sigma$ `Hubble tension'  between direct measurement of the Hubble parameter in the late-universe using cepheid-calibrated supernovae \cite{Riess:2021jrx} and the $\Lambda$CDM prediction when calibrated on {\it Planck} data  \cite{Planck:2018vyg}. 
There is in addition, a milder (2-3$\sigma$) `$S_8$ tension'  between galaxy weak-lensing \cite{KiDS:2020suj,DES:2021wwk,HSC:2018mrq} and  CMB \cite{Planck:2018vyg} determinations of the amplitude of matter fluctuations, parametrised as $S_8 = \sigma_8 \sqrt{\Omega_m/0.3}$, where $\Omega_m$ is the relative matter density today, and $\sigma_8$ corresponds to the root mean square of matter fluctuations on a $8 \ h^{-1}$Mpc scale, with $h = H_0/(100 \ \textrm{km/s/Mpc})$.  There is also a less known (yet potentially related) tension between determinations of the tilt of the matter power spectrum at redshift $z=3$ and around $k\sim 1\,h\,\Mpc^{-1}$ using Lyman-$\alpha$ (Ly-$\alpha$) data from eBOSS \cite{eBOSS:2018qyj} and the CMB under $\Lambda$CDM, which has reached a statistical level of $5\sigma$ \cite{Palanque-Delabrouille:2019iyz,Goldstein:2023gnw,Rogers:2023upm}. While the possibility of unknown systematic effects explaining these discrepancies remains extensively studied (see e.g.~Refs.~\cite{Freedman:2021ahq,Riess:2021jrx,Amon:2022ycy,Arico:2023ocu, Fernandez:2023grg,Walther:2024tcj,Freedman:2024eph,Riess:2024vfa} for discussion), they may also indicate the existence of physics beyond $\Lambda$CDM (see e.g.~Ref.~\cite{Abdalla:2022yfr} for a review). In that context, primordial features have been shown to potentially alleviate those tensions \cite{Liu:2019dxr, Keeley:2020rmo, Antony:2022ert, Hazra:2022rdl}, although more data are necessary to firmly confirm (or exclude) those models as a solution. 

\vskip 4pt
In addition to CMB data, large-volume photometric and spectroscopic surveys provide precise measurements of cosmic structures across a wide range of scales both in the linear and nonlinear regime of structure formation~\cite{Wang:1998gb, Zhan:2005rz, Huang:2012mr, Chen:2016vvw, Chen:2016zuu, Ballardini:2016hpi, Xu:2016kwz, Fard:2017oex, Palma:2017wxu, Ballardini:2017qwq, Ballardini:2019tuc, Debono:2020emh, Li:2021jvz, Euclid:2023shr, Ballardini:2024dto}. Analyses from BOSS and eBOSS have shown that constraints on primordial features derived from mildly nonlinear scales $k \sim (0.1 - 0.2)\,h\,\Mpc^{-1}$ are competitive with those obtained from {\it Planck} data~\cite{Beutler:2019ojk, Ballardini:2022vzh,Ballardini:2022wzu, Mergulhao:2023ukp}. Efforts to accurately model small-scale nonlinearities, extending up to $k\sim 0.3 \,h\, \Mpc^{-1}$, have used perturbative techniques, incorporating effective field theory contributions and infrared resummation~\cite{Vlah:2015zda, Vasudevan:2019ewf, Chen:2020ckc}, as well as the use of reconstruction algorithms~\cite{Li:2021jvz}. Nevertheless, an important open question is whether these primordial features can persist at smaller  scales ($k>1\,h\,\Mpc^{-1}$) without being erased by nonlinear processes, and whether they can be meaningfully constrained in this regime. Early studies investigating the potential of Ly-$\alpha$ forest, weak-lensing observations~\cite{Abolfathi_2018, DES:2017qwj}, and 21cm data have yielded encouraging forecasts, suggesting that such constraints may indeed be achievable~\cite{Munoz:2019hjh, Naik:2025mba}.

\vskip 4pt
In this paper, we present a proof of principle that some remnant signature of features in the primordial power spectrum can survive the nonlinear regime of structure formation. Specifically, starting from motivated concrete UV-complete inflationary models, we study for the first time the evolution of primordial features down to scales of $k \sim  10 \, h \, \Mpc^{-1}$ using N-body simulations. Although oscillatory patterns are significantly damped due to mode coupling, we demonstrate that characteristic bumps associated with sharp features persist, and may play a role in explaining tensions between the CMB and probes of the matter power spectrum at late times. These remnants could serve as signatures of a sudden, localised in time, breaking of slow-roll evolution during inflation~\cite{Starobinsky:1992ts, Adams:2001vc, Hunt:2004vt, Gong:2005jr, Joy:2007na}. While these preliminary results highlight a promising path toward constraining primordial sharp features at galactic scales, we find that the observed effects are significantly degenerate with other extensions of the standard cosmological model. Additionally, our work suggests that further advancements in modelling nonlinear processes are essential, as existing emulators are primarily trained on featureless simulated datasets.

\vskip 4pt
This paper is organised as follows. In Sec.~\ref{sec: Primordial Features from Inflationary UV Physics}, we introduce primordial sharp features from concrete UV-complete inflationary models with two distinct mechanisms and develop an effective description of features to derive simple templates for the primordial power spectrum. In Sec.~\ref{sec: Cosmological Evolution with Features}, we first constrain features from CMB data before performing N-body simulations in the nonlinear regime of structure formation. We then present the results for the nonlinear matter power spectrum and the halo mass function, before concluding in Sec.~\ref{sec: Conclusions and Outlook} with some future directions left for forthcoming works.

\section{Primordial Features from Inflationary Physics}
\label{sec: Primordial Features from Inflationary UV Physics}

Primordial features modify the primordial power spectrum of curvature perturbations $P_\zeta(k)$ by inducing an additional scale dependence to the slight red tilt well-constrained around $k_\star\sim 0.05\,\Mpc^{-1}$. In particular, sharp features emerge from strongly time-dependent dynamics encoded in the time evolution of parameters that dictate the behaviour of primordial fluctuations during inflation. In this section, we introduce sharp primordial features through concrete inflationary background models, which we will refer to as concrete UV models. Then, motivated by a unifying picture in which fluctuations either present a slight breaking of slow roll or propagate with a time-dependent speed of sound, we design templates for the primordial power spectrum $P_\zeta(k)$.

\subsection{Inflationary UV-complete Theories}
\label{subsec: Inflationary Theories}

For concreteness, we consider multi-field background models of inflation with a potentially curved field space, i.e.~general multi-field nonlinear sigma models (see e.g.~Ref.~\cite{Baumann:2014nda} for a review), described by the action
\begin{equation}
    S = \int \d^4x \sqrt{-g} \left(\frac{\Mpl^2}{2}R - \tfrac{1}{2}G_{IJ}\nabla_\mu \phi^I\nabla^\mu\phi^J - V\right)\,,
\end{equation}
with $\phi^I$ ($I = 1, \ldots, N$) are $N$ scalar fields, $G_{IJ}\equiv G_{IJ}(\phi^K)$ is the internal field-space metric, and $V\equiv V(\phi^K)$ is the (multi-field) potential. Here, $R$ denotes the Ricci scalar constructed out of the spacetime metric with determinant $g$. The spacetime background dynamics is then dictated by the homogenous parts of the fields $\bar{\phi}^I(t)$ whose evolutions are determined by their equations of motion. By specifying a concrete form for the field-space metric $G_{IJ}$ and the potential $V$, and supplementing these with the Friedmann equations, the entire system can be consistently solved provided appropriate initial conditions are given. At the level of fluctuations, the field and metric perturbations can be combined into entropic fluctuations orthogonal to the field-space background trajectory, and the gauge independent adiabatic curvature perturbation $\zeta$. In vanilla slow-roll models of inflation, in which the background trajectory follows a geodesic of the field-space manifold along an almost flat direction, the primordial power spectrum is almost scale-invariant $P_{\zeta}(k) = \tfrac{2\pi^2}{k^3} \P_{\zeta, 0}(k)$ where the dimensionless power spectrum $\P$ is given by
\begin{equation}
\label{eq: slow-roll dimensionless power spectrum}
    \P_{\zeta, 0}(k) = A_s \left(\frac{k}{k_\star}\right)^{n_s-1}\,,
\end{equation}
where $A_s$ and $n_s$ are the amplitude and the spectral index evaluated at the pivot scale $k_\star = 0.05\,\Mpc^{-1}$. In what follows, we review two well-known concrete realisations of background trajectories that generate sharp features, which will add corrections to~(\ref{eq: slow-roll dimensionless power spectrum}).

\subsubsection{Single-Field Step Potential}
\label{subsubsec: Single-field step potential}

\begin{figure}[h!]
    \centering    \includegraphics[width=1.1\columnwidth,trim={0pt 6pt 0pt 0}]{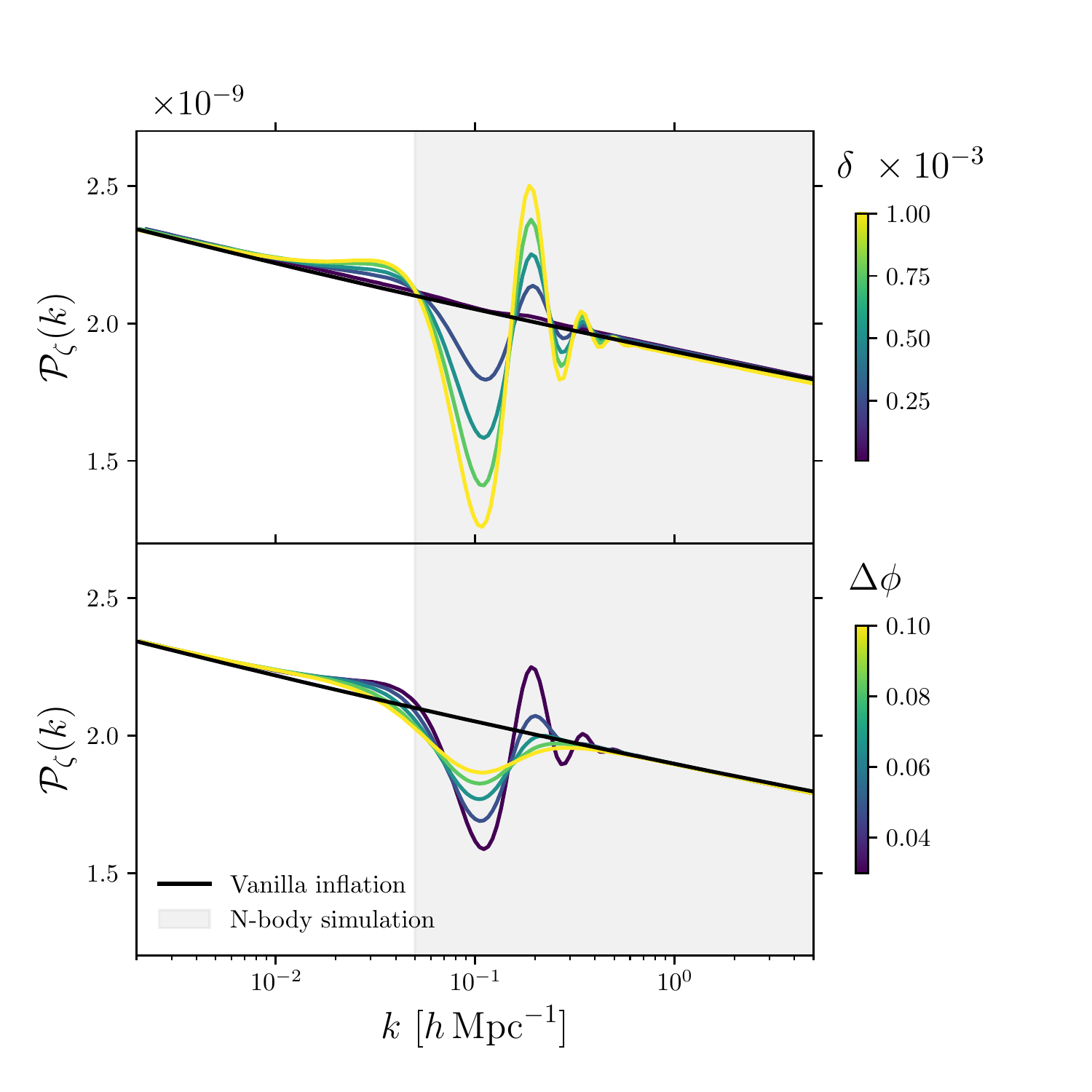}
    \caption{Primordial (dimensionless) power spectrum $\P_\zeta(k) \equiv k^3/2\pi^2 \, P_\zeta(k)$ for the single-field step potential model varying the amplitude of the feature $\delta$ with fixed $\phi_0=15.4$ and $\Delta\phi=0.03$ (\textit{top panel}), and varying the feature width $\Delta\phi$ with fixed $\phi_0=15.4$ and $\delta = 5\times10^{-4}$ (\textit{bottom panel}). The solid black line corresponds to the featureless vanilla inflation power spectrum $\P_{\zeta, 0}(k)$ in Eq.~(\ref{eq: slow-roll dimensionless power spectrum}). Note that $\delta \sim 10^{-3}$ produces a feature with an amplitude of approximately $20\%$, placing it at the threshold of compatibility with {\it Planck} data~\cite{Planck:2018jri}. The shaded grey region corresponds to the scales simulated in Sec.~\ref{subsec: N-body Simulations}.}
    \label{fig: Single-field step potential} 
 \end{figure}

The simplest model of inflation that exhibits a sharp feature is a single field rolling on a potential with a sudden step~\cite{Starobinsky:1992ts, Adams:2001vc}, see also~Ref.~\cite{Chen:2006xjb, Chen:2008wn} for the study of the corresponding non-Gaussianities. We consider the potential
\begin{equation}
\label{eq: single-field potential}
    V(\phi) = \frac{1}{2}m^2 \phi^2 \, \left[1 + \delta \tanh\left(\frac{\phi - \phi_0}{\Delta\phi}\right)\right]\,.
\end{equation}
In the absence of features, i.e.~setting $\delta = 0$, the amplitude and spectral index of the primordial power spectrum is related to the potential by the slow-roll condition
\begin{equation}
    \left(\frac{m}{\Mpl}\right)^2 = A_s \, \frac{(1-n_s)^2}{2}\,.
\end{equation}
As such, the mass parameter $m$ uniquely fixes the slow-roll background. The sharp feature we consider is small compared to the leading-order potential, and is described by its amplitude $\delta$, the location (in the field trajectory) $\phi_0$, and duration $\Delta\phi$. We have solved numerically the background and the dynamics of the fluctuations using \texttt{PyTransport} \href{https://github.com/jronayne/PyTransport}{\faGithub} \cite{Dias:2016rjq, Mulryne:2016mzv, Ronayne:2017qzn}. Fig.~\ref{fig: Single-field step potential} shows the primordial power spectrum $\P_\zeta(k)$ varying the parameters defining the feature. We have fixed $\phi_0$ that only dictates the characteristic scale $k_f$ of the feature without altering the overall shape.  The induced modification of the primordial power spectrum is characteristic of a sharp feature: a localised bump with oscillations linear in $k$. Notably, the profile of the feature is very sensitive to the potential parameters, and increasing the feature width washes out the oscillatory patterns.

\subsubsection{Turn in Multi-Field Space}
\label{subsubsec: Turn in multi-field space}

\begin{figure}[h!]
    \centering    \includegraphics[width=1.1\columnwidth,trim={0pt 6pt 0pt 0}]{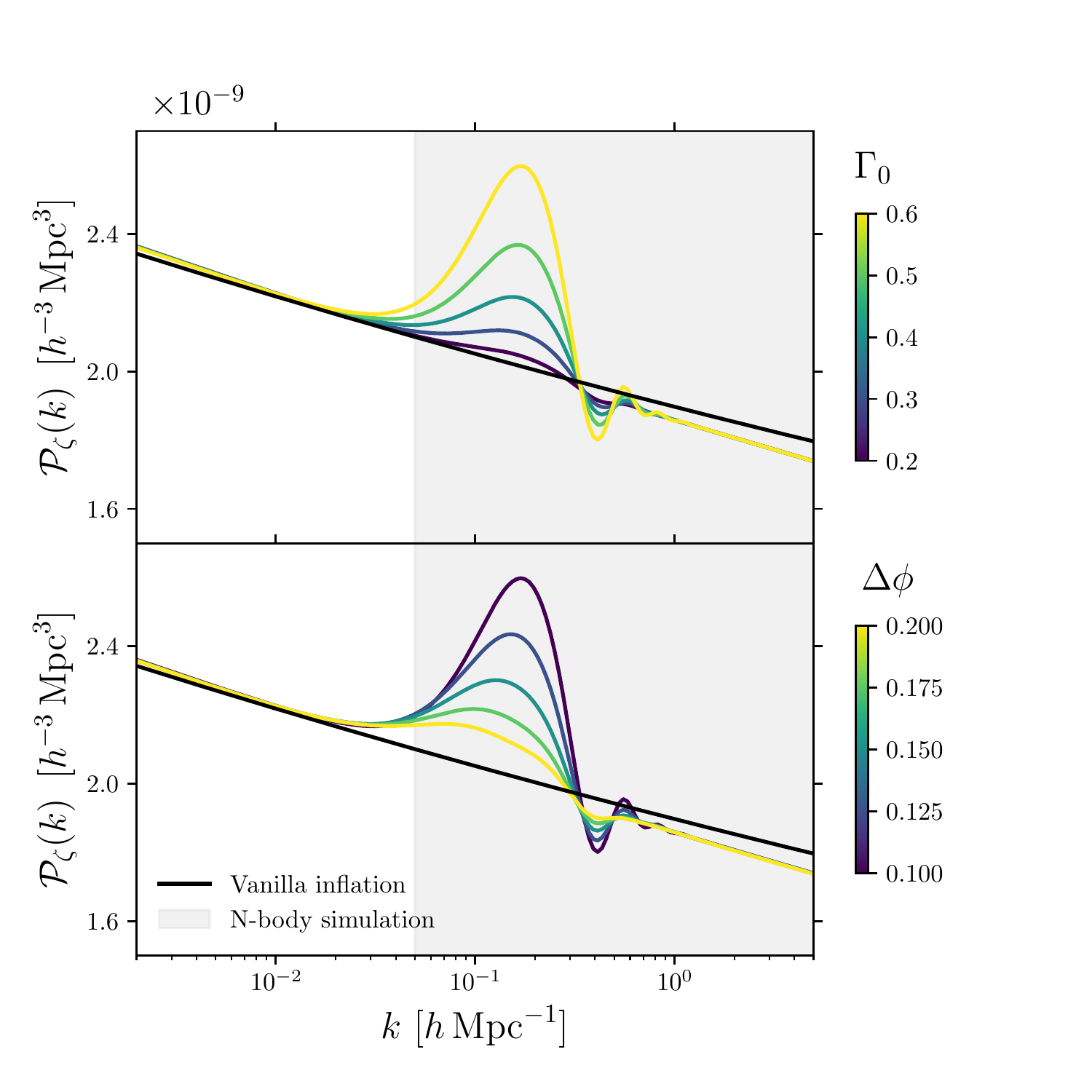}
    \caption{Primordial (dimensionless) power spectrum $\P_\zeta(k) \equiv k^3/2\pi^2 \, P_\zeta(k)$ for the multi-field model exhibiting a turn in field space varying the amplitude of the feature $\Gamma_0$ with fixed $\phi_0=8.8$ and $\Delta\phi=0.1$ (\textit{top panel}), and varying the feature width $\Delta\phi$ with fixed $\phi_0=8.8$ and $\Gamma_0 = 0.6$ (\textit{bottom panel}). The solid black line corresponds to the featureless vanilla inflation power spectrum $\P_{\zeta, 0}(k)$ in Eq.~(\ref{eq: slow-roll dimensionless power spectrum}).}
    \label{fig: Multi-field turn feature} 
 \end{figure}

Features notably also arise from non-trivial trajectories in a multi-dimensional field space, such as turns, that can lead to striking observational signatures. As models with two fields have been widely explored in the literature, see e.g.~Refs.~\cite{Langlois:1999dw, Wands:2002bn, Lesgourgues:1999uc, Peterson:2010np, Achucarro:2010da, Konieczka:2014zja, Garcia-Saenz:2019njm}, we present here a three-field model, which is an extension of the quasi-single field model of inflation~\cite{Chen:2009zp} with a non-unity field-space metric. Concretely, we consider the following potential 
\begin{equation}
    V = \frac{1}{2}\sum_{I=1}^3 m_I^2 \phi_I^2\,,
\end{equation}
and the following field-space metric
\begin{equation}
    G_{IJ} = 
    \begin{pmatrix}
        1      & \Gamma & 0 \\
        \Gamma & 1      & 0 \\
        0      & 0      & 1
    \end{pmatrix}\,,
\end{equation}
with
\begin{equation}
\label{eq: field-space component Gamma}
    \Gamma(\phi_1) = \frac{\Gamma_0}{\cosh^2\left[2 \left(\frac{\phi_1 - \phi_0}{\Delta \phi}\right)\right]}\,.
\end{equation}
This model, after fixing the mass ratios to be $m_2^2/m_1^2 = 300$ and $m_3^2/m_1^2 = 1/81$, was first explored in Ref.~\cite{Dias:2015rca} as a straightforward extension of the model proposed in Ref.~\cite{Achucarro:2010da} and was subsequently revisited in Ref.~\cite{Ronayne:2017qzn}. Physically, the two light fields $\phi_1$ and $\phi_3$ drive the inflationary background, and the heavy field $\phi_2$ interacts with the light ones through kinetic interactions. The characteristic trajectory then undertakes a fast turn in the plane of the two light fields resulting in a burst of particle production. This leads to a well-known feature boosting the power spectrum, see e.g.~Refs.~\cite{Achucarro:2010da, Gao:2012uq, Gao:2013ota, Achucarro:2013cva}. Similarly to the single-field model with a step in the potential, the multi-field feature is characterised by the three same parameters: the amplitude $\Gamma_0$, the location $\phi_0$, and the width $\Delta\phi$. Therefore, we expect to have qualitatively a similar feature profile. We illustrate the primordial power spectrum in Fig.~\ref{fig: Multi-field turn feature}. It is interesting to observe that both parameters $\Gamma_0$ and $\Delta\phi$ seem degenerate, as an increase of the feature amplitude can also be (almost perfectly) mimicked by decreasing the feature width. A key difference from the single-field model with a step in the potential is that, for the chosen range of parameters, the feature does not generate the characteristic and pronounced damped oscillations. This is due to the insufficient sharpness of the feature. As we will show in Sec.~\ref{subsec: N-body Simulations}, oscillatory features are naturally suppressed at the level of the matter power spectrum when entering the late-time nonlinear regime and nearly vanish entirely. However, a localised bump persists, providing a robust indicator of the primordial feature. One can also notice the slight decrease in power for scales smaller that the feature $k>k_f$ compared to the almost scale-invariant power spectrum before the feature. This behaviour arises from the double-stage inflationary dynamics, initially driven by the field $\phi_1$ and subsequently by $\phi_3$. This $\sim 1\%$-level decrease in power is irrelevant for all observational applications.
 
\subsection{Unified Effective Description of Features}
\label{subsec: Effective Description of Primordial Features}

The two concrete UV models we have presented, despite relying on different physical mechanisms, produce qualitatively similar small features in the power spectrum. This highlights the advantage of adopting a general, largely UV-independent framework for describing primordial features. In this section, we focus exclusively on the dynamics of fluctuations to derive general templates for the power spectrum with sharp features.

\subsubsection{Time-dependent Slow-roll Parameter or Sound Speed}
\label{subsubsec: Time-dependent Slow-roll Parameter or Sound Speed}

The background dynamics dictates the time evolution of the parameters in the action governing the fluctuations. A feature in the inflationary potential induces a time dependence in the first slow-roll parameter $\textcolor{pyblue}{\epsilon(t)} \equiv -\dot{H}/H^2$, which then translates to features in the primordial power spectrum, as in the model presented in Sec.~\ref{subsubsec: Single-field step potential}. Furthermore, when additional entropic fluctuations---orthogonal to the background trajectory---are sufficiently heavy, they can be integrated out under generalised adiabaticity conditions~\cite{Cespedes:2012hu, Achucarro:2012sm, Achucarro:2012yr, Achucarro:2014msa, Pinol:2023oux}, as in the UV-complete model of Sec.~\ref{subsubsec: Turn in multi-field space}. This results in a simplified action for the curvature perturbation $\zeta$, where the influence of UV physics and the effects of additional fluctuations are entirely hidden in the time dependence of the parameters entering the action, that reads
\begin{equation}
\label{eq: quadratic action}
    S = \Mpl^2 \int \d t\d^3x \, a^3\frac{\textcolor{pyblue}{\epsilon(t)}}{\textcolor{pyred}{c_s^2(t)}} \left[\dot{\zeta}^2 - \textcolor{pyred}{c_s^2(t)}\frac{(\partial_i \zeta)^2}{a^2}\right]\,,
\end{equation}
where $\textcolor{pyred}{c_s(t)}$ is the speed of sound of adiabatic fluctuations. Depending on the specific mechanism that generates features, the departure from vanilla near-scale invariant power spectrum is either hidden in the time dependence of the slow-roll parameter $\textcolor{pyblue}{\epsilon(t)}$, i.e.~in the Hubble parameter $H(t)$, or in the speed of sound $\textcolor{pyred}{c_s(t)}$. In fact, the action~(\ref{eq: quadratic action}) suggests a \textit{unified} effective description of features: every feature can be effectively described by a time-dependent function entering either the overall normalisation of the curvature perturbation $\zeta$ or the breaking of its light cone. 

\vskip 4pt
In what follows, in the case of features in $\epsilon(t)$, we will set $c_s = 1$, and for a time-dependent sound speed $c_s(t)$, we will assume an approximately constant slow-roll parameter, ensuring a stable slow-roll background.\footnote{An overall constant normalisation can be absorbed in $\epsilon$.} A localised feature is therefore entirely described by a transient enhancement of $\epsilon(t)$ or reduction in $c_s(t)$. Assuming this localised breaking of the slow-roll evolution is small enough, the power spectrum deviates from the vanilla one by 
\begin{equation}
    \P_\zeta(k) = \P_{\zeta, 0}(k) \left[1 + \delta\P_\zeta(k)\right]\,,
\end{equation}
with~\cite{Achucarro:2010jv, Achucarro:2010da, Bartolo:2013exa, Achucarro:2014msa, Palma:2014hra}
\begin{equation}
\label{eq: PPS feature}
    \delta\P_\zeta(k) = k \int_{-\infty}^0 \d \tau \, u(\tau) \times
        \left\{
            \begin{array}{ll}
                \textcolor{pyblue}{\frac{\sin(2k\tau)}{(k\tau)^2} - \frac{2\cos(2k\tau)}{k\tau}}\\
                \textcolor{pyred}{\sin(2k\tau)}
            \end{array}
        \right. \,,
\end{equation}
  where $u$ is either defined as the small (relative) deviation from a constant featureless slow-roll parameter $\epsilon \equiv \epsilon_0(1-u)$ or $u \equiv 1 - c_s^{-2}$, depending on the mechanism generating the feature, and $\tau$ is the conformal time defined by $\d\tau \equiv \d t/a$. We leave details of the derivation in App.~\ref{app: Details on PPS with Features}. Note that an enhanced/reduced but constant function $u(\tau)$ does not bring any new scale dependence to the power spectrum as oscillations average the power spectrum correction to zero after integration.

\begin{figure}[h!]
    \centering    \includegraphics[width=0.9\columnwidth,trim={0pt 6pt 0pt 0}]{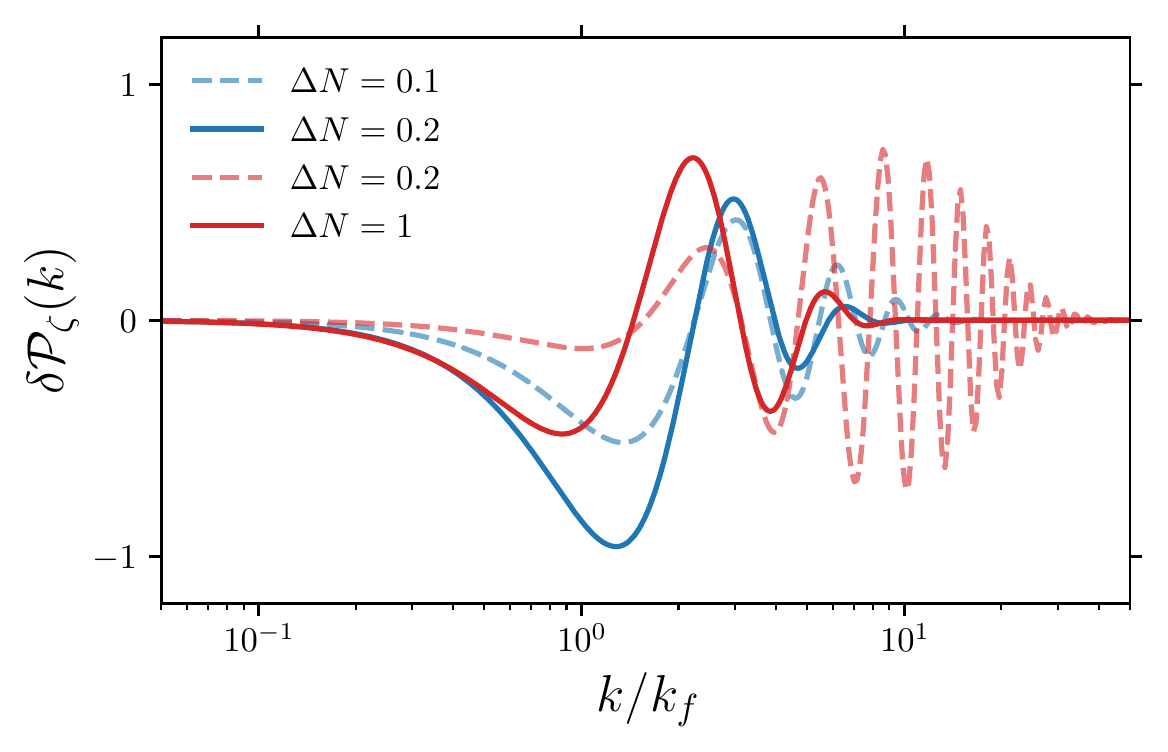}
    \caption{Relative correction to the dimensionless primordial power spectrum $\delta\P_\zeta(k)$ in Eq.~(\ref{eq: PPS feature}) for a feature in the slow-roll parameter with the Gaussian envelope~(\ref{eq: u Gaussian envelope}) (\textcolor{pyblue}{in blue}), and for a feature in the speed of sound with the time-dependence given in Eq.~(\ref{eq: u cosh envelope}) (\textcolor{pyred}{in red}). We have set $\delta\A=1$ and $N_0=0$.}
    \label{fig: Effective Feature PPS} 
 \end{figure}

\vskip 4pt
\textbf{Single-field step potential.} In the single-field model considered in Sec.~\ref{subsubsec: Single-field step potential}, we effectively consider the following Gaussian feature in the slow-roll parameter 
\begin{equation}
\label{eq: u Gaussian envelope}
    u(N) = \delta\A \, \exp\left[-\frac{(N-N_0)^2}{2\Delta N^2}\right]\,,
\end{equation}
where $N\equiv \log(a)$ is the number of $e$-folds. Fig~\ref{fig: Effective Feature PPS} shows the correction to the dimensionless power spectrum after numerically integrating~(\ref{eq: PPS feature}). The profile qualitatively reproduces the one found from the concrete UV-complete model~(\ref{eq: single-field potential}), and a sharper feature (i.e.~decreasing the width $\Delta N$) naturally makes the oscillatory pattern more pronounced.

\vskip 4pt
\textbf{Turn in multi-field space.} For the multi-field model considered in Sec.~\ref{subsubsec: Turn in multi-field space}, we consider the following effective time-dependent correction to the speed of sound 
\begin{equation}
\label{eq: u cosh envelope}
    u(N) = \frac{\delta\A}{\cosh^4\left[2 \left(\frac{N - N_0}{\Delta N}\right)\right]}\,,
\end{equation}
which is motivated by the form of field-space metric component~(\ref{eq: field-space component Gamma}).\footnote{At the level of the fluctuations, provided the entropic fields are heavy enough, the correction to the sound speed is proportional to the bending of the background trajectory squared $u \propto \eta_\perp^2$, which is itself in one-to-one correspondence with the field-space metric component $\Gamma$, hence explaining the exponent in~(\ref{eq: u cosh envelope}).} Fig.~\ref{fig: Effective Feature PPS} shows the corresponding power spectrum correction. Interestingly, even though the feature time dependence encoded in the $u$ function is qualitatively similar to that in~(\ref{eq: u Gaussian envelope}), the kernel modifies the residual oscillatory pattern at $k>k_f$. Nevertheless, the characteristic bump of the feature---which survives the late-time nonlinear regime of structure formation---still remains similar. A relatively wide feature in the speed of sound, with $\Delta N\approx 2$, reproduces well the power spectrum profile found in Fig.~\ref{fig: Multi-field turn feature}.

\subsubsection{Primordial Feature Templates} 
\label{subsubsec: Primordial Feature Template}

\begin{figure*}[t]
    \centering
    \includegraphics[width=1\textwidth]{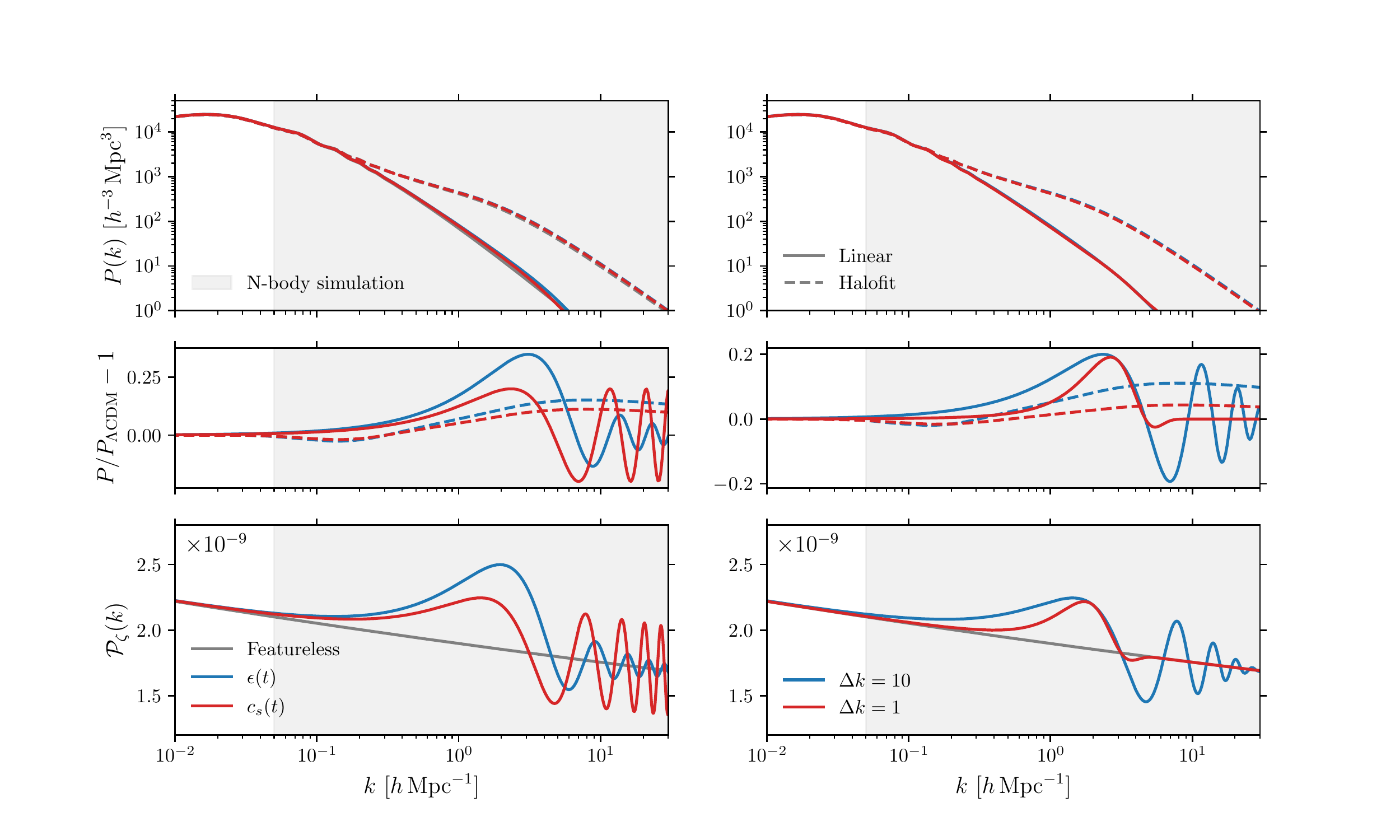}
    \caption{
    Dimensionless primordial power spectrum $\P_\zeta(k)$ (\textit{bottom panel}), matter power spectrum $P(k)$ (\textit{top panel}) and relative difference between $P(k)$ and the featureless matter power spectrum, for features in the \textcolor{pyblue}{slow-roll parameter} and in the \textcolor{pyred}{speed of sound} using the templates~(\ref{eq: PPS template epsilon cs}) (\textit{left panel}), 
    and using the template~(\ref{eq: final template}) with $\Delta k = 10\,\Mpc^{-1}$ and $\Delta k = 1\,\Mpc^{-1}$ (\textit{right panel}). 
    We have set $\delta \A = 0.2$ and $k_f = 2\,\Mpc^{-1}$. The solid lines corresponds to a linear evolution of primordial fluctuations whereas the dashed line corresponds to a nonlinear evolution using \texttt{Halofit}. The shaded grey region corresponds to the the scales simulated in Sec.~\ref{subsec: N-body Simulations}.
    }
    \label{fig: PPS template} 
 \end{figure*}
 
As seen previously, the feature precise characteristic significantly depends on the details of the inflationary model. However, its overall shape is only determined by the characteristic time of the feature. In order to derive an analytical template for the primordial power spectrum, we first consider a pulse sharp feature of the form $u(\tau) = \delta\A \, k^{-1}\delta_D(\tau - \tau_f)$, with $\delta_D$ the Dirac-delta distribution. The power spectrum correction reads
\begin{equation}
\label{eq: PPS template epsilon cs}
    \delta \P_\zeta^{\textcolor{pyblue}{\epsilon}, \textcolor{pyred}{c_s}}(k) = \delta\A \times
        \left\{
            \begin{array}{ll}
                \textcolor{pyblue}{\frac{\sin(2k/k_f)}{(k/k_f)^2} - \frac{2\cos(2k/k_f)}{k/k_f}}\\
                \textcolor{pyred}{\sin(2k/k_f)}
            \end{array}
        \right. \,,
\end{equation}
where $k_f$ is the scale of the sharp feature corresponding to the conformal time $\tau_f$. The bottom left panel of Fig.~\ref{fig: PPS template} shows both templates. It is interesting to notice that while a pulse induces oscillations of constant amplitude across all scales when originating from a feature in the sound speed, the oscillatory kernel associated with a feature in the slow-roll parameter naturally attenuates these oscillations for $k>k_f$. 
 
\vskip 4pt
These two behaviours can be described by a unified template by adding instead of a Dirac-delta a Gaussian envelope (wave packet) to the simple oscillatory pattern
\begin{equation}
\label{eq: final template}
    \delta\P_\zeta(k) = \delta\A \, \exp\left[-\frac{(k - k_f)^2}{2\Delta k^2}\right]\, \sin(2k/k_f)\,,
\end{equation}
which effectively adds the third parameter $\Delta k$ describing the width of the feature in momentum space. We note that this shape was used in Refs.~\cite{Planck:2018jri, Domenech:2019cyh, Ballardini:2022vzh, Ballardini:2024dto} in an ad hoc manner in the context of the $A_L$ anomaly and time-sliced perturbation theory. However, we independently rederived it on a more robust theoretical basis. 
To model localised feature, it is also necessary to enforce $\Delta k \sim {\cal O}(k_f)$, and we will comment on this further when performing dedicated data analysis. We illustrate this template in the bottom right panel of Fig.~\ref{fig: PPS template}, for $\delta A = 0.2$, $k_f = 1\,\Mpc^{-1}$ and two different values of $\Delta k = 1\,\Mpc^{-1}$ and $\Delta k = 10\,\Mpc^{-1}$. This universal template can be easily understood physically. The localised feature provides enough energy to classically excite a range of modes inside the horizon, which then develop a negative frequency component. The mixing between these positive and negative modes creates the oscillations linear in the wavenumber $k$ and the generic factor $2$ entering the oscillation frequency. Notably, as the feature excites modes inside the horizon, the scale of the feature can in principle be decorrelated from the oscillatory frequency such that $k\ge k_f$. However, a too large hierarchy would violate perturbativity,\footnote{If the feature is excessively sharp, the fluctuations would become strongly coupled, leading to a breakdown of perturbative unitarity.} and inspecting concrete UV models in Sec.~\ref{subsec: Inflationary Theories} shows that $k\sim k_f$ with only a few visible oscillations. We thus fix the scale of the feature to the oscillation frequency in Eq.~\eqref{eq: final template}.
The free parameter $\Delta k$ entering the template~(\ref{eq: final template}) governs the number of residual oscillations in the primordial power spectrum. Varying $\Delta k$ enables us to design bump-like primordial templates with and without oscillations.\footnote{Obtaining an enhancement or a decrease in power at small scales is related to the sign of the amplitude of the feature $\delta\A$.} This template thus recovers the behaviour illustrated on the left for realistic models. In the following, Eq.~(\ref{eq: final template}) will be considered as the template to propagate fluctuations through the late-time Universe and initialise N-body simulations.

\section{Cosmological Evolution with Features}
\label{sec: Cosmological Evolution with Features}

We are now ready to evolve Gaussian primordial fluctuations whose statistics is governed by the primordial power spectrum with sharp features. In this section, we first constrain the template~(\ref{eq: final template}) considering linear physics. Then, we focus on the nonlinear regime of structure formation and perform N-body simulations using initial conditions that incorporate these sharp features.

\vskip 4pt
In Fig.~\ref{fig: PPS template}, we show the linear matter power spectrum and the correction for nonlinearities using the \texttt{Halofit} emulator \cite{Takahashi:2012em} in $\Lambda$CDM and for our primordial feature templates with parameters specified above (top panel shows the matter power spectrum in both models, while the middle panel show the relative difference between our model and $\Lambda$CDM). The \texttt{Halofit} emulator reveals the suppression of oscillations by nonlinear effects, while the characteristic enhancement of the nonlinear matter power spectrum remains largely unchanged. However, \texttt{Halofit} was not specifically design to handle oscillatory features in the primordial power spectrum. Using dedicated simulations, we will be able to check this prediction from \texttt{Halofit}.

\subsection{Constraints from Linear CMB Physics}
\label{subsec:Linear_Evolution}
We now turn to a detailed analysis of constraints from the CMB data on large scales. 
It is indeed important to make sure that primordial sharp features at scales $k \sim 1 \,\Mpc^{-1}$, which can extend down to $k\sim 0.1 \,\Mpc^{-1}$, do not spoil the well-constrained CMB scales.

\vskip 4pt
To that purpose, we modify the Boltzmann code \texttt{CLASS} \href{https://lesgourg.github.io/class_public/class.html}{\faGithub} \cite{Blas:2011rf} to read in our template given by Eq.~\eqref{eq: final template} using the `\texttt{External Pk}' module. We perform Monte Carlo Markov Chains (MCMC) analysis using the code \texttt{MontePython-v3} \href{https://github.com/brinckmann/montepython_public}{\faGithub} \cite{Audren:2012wb,Brinckmann:2018cvx} interfaced with our modified \texttt{CLASS}. We analyse the $\Lambda$CDM model with a modified primordial power spectrum in light of {\it Planck} 2018 TTTEEE and lensing data with the \texttt{Plik} likelihood. As is conventional, we make use of large flat prior on the baryon density contrast $\Omega_b$, dark matter density contrast $\Omega_m$, angular size of the sound horizon $\theta_s$, optical depth to reionization $\tau$, amplitude $A_s$ and tilt $n_s$ of the primordial power spectrum with pivot scale $k_* = 0.05\,\Mpc^{-1}$ (describing the canonical power law part). We also include parameters describing the modified power spectrum $k_f$, $\Delta k$ and $\delta A$. For the position of the feature $k_f$, we make use of the uniform logarithmic prior $\log_{10}(k_f/\Mpc^{-1})\in [-2,1]$ to ensure a correct exploration of the parameter space, as it spans several orders of magnitudes.  For the Gaussian width, we run with a flat prior on the parameter combination $\Delta k/ k_f\in[0.5,5]$ to ensure that $\Delta k$ always remains of the same order as $k_f$ while $k_f$ varies across several orders of magnitudes. Finally, we decide to also use a logarithmic prior on the amplitude $\delta A$ due to difficulties in exploring the highly non-Gaussian posteriors with a linear prior. To that end, we perform two separate runs for positive and negative values of $\delta A$, varied according to the same flat prior on the absolute value $\log_{10}{|\delta A|}\in [-3,\log_{10}(0.4)]$. More details about this choice are given in App.~\ref{app:deltaA}.
We also include the numerous nuisance parameters necessary for {\it Planck} (see Ref.~\cite{Planck:2018vyg} for details) and make use of a Choleski `fast-slow' decomposition to better handle those \cite{Lewis:2013hha}. We consider our chains to be converged when the Gelman-Rubin criterion \cite{Gelman:1992zz} reaches $R-1 < 0.03$ for all parameters.

\begin{figure}[h!]
     \centering
     \includegraphics[width=\columnwidth]{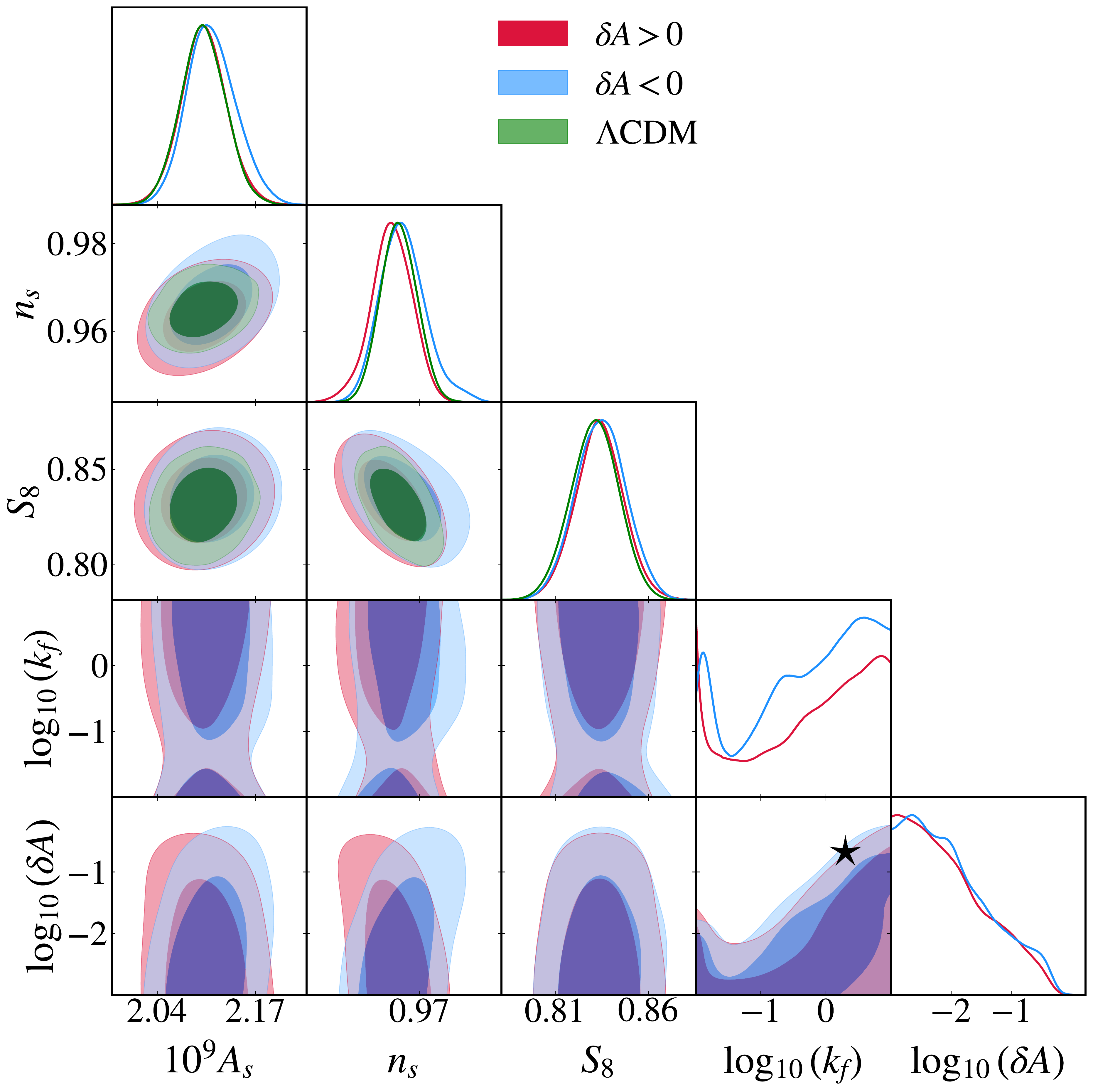}
     \caption{\justifying 2D posterior distributions of the most relevant parameters reconstructed from analysis of {\it Planck} 2018 TTTEEE+lensing data in the $\Lambda$CDM model without (\textcolor{pygreen}{green}) or with features in the primordial power spectra, with either a positive (\textcolor{pyred}{red}) or negative (\textcolor{pyblue}{blue}) amplitude. The black five-branched star corresponds to the parameters' values used for the N-body simulations described in section \ref{subsec: N-body Simulations}.}
     \label{fig: CMB constraints}
 \end{figure}
 
\vskip 4pt
Our results are shown in Fig.~\ref{fig: CMB constraints}. We do not show the parameter $\Delta k/ k_f$ that is fully unconstrained to avoid cluttering the plot.  One can see that, as expected, constraints from {\it Planck} are significant below $k_f \lesssim 0.1\,\Mpc^{-1}$, where they enforce $\delta A\lesssim {\cal O}(1\%)$ (as expected from the error on $A_s$, $\sigma(A_s)=0.03$) and then rapidly degrades. This suggests that a large feature $\delta\A \sim {\cal O}(1)$ at $k_f \gtrsim 1\,\Mpc^{-1}$ is acceptable and can be explored with N-body simulations. Note that we anticipate better constraints could be obtained by including further small scales data, in particular from CMB lensing measurements. However, we do not expect significantly different constraints at $k_f \gtrsim 1\,\Mpc^{-1}$ given current measurements (by ACT DR6 \cite{ACT:2023dou} and SPT-3G \cite{SPT-3G:2024atg}) and including those data at this stage would require improving the nonlinear prediction. We thus defer this to future work. 
We also note some mild degeneracies between $\delta A$ and $n_s$ (and even milder with $A_s$) for the larger values of $\delta A$. This in turn leads to slightly weaker bound on $n_s$, suggesting some interplay between constraints on the inflationary potential's slope coming from $n_s$ and the presence of a feature in the spectrum. Indeed, one might intuitively expect that a decrease (increase) in the power spectrum at small scales can be mimicked by a slight running red (blue) tilt on large scales.
However, we find no impact on the parameter $S_8$, suggesting that such features cannot resolve the $S_8$ tension purely from the linear behaviour. Following Refs.~\cite{Amon:2022azi,Preston:2023uup}, we now explore whether deviations in the nonlinear regime can help.

\subsection{N-body Simulations of the Inflationary Template}
\label{subsec: N-body Simulations}

While CMB physics is linear, the gravitational clustering of large scale structure and galaxies is a nonlinear process and its modelling for $k > 0.3\,h\,\Mpc^{-1}$ requires the use of cosmological simulations \cite{Angulo:2021kes}. Based on CMB constraints discussed in Sec.~\ref{subsec:Linear_Evolution}, we will now run four simulations: a benchmark featureless simulation compatible with {\it Planck} 2018 (cf.~Tab.~\ref{tab: cosmo parameters}) referred to as $\Lambda{\rm CDM}$ and three simulations including the sharp features, denoted sim-1, sim-2 and sim-3 respectively. They will be representative of the phenomenology of the features discussed in Sec.~\ref{sec: Primordial Features from Inflationary UV Physics}. We gather parameters entering their primordial power spectrum in Tab.~\ref{tab: simu init parameters}. Using a basic reconstruction method, we also relate \{$\delta\A$,$\Delta k$,$k_f$\} to their corresponding parameters for UV-complete theories discussed in Sec.~\ref{sec: Primordial Features from Inflationary UV Physics}. 

\begin{table}[h!]
\begin{center}
\begin{tabular}{c|c|c|c|c|c}
\hline
    $\Omega_{\text{m}}$ & $\Omega_{\rm b}$ & $\Omega_{\Lambda}$ & $H_0$  & $A_s$ & $n_s$ \\
\hline 
    0.31 & 0.049 & 0.69 & 67.32 & $2.10 \times 10^{-9}$ & 0.966 \\
\hline
\end{tabular}
\caption{Cosmological parameters adopted for the N-body simulations, compatible with the results from our analyses of \textit{Planck} 2018 data of section \ref{subsec:Linear_Evolution}.}
\label{tab: cosmo parameters}
\vspace*{-0.5cm}
\end{center}
\end{table}

\vskip 4pt
For all simulations, we take $|\delta \mathcal{A}|=0.2$ and $k_f=2\,\Mpc^{-1}$ to have a $\sim 20\%$ correction to $\Lambda$CDM at scales free from CMB constraints, and we compare two different values of $\Delta k = 1\,\Mpc^{-1}$ and $\Delta k = 10\,\Mpc^{-1}$. With our choice of parameters, sim-1 will have one main oscillation boosting the linear power spectrum ($\delta \mathcal{A}>0$) compared to $\Lambda$CDM. Conversely sim-2 will have the same setup but with a negative $\delta \mathcal{A}$. Finally, sim-3 that has a larger $\Delta k$, will have oscillations  over a wider range of $k$ (specifically it will have $\sim \Delta k / k_f = 5$ oscillations).

\vskip 4pt
Our choice of parameters is motivated by both theoretical and phenomenological grounds. On the one hand, a choice of $\delta \mathcal{A}>1$ would break the perturbative scheme used in Sec.~\ref{sec: Primordial Features from Inflationary UV Physics} while a too small value of $\delta \mathcal{A}$ would not lead to any sizeable effect due to primordial features. On the other hand, the choice of $k_f$ is made to introduce a feature at a range of modes relevant for the $S_8$ and Ly-$\alpha$ tensions discussed previously. With these simulations, we will be able to check whether such features can survive nonlinear evolution and play a role in those tensions.

\begin{table}
\begin{center}
\begin{tabular}{c|c|c|c|c}
\hline
    $L \, [\Mpc/h]$ & $N_{\text{part}}$ & $z_{\text{start}}$ & $m_{\text{part}}\,[M_\odot/h]$  & $\epsilon\, [\kpc/h]$\\
\hline 
    100 & $1024^3$ & 32 & $8.1\times 10^7$ & 5 \\
\hline
\end{tabular}
\caption{Numerical parameters for the N-body simulations.}
\label{tab: simu parameters}
\vspace*{-0.5cm}
\end{center}
\end{table}

\vskip 4pt
We initialise $1024^3$ particles' positions of the simulations using \texttt{S-GenIC} \href{https://github.com/sbird/S-GenIC/tree/master}{\faGithub} in order to compute the displacement field calculated with Lagrangian perturbation theory at second order (2LPT) at redshift $z=32$ on a $1024^3$ grid. We use the same random seed for each simulation to ease the direct comparisons between simulations. To track the nonlinear evolution of the cosmological perturbation down to $z=0$, we use the Tree-PM code \texttt{Gadget-4} \href{https://wwwmpa.mpa-garching.mpg.de/gadget4/}{\faGithub} \cite{Springel:2020plp}. It numerically follows the gravitational evolution of the particules' positions with a gravitational softening length of $\epsilon=0.05\,\ell = 0.05\,(L/N_{\rm part})$, where $\ell$ is the mean inter-particle separation, $L$ the length of the box simulated and $N_{\text{part}}$ the number of particles per linear dimension. We have stored snapshots at redshifts $z=32, 3, 1$ and $0$.

\begin{table*}
\begin{center}
\begin{tabular}{c|>{\centering\arraybackslash}p{1.5cm}|c|c||c|c} 
\hline
 & $\delta\A$ & $\Delta k\,[\,\Mpc^{-1}]$ & $k_f\,[\,\Mpc^{-1}]$ & $(\delta, \phi_0, \Delta\phi)$  & $(\Gamma_0, \phi_0, \Delta\phi)$\\
\hline
\textcolor{pyblue}{sim-1} & 0.2 & 1 & 2 & $(-3.7\times 10^{-4}, 15.04, 0.043)$ & $ (0.57, 8.47, 0.12)$\\ 
\hline
\textcolor{pygreen}{sim-2} & $-0.2$ & 1 & 2 & $(6.2\times 10^{-4}, 15.02, 0.054)$ & \xmark \\ 
\hline
\textcolor{pyred}{sim-3} & $-0.2$ & 10 & 2 & $(5.5\times 10^{-4}, 15.04, 0.042)$ & \xmark \\ 
\hline
\end{tabular}
\caption{Numerical parameters for the primordial power spectrum template with feature~(\ref{eq: final template}) used for the N-body simulations, along with the corresponding values of the parameters of the single-field potential (see Sec.~\ref{subsubsec: Single-field step potential}) and of the model with a turn in multi-field space (see Sec.~\ref{subsubsec: Turn in multi-field space}). Note that this model cannot predict a dip in power (i.e.~negative $\delta\A$) as a turn in field space is accompanied by particle production that always boosts the primordial power spectrum.}
\label{tab: simu init parameters}
\end{center}
\end{table*}

\subsubsection{Nonlinear Matter Power Spectrum}

\begin{figure*}[t]
    \centering
    \includegraphics[width=1\textwidth]{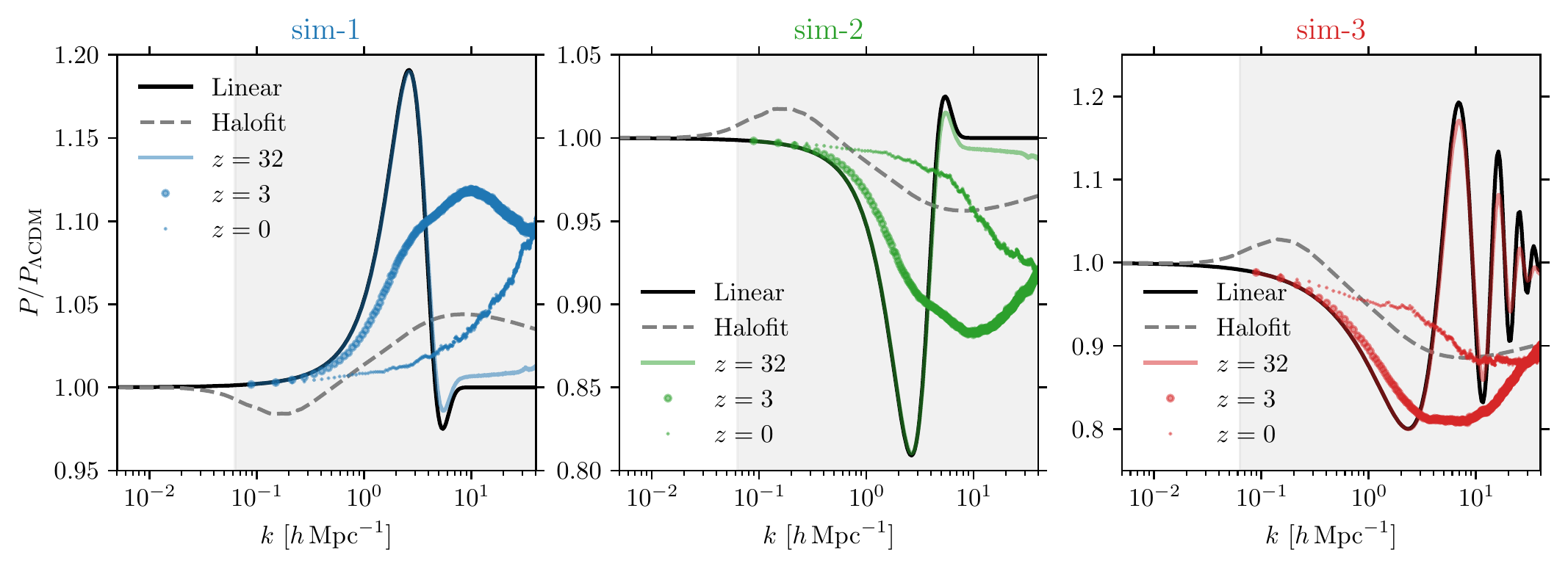}
    \caption{Ratio of the matter power spectra between the run with and without features, for the three simulations with parameters summarised in Tab.~\ref{tab: simu init parameters}, at redshifts $z=32, 3$ and $0$. The shaded gray region corresponds to the simulated scales between the fundamental mode of the simulated box and of the Nyquist mode.
    The black lines correspond to the linear power spectrum (\textit{solid}) and the nonlinear one computed with \texttt{Halofit} (\textit{dashed}).
    }
    \label{fig: PS} 
 \end{figure*}

In Fig.~\ref{fig: PS}, we show the ratio of the matter power spectra with and without features. The linear power spectrum generated with \texttt{CLASS} and the initial $z=32$ power spectrum of the simulation agree below $2\%$ thus validating our initial condition implementation. 

\vskip 4pt
We first notice that, as \texttt{Halofit} predicted, gravitational nonlinearities effectively suppress oscillatory pattern over time (see also a similar result in different contexts~\cite{Seo:2008yx, Buckley:2014hja, Schaeffer:2021qwm}). We expect that resonant features—arising when a background component oscillates with a frequency exceeding the Hubble scale during inflation, thereby imprinting persistent oscillations on the primordial power spectrum—are not accessible in the nonlinear regime of structure formation. This is particularly visible on the right panel of Fig.~\ref{fig: PS} for $\Delta k = 10\,\Mpc^{-1}$. However, even at $z=0$, all simulations show that the nonlinear matter power spectrum carry some information about the primordial feature, imprinted in the form of a characteristic enhancement/decrease in power above $k=2\,\Mpc^{-1}$, and reaching up to $\sim\pm 10\%$ at the smallest scales accessible with our simulations, with a sign that depends on that of the primordial feature amplitude $\delta \A$. This suggests that primordial features may indeed play a role in the $S_8$ tension, although the specific choice of $\delta \A$ and $k_f$ we have made may not be optimal.  
Interestingly, at $z=3$, while the nonlinearities have already washed away the oscillatory pattern, the suppression is similar to that needed to solve tensions between Ly-$\alpha$ forest and CMB data~\cite{Rogers:2023upm}. Firmly establishing the parameter space that may resolve tension requires more work and could be done using e.g.~an emulator of the nonlinear matter power spectrum tuned on simulations made for our model. Indeed, the \texttt{Halofit} emulator which has been trained only on featureless simulations (predictions are shown in grey in Fig.~\ref{fig: PS})  fails to mimic the effect of the feature at redshift $z=0$.  This shows that current emulators require significant improvements to accurately disentangle nonlinear clustering from primordial effects.

\vskip 4pt
We also note that the amplitude of the dump/dip in the matter power spectrum is non-trivially related to the template parameters, in particular to the feature width $\Delta k$. Although there is a generic trend showing that the broader the primordial feature, the broader the nonlinear response, it does not follow a simple scaling behaviour. Indeed, due to the gravitational mode coupling, we would naively expect that the nonlinear correction due to the feature is roughly dictated by the integrated power spectrum (at least at first order in perturbation theory): $P/P_{\Lambda{\rm CDM}}-1 \propto \int \tfrac{\d^3k}{(2\pi)^3} \Delta P_\zeta(k)$, where $\Delta P_\zeta(k) = P_\zeta(k) - P_{\zeta, 0}(k) = A_s \tfrac{2\pi^2}{k^3} \delta\P_\zeta(k)$. Performing the integral over a wave packet of width $\Delta k$ and assuming that nonlinear effects come from the integrated long-wavelength modes (as the localised feature suppresses small-scale physics), we naturally obtain the simple scaling behaviour $P/P_{\Lambda{\rm CDM}}-1 \propto \delta\A\,\tfrac{\Delta k}{k_f}$. However, this scaling, in particular linearly proportional to $\Delta k$, fails to match the results obtained from simulations, showing that this regime can only be reached through dedicated simulations. 
We leave a detailed analytical study of the effect of nonlinearities onto primordial features characterised by Eq.~\eqref{eq: final template} using perturbation theory and effective field theory technics, e.g. following Refs.~\cite{Vlah:2015zda, Vasudevan:2019ewf, Chen:2020ckc,Ballardini:2022vzh}, to future work. We nevertheless note that those analytical technics will be limited in the range of scales they can study compared to our N-body simulations.

\subsubsection{Halo Mass Function}

Next, we check the impact of the primordial features on the abundance of virialised objects. We detected halos using \texttt{Subfind} that is included in the public version of \texttt{Gadget-4} and adopt for the definition of mass $M_{200}$: a halo at $z=0$ has a boundary when its density is 200 times the critical density of the Universe. We consider only halos of at least 100 particles. 

\begin{figure}[h!]
    \centering 
    \includegraphics[width=0.9\columnwidth,trim={0pt 6pt 0pt 0}]{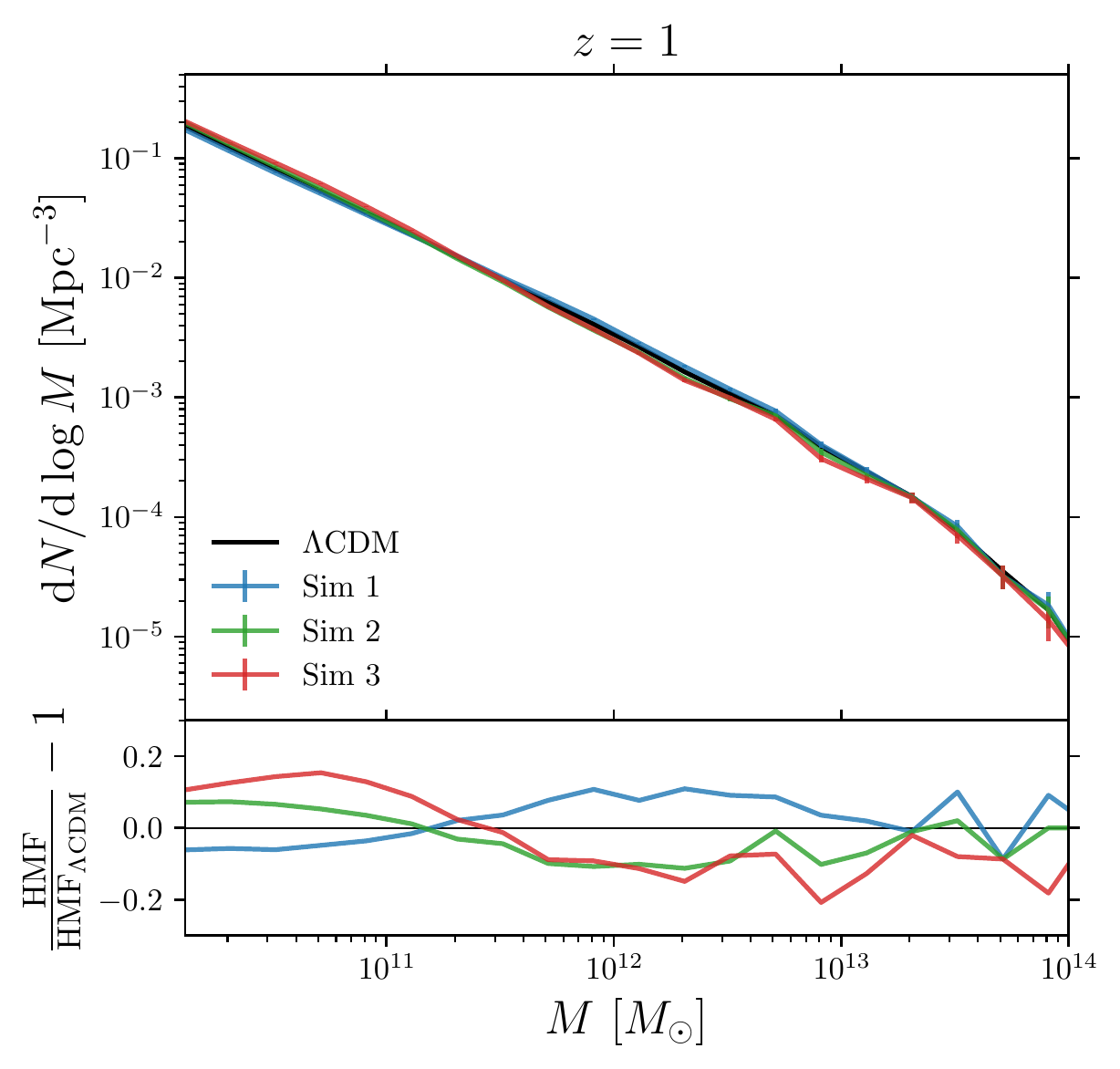}
    \caption{Halo mass function for the three simulations of Tab.~\ref{tab: simu init parameters} and its ratios with the featureless benchmark $\Lambda$CDM at $z=1$.}
    \label{fig: HMF} 
 \end{figure}

\vskip 4pt
Intuitively, one expects that an increase (decrease) in the matter power spectrum leads to more (less) massive halos around a mass scale that corresponds to the scale of the feature. 
To support this, we present the $z=1$ halo mass function ($\text{HMF} \equiv \d N / \d \log M$) for the simulations featuring oscillatory patterns in Fig.~\ref{fig: HMF}. At high masses, the presence of the features reduces/enhances the formation of dark matter halos by roughly $10\%$. Interestingly, this trend reverses for low-mass halos with the HMF unaffected by the feature for $M_h \sim 10^{11} M_\odot$. This ``oscillation'' was  also noticed in Ref.~\cite{Schaeffer:2021qwm}. It is expected that adding primordial non-Gaussianities further modifies the HMF \cite{Stahl:2024stz, Fiorino:2024ncx}. 

\subsubsection{Discussion}
The influence of the primordial features on the power spectrum and the HMF is similar to other extensions of the standard model of cosmology. To illustrate this, we relate the dip in the \textit{linear} matter power spectrum of the features with $\delta \mathcal{A}<0$ to a phenomenological parametrisation of the impact of dark matter candidates on the linear power spectrum, following the analytical fitting formula~\cite{Murgia:2017lwo, Stucker:2021vyx}
\begin{equation}
\label{eq:abg}
    \sqrt{\frac{P}{P_{\Lambda\rm CDM}}}=\left[ 1+(\alpha k)^{\beta}\right]^{-\gamma} \,.
\end{equation}
Such a parametrisation allows to explore dark matter candidates such as thermal relics, sterile neutrinos, fuzzy dark matter, and effective theory of structure formation. In Eq.~\eqref{eq:abg}, $\alpha, \beta$ and $\gamma$ are related to the microphysics of particle dark matter~\cite{Cyr-Racine:2015ihg}. We stress that the physical processes leading to primordial feature have {\it nothing to do} with particle dark matter candidates but we show here how features mimic the effect of dark matter on large scales. To do, so we will first take $\gamma=5$ without loss of generality~\cite{Stucker:2021vyx}. The left hand side of Eq.~\eqref{eq:abg} being equal to $1/2$ defines the characteristic half-mode
\begin{equation}
\label{eq:khm}
    k_{1/2} = \frac{1}{\alpha}\left(2^{1/\gamma}-1 \right)^{1/\beta}\, .
\end{equation}
A direct fit of the main bump-like feature for the simulations ($k<1.5\,h\,\Mpc^{-1}$) gives $(\alpha, \beta) = (0.04 \,\Mpc/h, 1.7)$ for sim-2, and $(0.01\,\Mpc/h, 1.0)$ for sim-3. This corresponds to a half mode of $k_{1/2} \approx 7.2 \,h\,\Mpc^{-1}$ and  $12 \,h\,\Mpc^{-1}$ and would correspond to a thermal relics of mass $m_{\rm WDM} \sim$ $1.1$ eV and $3.5$ eV. Such warm dark matter model would be totally degenerate with sim-2 and sim-3 at large and intermediate scales but would be ruled out by observations, e.g.~Refs.~\cite{Enzi:2020ieg,Das:2021pof}, due to the stark suppression of power for $k>10\,h\,\Mpc^{-1}$. The bump-like behaviour presented in sim-2 and sim-3 prevents such constraints to hold and models such as sim-2 and sim-3 may share the desirable properties of non-cold dark matter models without being ruled out. Moreover, this degeneracy between primordial features and particle dark matter at the level of the matter power spectrum is also broken when considering the HMF: features do not present the typical drop at low mass present in simulations of warm dark matter, see e.g.~Ref.~\cite{Stahl:2025nta} for more discussions on degeneracy between primordial physics and dark matter candidates.

\vskip 4pt
The models studied in this work also share similarities with Dark Acoustic Oscillations (DAO)~\cite{Cyr-Racine:2015ihg,Bohr:2020yoe,Bohr:2021bdm}, in particular the models ODM2 of Ref.~\cite{Schaeffer:2021qwm}. Our findings align with their conclusions: the oscillations present in the linear power spectrum do {\it not} transfer to the low-$z$ nonlinear power spectrum, but the overall shape and, in particular, the typical scale of the feature survives the nonlinear regime. The HMF also retains a characteristic behaviour related to the sign of $\delta \mathcal{A}$. The main differences between the primordial features studied in this work and DAO are that (i) features can boost clustering ($\delta \mathcal{A}>0$) and (ii) features do not impact (at the linear level) the smallest scales whereas DAO do. Moreover, features in the primordial power spectrum naturally come along features in the primordial bispectrum whereas, to the best of our knowledge, the impact of particle dark matter on the bispectrum was not discussed in the context of DAO.

\vskip 4pt
Neutrinos and baryons are also known to decrease the nonlinear matter power spectrum and their impact is also mutually non-exclusive with primordial features with negative amplitude $\delta \mathcal{A}$. Moreover, modified gravity, e.g.~\cite{Euclid:2024xfd}, and extra scalar field to describe dark energy, e.g.~\cite{Smith:2024ibv}, typically boost the matter power spectrum thus inducing possible degeneracies with positive $\delta \mathcal{A}$. Breaking these degeneracies might be possible by combining different probes such as the HMF and the power spectrum at different redshift as the physical origin of these extensions of the cosmological model is different. We note that 21cm surveys appear as a promising probe in this context~\cite{Munoz:2019hjh}.

\subsection{Case Study: the Phenomenological Feature Favoured by {\it Planck}}

As discussed previously, a localised feature similar to the one we study here as been shown to potentially explain the `$A_L$ anomaly' observed within {\it Planck} 2018 data \cite{Planck:2018jri,Domenech:2019cyh,Ballardini:2022vzh}.
More precisely, the feature that is favoured by {\it Planck} is parametrised according to the template
\begin{equation}
    \label{eq:planck_template}
    \delta\P_\zeta(k) = \delta\A \, \exp\left[-\frac{(k - k_f)^2}{2\Delta k^2}\right]\, \cos\big(\omega(k/k_*)+\phi\big)\,.
\end{equation}
From the analysis of {\it Planck} TT 2018 data, it has been found that the following parameter values are favoured\footnote{Note that EE and TE data, that are out of phase with respect to TT, do not favour that oscillation \cite{Planck:2018jri}.}:  $\delta\A = 0.16$, $\omega = 10^{1.158}\simeq 14.4$, $k_f = 0.2$ Mpc$^{-1}$, $\Delta k = 0.057$ Mpc$^{-1}$ and $\phi =\pi$. 
This template can be compared with that derived in Eq.~\eqref{eq: final template}. The main difference is that the frequency of the oscillatory feature $\omega$ is explicitly decoupled from its position, set by $k_f$, and out of phase by $\pi/2$.
Inserting the values found by {\it Planck} for $\omega$ in Eq.~\eqref{eq: final template} would predict that this template describes a feature that excites modes $k$ well inside the horizon such that $k/k_f \approx 30 \gg 1$. Even though this scenario is under perturbative control, i.e.~$(k/k_f)^4 \P_\zeta \ll 1$, see~\cite{Bartolo:2013exa, Cannone:2014qna}, it would be a highly non-trivial task to design a concrete inflationary UV model to reproduce such a feature.
Hence, it seems unlikely that the $A_L$ anomaly is tied to an inflationary feature.\footnote{It is also worth mentioning that the significance of the $A_L$ anomaly has decreased with updated data and alternative {\it Planck} likelihood \cite{Rosenberg:2022sdy,Tristram:2023haj}, and may simply be an artifact from a statistical fluke or unmitigated systematic error.}

\begin{figure*}[t]
    \centering
    \includegraphics[width=1\textwidth]{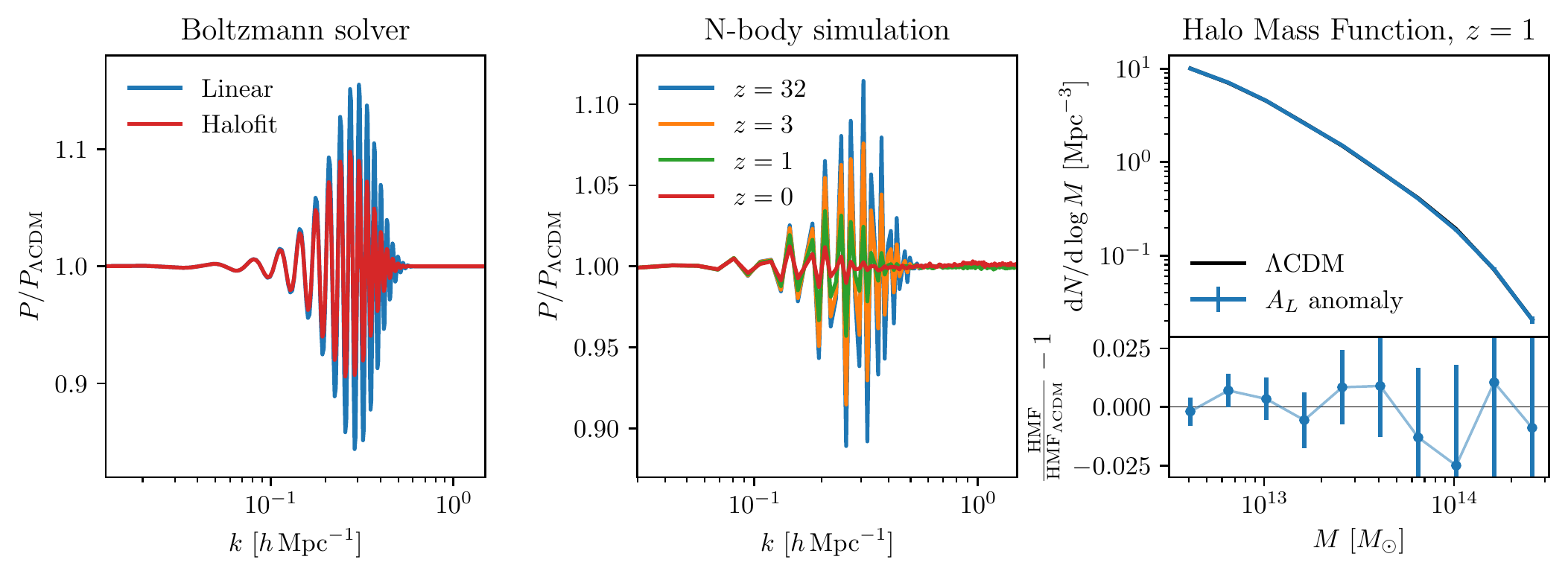}
    \caption{Impact of the phenomenological feature favoured by {\it Planck} on the $z=0$ matter power spectrum inferred from linear evolution using \texttt{CLASS} and the nonlinear emulator \texttt{Halofit} (\textit{left panel}), extracted from N-body simulations at z=32, 3, 1 and 0 (\textit{middle panel}), and on the corresponding $z=1$ halo mass function (\textit{right panel}). \texttt{Halofit} manages well to describe the damping of the oscillations at these scales, which contrasts with the sharp feature deep in the nonlinear regime in Sec.~\ref{subsec: N-body Simulations}. The nonlinear power spectra are in reasonable agreement with the simulations performed in Ref.~\cite{Ballardini:2022vzh}. The HMFs agree at the $\sim 2\%$ level, of the order of the Poisson errors.}
    \label{fig: Al} 
 \end{figure*}

\vskip 4pt
Nevertheless, for completeness, we run a final N-body simulation using the template given by Eq.~\eqref{eq:planck_template}, and the parameter values favoured by {\it Planck} (Tab.~\ref{tab: cosmo parameters}). As the feature is at slightly larger scales, we simulate a $L=500$ Mpc/$h$ box and use a grid of $512^3$. We present in Fig.~\ref{fig: Al} the ratio of the primordial power spectrum with and without features (left), the matter power spectra between the run with and without features (middle), and the halo mass function and its ratio with a featureless benchmark (right).

\vskip 4pt
Our results can be explicitly compared with those of Figure 4 of Ref.~\cite{Ballardini:2022vzh}. We find, as expected, that the oscillatory behaviour gets damped by the nonlinear evolution but as they are in the mildly nonlinear regime, their position does not get shifted toward smaller scales. At $z=0$, only a few per-cents effect remains but as in Sec.~\ref{subsec: N-body Simulations}, the primordial feature leaves a more pronounced effect at higher redshift. However, in contrast to the previous section, \texttt{Halofit} reasonably grasps the mildly nonlinear evolution of the feature. 
At the level of the HMF, no specific effect of the feature is found for the clustering of virialised objects with and without feature. 
From the model builder perspective, it might be possible to combine this resolution of the $A_L$ anomaly with features potentially being able to solve the $S_8$ and/or Ly-$\alpha$ tension as we have studied in Sec.~\ref{subsec: N-body Simulations}. However, it is not clear how motivated such a template is from the point of view of  inflationary model building. We leave the explicit construction and constraint of such models for follow-up works.
\section{Conclusions and Outlook}
\label{sec: Conclusions and Outlook}

In this work, we explored the impact of primordial features on the nonlinear regime of structure formation, motivated by UV completions of inflationary models. We began by deriving general templates for the primordial power spectrum that capture sharp features, reproducing those predicted by specific inflationary models. In particular, we solved the background and fluctuation dynamics for two distinct inflationary scenarios: a single-field model with a step in the potential and a multi-field model with a turn in the multi-field space trajectory.
We have shown that these features can be described within a unified picture, based on the time-dependence of either the slow-roll parameter or the speed of sound of curvature perturbations. This leads to a simple template characterised by three free parameters: the feature's amplitude, width, and scale.
Using linear cosmological data, we placed constraints on these features. Then, using N-body simulations, we demonstrated that sharp features consistent with CMB data can induce, in the nonlinear power spectrum, enhancements or suppressions proportional to its primordial amplitude. A broader primordial feature induces a broader and larger nonlinear response. In the HMF, we have also found a specific ``oscillation''. These characteristic corrections provide a potential means to recover the scale, the amplitude and the sign of the feature, and consequently the parameters governing the UV-complete models of inflation.
While our results highlight the promising potential of small-scale nonlinear structures to constrain primordial features, this study remains preliminary. Indeed, we also found that the observed effects may be significantly degenerate with other extensions of the standard cosmological model, such as alternative particle dark matter candidates. Moving forward,  improved modelling of nonlinear physics will be essential to robustly reconstruct primordial sharp features from the nonlinear regime of structure formation.

\vskip 4pt
Looking ahead, our work opens several avenues for future research. First, it will be essential to investigate the role of primordial non-Gaussianities. Since scale-dependent \textit{local} non-Gaussianities are predominantly gauge artifacts and hence non-physical, see e.g.~\cite{Pajer:2013ana}, simulating scale-dependent \textit{equilateral} non-Gaussianities will be crucial to definitively understand how features influence structure formation.
Second, our findings suggest that primordial features could provide a viable explanation for recent hints of suppressed power in the nonlinear regime. Examples include the power deficit at $k \sim 1 \,h\,\Mpc^{-1}$ inferred from Ly-$\alpha$ forest data at $z = 3$~\cite{Palanque-Delabrouille:2019iyz,Goldstein:2023gnw,Fernandez:2023grg, Rogers:2023upm} and the ongoing $S_8$ tension~\cite{KiDS:2020suj,DES:2021wwk,HSC:2018mrq,Amon:2022azi, Preston:2023uup, Stahl:2024stz}. 
Lastly, we have highlighted the necessity of training emulators on simulations that include primordial features to enable accurate reconstruction of primordial signals in the nonlinear regime. Given that the precise shape of these features depends sensitively on the microphysical details of inflationary models, this task presents significant challenges.
Constraining the physics of inflation, as always, proves to be a more formidable challenge than anticipated. Yet, the potential insights into the early Universe make the effort profoundly worthwhile.

\begin{acknowledgments}
We thank Sébastien Renaux-Petel for insightful discussions and comments on the draft. VP is supported by funding from the European Research Council (ERC) under the European Union’s HORIZON-ERC-2022 (grant agreement No 101076865). DW is supported by the European Research Council under the European Union’s Horizon 2020 research and innovation programme (grant agreement No 758792, Starting Grant project GEODESI). 
This article is distributed under the Creative Commons Attribution International Licence (\href{https://creativecommons.org/licenses/by/4.0/}{CC-BY 4.0}). This work has made use of the Infinity Cluster hosted by the Institut d'Astrophysique de Paris. The analysis was partially made using \texttt{CLASS} \href{https://lesgourg.github.io/class_public/class.html}{\faGithub}~\cite{Blas:2011rf}, \texttt{YT} \href{https://yt-project.org/}{\faGithub}~\cite{Turk:2010ah}, \texttt{Pylians} \href{https://pylians3.readthedocs.io/en/master/index.html}{\faGithub}~\cite{Pylians} as well as IPython~\cite{Perez:2007emg}, Matplotlib~\cite{Hunter:2007ouj} and NumPy~\cite{vanderWalt:2011bqk}.
\end{acknowledgments}

\subsection*{Carbon Footprint} 
Based on the methodology of Ref.~\cite{berthoud} to convert the CPU hours used for the simulations presented in this work, we estimate that we have emitted 3 TCO2eq.\footnote{Including the global utilisation of the cluster and the pollution due to the electrical source, the conversion factor is 4.7 gCO2e/(h.core).}

\appendix
\section{Details on Primordial Power Spectrum with Features}
\label{app: Details on PPS with Features}

In this appendix, we derive and justify the integral~(\ref{eq: PPS feature}) giving the deviation of the primordial power spectrum from the vanilla slow-roll one due to a feature either in the slow-roll parameter $\epsilon(t)$ or in the speed of sound $c_s(t)$. We begin with a small reduction in the sound speed assuming a constant $\epsilon$. The quadratic action~(\ref{eq: quadratic action}) can be written
\begin{equation}
\label{eq: sound speed perturbation}
    S = S_0 - \Mpl^2 \int \d t \d^3x \, a^3 \epsilon \left(1 - \frac{1}{c_s^2}\right)\dot{\zeta}^2\,,
\end{equation}
where $S_0$ is the free action obtained from~(\ref{eq: quadratic action}) after setting $c_s = 1$. The second term in~(\ref{eq: sound speed perturbation}) can be treated as a small perturbation to the free action as long as $|1 - c_s^{-2}| \ll 1$. Using the in-in formalism~\cite{Weinberg:2005vy}, the correction to the power spectrum can be written (using notations of Ref.~\cite{Chen:2017ryl})
\begin{equation}
    \begin{aligned}
    \raisebox{0pt}{
\begin{tikzpicture}[line width=1. pt, scale=2]
\draw[fill=white] ([xshift=0pt,yshift=-1.5pt]-0.1, 0) rectangle ++(3pt,3pt);
\draw[fill=black] (0.5, 0) circle (.05cm) node[above=0.5mm] {};
\draw[black] (0, 0) -- (1, 0);
\draw[fill=white] ([xshift=0pt,yshift=-1.5pt]1, 0) rectangle ++(3pt,3pt);
\end{tikzpicture} 
} = -&i \int_{-\infty}^0 \d t \, a^3(t) u_{c_s}(t)\\ &\times[\partial_t G_+(k; t)][\partial_t G_+(k; t)]\,,
    \end{aligned}
\end{equation}
where $u_{c_s} \equiv 1 - c_s^{-2}$ and $G_+(k; t) = \zeta_k^*(t)\zeta_k(0)$ with $\zeta_k(\tau) = \tfrac{H}{\sqrt{2k^3}}(1+ik\tau)e^{-ik\tau}$ the mode function of the (canonically normalised) curvature perturbation field.\footnote{We have removed the overall normalisation $\Mpl^2\epsilon$ as it does not enter the relative correction to the featureless power spectrum. Note that the power spectrum of a canonically normalised field reads $H^2/2k^3$.} Expressing this integral in terms of conformal time $\tau$, setting $a(\tau) = -1/H\tau$, and adding the complex conjugate expression, we obtain~(\ref{eq: PPS feature}).

\vskip 4pt
For a small feature in the slow-roll parameter $\epsilon(t) = \epsilon_0[1 -u_\epsilon(t)]$, the action now reads
\begin{equation}
    S = S_0 - \Mpl^2 \int \d t \d^3x\, a^3 \epsilon_0 u_\epsilon \left(\dot{\zeta}^2 - \frac{(\partial_i \zeta)^2}{a^2}\right)\,,
\end{equation}
where we have set $c_s=1$, and $S_0$ is the quadratic action~(\ref{eq: quadratic action}) after setting $\epsilon = \epsilon_0$. The first interaction term $\propto \dot{\zeta}^2$ gives the same correction as a feature in the sound speed
\begin{equation}
    \delta\P_\zeta^{\dot{\zeta}^2}(k) = k \int_{-\infty}^0\d\tau \, u_\epsilon(\tau) \sin(2k\tau)\,.
\end{equation}
The correction due to the second term $\propto (\partial_i\zeta)^2$ gives the following integral
\begin{equation}
    \begin{aligned}
    \raisebox{0pt}{
\begin{tikzpicture}[line width=1. pt, scale=2]
\draw[fill=white] ([xshift=0pt,yshift=-1.5pt]-0.1, 0) rectangle ++(3pt,3pt);
\draw[fill=black] (0.5, 0) circle (.05cm) node[above=0.5mm] {};
\draw[black] (0, 0) -- (1, 0);
\draw[fill=white] ([xshift=0pt,yshift=-1.5pt]1, 0) rectangle ++(3pt,3pt);
\end{tikzpicture} 
} = -&i \int_{-\infty}^0 \d t \, a^3(t) u_{\epsilon}(t)\\ &\times \frac{k^2}{a^2} \, G_+(k; t) G_+(k; t)\,,
    \end{aligned}
\end{equation}
which, after adding the complex conjugate contribution, yields
\begin{equation}
    \begin{aligned}
        &\delta\P_\zeta^{(\partial_i \zeta)^2}(k) = \frac{1}{k} \int_{-\infty}^0 \frac{\d\tau}{\tau^2}\, u_\epsilon(\tau) \sin(2k\tau) \\
        &- \int_{-\infty}^0 \frac{\d\tau}{\tau}\, u_\epsilon(\tau)\left[2\cos(2k\tau) + k\tau \sin(2k\tau)\right]\,.
    \end{aligned}
\end{equation}
The last term in $\delta\P_\zeta^{(\partial_i \zeta)^2}$ cancels with $\delta\P_\zeta^{\dot{\zeta}^2}$. In the end, we recover Eq.~(\ref{eq: PPS feature}).

\section{On the Use of a Logarithmic Prior on $\delta A$ in the Bayesian Analysis}
\label{app:deltaA}

\begin{figure}[h!]
    \centering
    \includegraphics[width=0.8\columnwidth]{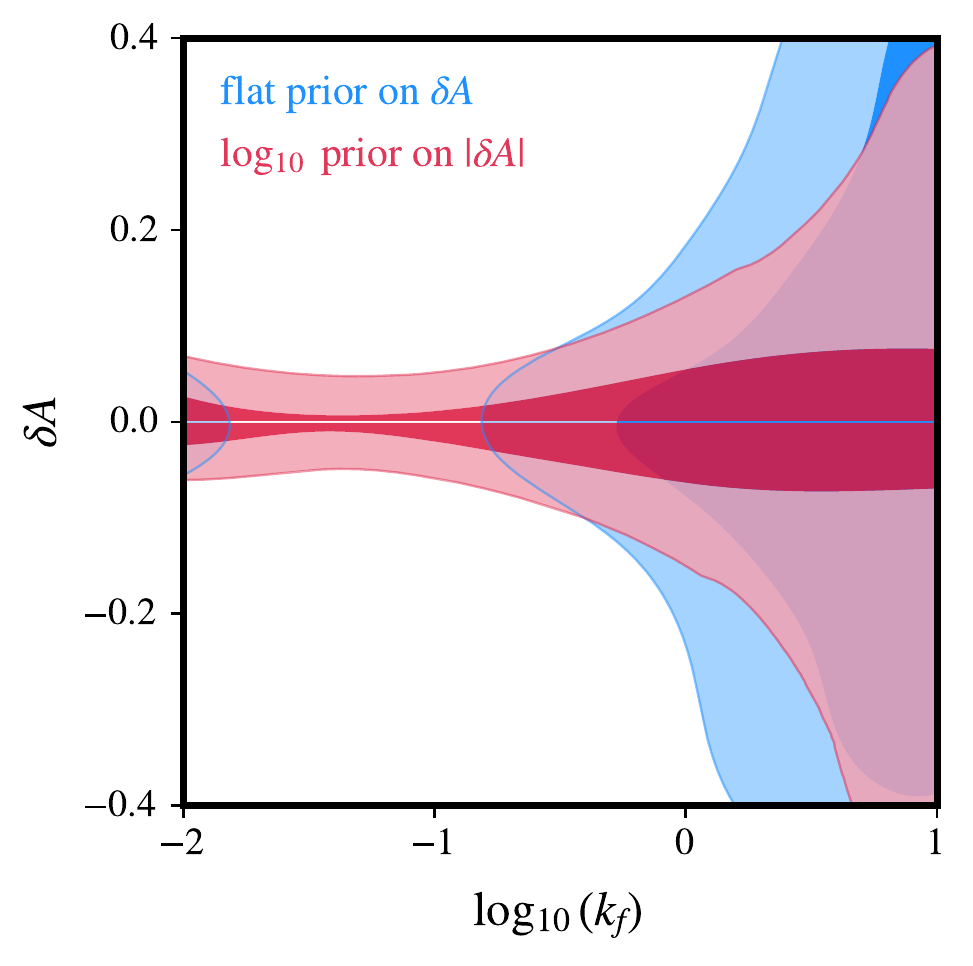}
    \caption{Comparison of the 2D posteriors in the plane $\{\log_{10}(k_f),\delta A\}$ when using either a logarithmic or a linear prior on $\delta A$. Note that the logarithmic prior correspond to two different runs, considering $\delta A > 0$ and $\delta A < 0$ separately. The use of a logarithmic prior is warranted to prevent difficulties in exploring the parameter space due to highly non-Gaussian posteriors.}
    \label{fig:log-vs-lindeltaA}
\end{figure}

In the main text, we have presented results that made use of logarithmic prior on $\delta A$, though this is not necessarily justified theoretically (our template suggests a linear prior may be more suited). This choice avoids spurious constraints due to difficulties in exploring a highly non-Gaussian parameter space. This is illustrated in Fig.~\ref{fig:log-vs-lindeltaA}, where we compare the 2D posteriors in the plane $\{\log_{10} k_f,\delta A\}$ when using either a logarithmic or a linear prior on $\delta A$.  One can see that the region of small $\log_{10}(k_f)$ appears artificially disfavoured with the linear prior, while it is correctly recovered with a logarithmic one. However, one should bear in mind that the logarithmic prior instead leads to stronger constraints in the region of large $\log_{10}(k_f)$. Alternative sampling methods (e.g. using nested sampling \cite{Feroz:2008xx}), a re-parametrisation of $\delta A$, or the use of frequentist methods (as e.g. in Ref.~\cite{Karwal:2024qpt}) may help better explore the parameter space. 

\let\oldaddcontentsline\addcontentsline
\renewcommand{\addcontentsline}[3]{}

\twocolumngrid
\bibliography{References}

\begin{thebibliography}{171}%
\makeatletter
\providecommand \@ifxundefined [1]{%
 \@ifx{#1\undefined}
}%
\providecommand \@ifnum [1]{%
 \ifnum #1\expandafter \@firstoftwo
 \else \expandafter \@secondoftwo
 \fi
}%
\providecommand \@ifx [1]{%
 \ifx #1\expandafter \@firstoftwo
 \else \expandafter \@secondoftwo
 \fi
}%
\providecommand \natexlab [1]{#1}%
\providecommand \enquote  [1]{``#1''}%
\providecommand \bibnamefont  [1]{#1}%
\providecommand \bibfnamefont [1]{#1}%
\providecommand \citenamefont [1]{#1}%
\providecommand \href@noop [0]{\@secondoftwo}%
\providecommand \href [0]{\begingroup \@sanitize@url \@href}%
\providecommand \@href[1]{\@@startlink{#1}\@@href}%
\providecommand \@@href[1]{\endgroup#1\@@endlink}%
\providecommand \@sanitize@url [0]{\catcode `\\12\catcode `\$12\catcode `\&12\catcode `\#12\catcode `\^12\catcode `\_12\catcode `\%12\relax}%
\providecommand \@@startlink[1]{}%
\providecommand \@@endlink[0]{}%
\providecommand \url  [0]{\begingroup\@sanitize@url \@url }%
\providecommand \@url [1]{\endgroup\@href {#1}{\urlprefix }}%
\providecommand \urlprefix  [0]{URL }%
\providecommand \Eprint [0]{\href }%
\providecommand \doibase [0]{http://dx.doi.org/}%
\providecommand \selectlanguage [0]{\@gobble}%
\providecommand \bibinfo  [0]{\@secondoftwo}%
\providecommand \bibfield  [0]{\@secondoftwo}%
\providecommand \translation [1]{[#1]}%
\providecommand \BibitemOpen [0]{}%
\providecommand \bibitemStop [0]{}%
\providecommand \bibitemNoStop [0]{.\EOS\space}%
\providecommand \EOS [0]{\spacefactor3000\relax}%
\providecommand \BibitemShut  [1]{\csname bibitem#1\endcsname}%
\let\auto@bib@innerbib\@empty
\bibitem [{\citenamefont {Starobinsky}(1980)}]{Starobinsky:1980te}%
  \BibitemOpen
  \bibfield  {author} {\bibinfo {author} {\bibfnamefont {A.~A.}\ \bibnamefont {Starobinsky}},\ }\href {\doibase 10.1016/0370-2693(80)90670-X} {\bibfield  {journal} {\bibinfo  {journal} {Phys. Lett. B}\ }\textbf {\bibinfo {volume} {91}},\ \bibinfo {pages} {99} (\bibinfo {year} {1980})}\BibitemShut {NoStop}%
\bibitem [{\citenamefont {Guth}(1981)}]{Guth:1980zm}%
  \BibitemOpen
  \bibfield  {author} {\bibinfo {author} {\bibfnamefont {A.~H.}\ \bibnamefont {Guth}},\ }\href {\doibase 10.1103/PhysRevD.23.347} {\bibfield  {journal} {\bibinfo  {journal} {Phys. Rev. D}\ }\textbf {\bibinfo {volume} {23}},\ \bibinfo {pages} {347} (\bibinfo {year} {1981})}\BibitemShut {NoStop}%
\bibitem [{\citenamefont {Linde}(1982)}]{Linde:1981mu}%
  \BibitemOpen
  \bibfield  {author} {\bibinfo {author} {\bibfnamefont {A.~D.}\ \bibnamefont {Linde}},\ }\href {\doibase 10.1016/0370-2693(82)91219-9} {\bibfield  {journal} {\bibinfo  {journal} {Phys. Lett. B}\ }\textbf {\bibinfo {volume} {108}},\ \bibinfo {pages} {389} (\bibinfo {year} {1982})}\BibitemShut {NoStop}%
\bibitem [{\citenamefont {Albrecht}\ and\ \citenamefont {Steinhardt}(1982)}]{Albrecht:1982wi}%
  \BibitemOpen
  \bibfield  {author} {\bibinfo {author} {\bibfnamefont {A.}~\bibnamefont {Albrecht}}\ and\ \bibinfo {author} {\bibfnamefont {P.~J.}\ \bibnamefont {Steinhardt}},\ }\href {\doibase 10.1103/PhysRevLett.48.1220} {\bibfield  {journal} {\bibinfo  {journal} {Phys. Rev. Lett.}\ }\textbf {\bibinfo {volume} {48}},\ \bibinfo {pages} {1220} (\bibinfo {year} {1982})}\BibitemShut {NoStop}%
\bibitem [{\citenamefont {Akrami}\ \emph {et~al.}(2020)\citenamefont {Akrami} \emph {et~al.}}]{Planck:2018jri}%
  \BibitemOpen
  \bibfield  {author} {\bibinfo {author} {\bibfnamefont {Y.}~\bibnamefont {Akrami}} \emph {et~al.} (\bibinfo {collaboration} {Planck}),\ }\href {\doibase 10.1051/0004-6361/201833887} {\bibfield  {journal} {\bibinfo  {journal} {Astron. Astrophys.}\ }\textbf {\bibinfo {volume} {641}},\ \bibinfo {pages} {A10} (\bibinfo {year} {2020})},\ \Eprint {http://arxiv.org/abs/1807.06211} {arXiv:1807.06211 [astro-ph.CO]} \BibitemShut {NoStop}%
\bibitem [{\citenamefont {Baumann}\ and\ \citenamefont {McAllister}(2015)}]{Baumann:2014nda}%
  \BibitemOpen
  \bibfield  {author} {\bibinfo {author} {\bibfnamefont {D.}~\bibnamefont {Baumann}}\ and\ \bibinfo {author} {\bibfnamefont {L.}~\bibnamefont {McAllister}},\ }\href {\doibase 10.1017/CBO9781316105733} {\emph {\bibinfo {title} {{Inflation and String Theory}}}},\ Cambridge Monographs on Mathematical Physics\ (\bibinfo  {publisher} {Cambridge University Press},\ \bibinfo {year} {2015})\ \Eprint {http://arxiv.org/abs/1404.2601} {arXiv:1404.2601 [hep-th]} \BibitemShut {NoStop}%
\bibitem [{\citenamefont {Chen}(2010)}]{Chen:2010xka}%
  \BibitemOpen
  \bibfield  {author} {\bibinfo {author} {\bibfnamefont {X.}~\bibnamefont {Chen}},\ }\href {\doibase 10.1155/2010/638979} {\bibfield  {journal} {\bibinfo  {journal} {Adv. Astron.}\ }\textbf {\bibinfo {volume} {2010}},\ \bibinfo {pages} {638979} (\bibinfo {year} {2010})},\ \Eprint {http://arxiv.org/abs/1002.1416} {arXiv:1002.1416 [astro-ph.CO]} \BibitemShut {NoStop}%
\bibitem [{\citenamefont {Chluba}\ \emph {et~al.}(2015)\citenamefont {Chluba}, \citenamefont {Hamann},\ and\ \citenamefont {Patil}}]{Chluba:2015bqa}%
  \BibitemOpen
  \bibfield  {author} {\bibinfo {author} {\bibfnamefont {J.}~\bibnamefont {Chluba}}, \bibinfo {author} {\bibfnamefont {J.}~\bibnamefont {Hamann}}, \ and\ \bibinfo {author} {\bibfnamefont {S.~P.}\ \bibnamefont {Patil}},\ }\href {\doibase 10.1142/S0218271815300232} {\bibfield  {journal} {\bibinfo  {journal} {Int. J. Mod. Phys. D}\ }\textbf {\bibinfo {volume} {24}},\ \bibinfo {pages} {1530023} (\bibinfo {year} {2015})},\ \Eprint {http://arxiv.org/abs/1505.01834} {arXiv:1505.01834 [astro-ph.CO]} \BibitemShut {NoStop}%
\bibitem [{\citenamefont {Slosar}\ \emph {et~al.}(2019)\citenamefont {Slosar} \emph {et~al.}}]{Slosar:2019gvt}%
  \BibitemOpen
  \bibfield  {author} {\bibinfo {author} {\bibfnamefont {A.}~\bibnamefont {Slosar}} \emph {et~al.},\ }\href@noop {} {\bibfield  {journal} {\bibinfo  {journal} {Bull. Am. Astron. Soc.}\ }\textbf {\bibinfo {volume} {51}},\ \bibinfo {pages} {98} (\bibinfo {year} {2019})},\ \Eprint {http://arxiv.org/abs/1903.09883} {arXiv:1903.09883 [astro-ph.CO]} \BibitemShut {NoStop}%
\bibitem [{\citenamefont {Ach\'ucarro}\ \emph {et~al.}(2022)\citenamefont {Ach\'ucarro} \emph {et~al.}}]{Achucarro:2022qrl}%
  \BibitemOpen
  \bibfield  {author} {\bibinfo {author} {\bibfnamefont {A.}~\bibnamefont {Ach\'ucarro}} \emph {et~al.},\ }\href@noop {} {\  (\bibinfo {year} {2022})},\ \Eprint {http://arxiv.org/abs/2203.08128} {arXiv:2203.08128 [astro-ph.CO]} \BibitemShut {NoStop}%
\bibitem [{\citenamefont {Chen}\ \emph {et~al.}(2007)\citenamefont {Chen}, \citenamefont {Easther},\ and\ \citenamefont {Lim}}]{Chen:2006xjb}%
  \BibitemOpen
  \bibfield  {author} {\bibinfo {author} {\bibfnamefont {X.}~\bibnamefont {Chen}}, \bibinfo {author} {\bibfnamefont {R.}~\bibnamefont {Easther}}, \ and\ \bibinfo {author} {\bibfnamefont {E.~A.}\ \bibnamefont {Lim}},\ }\href {\doibase 10.1088/1475-7516/2007/06/023} {\bibfield  {journal} {\bibinfo  {journal} {JCAP}\ }\textbf {\bibinfo {volume} {06}},\ \bibinfo {pages} {023} (\bibinfo {year} {2007})},\ \Eprint {http://arxiv.org/abs/astro-ph/0611645} {arXiv:astro-ph/0611645} \BibitemShut {NoStop}%
\bibitem [{\citenamefont {Chen}\ \emph {et~al.}(2008)\citenamefont {Chen}, \citenamefont {Easther},\ and\ \citenamefont {Lim}}]{Chen:2008wn}%
  \BibitemOpen
  \bibfield  {author} {\bibinfo {author} {\bibfnamefont {X.}~\bibnamefont {Chen}}, \bibinfo {author} {\bibfnamefont {R.}~\bibnamefont {Easther}}, \ and\ \bibinfo {author} {\bibfnamefont {E.~A.}\ \bibnamefont {Lim}},\ }\href {\doibase 10.1088/1475-7516/2008/04/010} {\bibfield  {journal} {\bibinfo  {journal} {JCAP}\ }\textbf {\bibinfo {volume} {04}},\ \bibinfo {pages} {010} (\bibinfo {year} {2008})},\ \Eprint {http://arxiv.org/abs/0801.3295} {arXiv:0801.3295 [astro-ph]} \BibitemShut {NoStop}%
\bibitem [{\citenamefont {Adshead}\ \emph {et~al.}(2012)\citenamefont {Adshead}, \citenamefont {Dvorkin}, \citenamefont {Hu},\ and\ \citenamefont {Lim}}]{Adshead:2011jq}%
  \BibitemOpen
  \bibfield  {author} {\bibinfo {author} {\bibfnamefont {P.}~\bibnamefont {Adshead}}, \bibinfo {author} {\bibfnamefont {C.}~\bibnamefont {Dvorkin}}, \bibinfo {author} {\bibfnamefont {W.}~\bibnamefont {Hu}}, \ and\ \bibinfo {author} {\bibfnamefont {E.~A.}\ \bibnamefont {Lim}},\ }\href {\doibase 10.1103/PhysRevD.85.023531} {\bibfield  {journal} {\bibinfo  {journal} {Phys. Rev. D}\ }\textbf {\bibinfo {volume} {85}},\ \bibinfo {pages} {023531} (\bibinfo {year} {2012})},\ \Eprint {http://arxiv.org/abs/1110.3050} {arXiv:1110.3050 [astro-ph.CO]} \BibitemShut {NoStop}%
\bibitem [{\citenamefont {Ach\'ucarro}\ \emph {et~al.}(2013)\citenamefont {Ach\'ucarro}, \citenamefont {Gong}, \citenamefont {Palma},\ and\ \citenamefont {Patil}}]{Achucarro:2012fd}%
  \BibitemOpen
  \bibfield  {author} {\bibinfo {author} {\bibfnamefont {A.}~\bibnamefont {Ach\'ucarro}}, \bibinfo {author} {\bibfnamefont {J.-O.}\ \bibnamefont {Gong}}, \bibinfo {author} {\bibfnamefont {G.~A.}\ \bibnamefont {Palma}}, \ and\ \bibinfo {author} {\bibfnamefont {S.~P.}\ \bibnamefont {Patil}},\ }\href {\doibase 10.1103/PhysRevD.87.121301} {\bibfield  {journal} {\bibinfo  {journal} {Phys. Rev. D}\ }\textbf {\bibinfo {volume} {87}},\ \bibinfo {pages} {121301} (\bibinfo {year} {2013})},\ \Eprint {http://arxiv.org/abs/1211.5619} {arXiv:1211.5619 [astro-ph.CO]} \BibitemShut {NoStop}%
\bibitem [{\citenamefont {Ach\'ucarro}\ \emph {et~al.}(2014)\citenamefont {Ach\'ucarro}, \citenamefont {Atal}, \citenamefont {Ortiz},\ and\ \citenamefont {Torrado}}]{Achucarro:2013cva}%
  \BibitemOpen
  \bibfield  {author} {\bibinfo {author} {\bibfnamefont {A.}~\bibnamefont {Ach\'ucarro}}, \bibinfo {author} {\bibfnamefont {V.}~\bibnamefont {Atal}}, \bibinfo {author} {\bibfnamefont {P.}~\bibnamefont {Ortiz}}, \ and\ \bibinfo {author} {\bibfnamefont {J.}~\bibnamefont {Torrado}},\ }\href {\doibase 10.1103/PhysRevD.89.103006} {\bibfield  {journal} {\bibinfo  {journal} {Phys. Rev. D}\ }\textbf {\bibinfo {volume} {89}},\ \bibinfo {pages} {103006} (\bibinfo {year} {2014})},\ \Eprint {http://arxiv.org/abs/1311.2552} {arXiv:1311.2552 [astro-ph.CO]} \BibitemShut {NoStop}%
\bibitem [{\citenamefont {Bartolo}\ \emph {et~al.}(2013)\citenamefont {Bartolo}, \citenamefont {Cannone},\ and\ \citenamefont {Matarrese}}]{Bartolo:2013exa}%
  \BibitemOpen
  \bibfield  {author} {\bibinfo {author} {\bibfnamefont {N.}~\bibnamefont {Bartolo}}, \bibinfo {author} {\bibfnamefont {D.}~\bibnamefont {Cannone}}, \ and\ \bibinfo {author} {\bibfnamefont {S.}~\bibnamefont {Matarrese}},\ }\href {\doibase 10.1088/1475-7516/2013/10/038} {\bibfield  {journal} {\bibinfo  {journal} {JCAP}\ }\textbf {\bibinfo {volume} {10}},\ \bibinfo {pages} {038} (\bibinfo {year} {2013})},\ \Eprint {http://arxiv.org/abs/1307.3483} {arXiv:1307.3483 [astro-ph.CO]} \BibitemShut {NoStop}%
\bibitem [{\citenamefont {{Ach{\'u}carro}}\ \emph {et~al.}(2014)\citenamefont {{Ach{\'u}carro}}, \citenamefont {{Atal}}, \citenamefont {{Hu}}, \citenamefont {{Ortiz}},\ and\ \citenamefont {{Torrado}}}]{Achucarro:2014msa}%
  \BibitemOpen
  \bibfield  {author} {\bibinfo {author} {\bibfnamefont {A.}~\bibnamefont {{Ach{\'u}carro}}}, \bibinfo {author} {\bibfnamefont {V.}~\bibnamefont {{Atal}}}, \bibinfo {author} {\bibfnamefont {B.}~\bibnamefont {{Hu}}}, \bibinfo {author} {\bibfnamefont {P.}~\bibnamefont {{Ortiz}}}, \ and\ \bibinfo {author} {\bibfnamefont {J.}~\bibnamefont {{Torrado}}},\ }\href {\doibase 10.1103/PhysRevD.90.023511} {\bibfield  {journal} {\bibinfo  {journal} {Physics Review D}\ }\textbf {\bibinfo {volume} {90}},\ \bibinfo {eid} {023511} (\bibinfo {year} {2014})},\ \Eprint {http://arxiv.org/abs/1404.7522} {arXiv:1404.7522 [astro-ph.CO]} \BibitemShut {NoStop}%
\bibitem [{\citenamefont {{Torrado}}\ \emph {et~al.}(2017)\citenamefont {{Torrado}}, \citenamefont {{Hu}},\ and\ \citenamefont {{Ach{\'u}carro}}}]{Torrado:2016sls}%
  \BibitemOpen
  \bibfield  {author} {\bibinfo {author} {\bibfnamefont {J.}~\bibnamefont {{Torrado}}}, \bibinfo {author} {\bibfnamefont {B.}~\bibnamefont {{Hu}}}, \ and\ \bibinfo {author} {\bibfnamefont {A.}~\bibnamefont {{Ach{\'u}carro}}},\ }\href {\doibase 10.1103/PhysRevD.96.083515} {\bibfield  {journal} {\bibinfo  {journal} {Physics Review D}\ }\textbf {\bibinfo {volume} {96}},\ \bibinfo {eid} {083515} (\bibinfo {year} {2017})},\ \Eprint {http://arxiv.org/abs/1611.10350} {arXiv:1611.10350 [astro-ph.CO]} \BibitemShut {NoStop}%
\bibitem [{\citenamefont {Chen}\ \emph {et~al.}(2022)\citenamefont {Chen}, \citenamefont {Ebadi},\ and\ \citenamefont {Kumar}}]{Chen:2022vzh}%
  \BibitemOpen
  \bibfield  {author} {\bibinfo {author} {\bibfnamefont {X.}~\bibnamefont {Chen}}, \bibinfo {author} {\bibfnamefont {R.}~\bibnamefont {Ebadi}}, \ and\ \bibinfo {author} {\bibfnamefont {S.}~\bibnamefont {Kumar}},\ }\href {\doibase 10.1088/1475-7516/2022/08/083} {\bibfield  {journal} {\bibinfo  {journal} {JCAP}\ }\textbf {\bibinfo {volume} {08}},\ \bibinfo {pages} {083} (\bibinfo {year} {2022})},\ \Eprint {http://arxiv.org/abs/2205.01107} {arXiv:2205.01107 [hep-ph]} \BibitemShut {NoStop}%
\bibitem [{\citenamefont {Werth}\ \emph {et~al.}(2024)\citenamefont {Werth}, \citenamefont {Pinol},\ and\ \citenamefont {Renaux-Petel}}]{Werth:2023pfl}%
  \BibitemOpen
  \bibfield  {author} {\bibinfo {author} {\bibfnamefont {D.}~\bibnamefont {Werth}}, \bibinfo {author} {\bibfnamefont {L.}~\bibnamefont {Pinol}}, \ and\ \bibinfo {author} {\bibfnamefont {S.}~\bibnamefont {Renaux-Petel}},\ }\href {\doibase 10.1103/PhysRevLett.133.141002} {\bibfield  {journal} {\bibinfo  {journal} {Phys. Rev. Lett.}\ }\textbf {\bibinfo {volume} {133}},\ \bibinfo {pages} {141002} (\bibinfo {year} {2024})},\ \Eprint {http://arxiv.org/abs/2302.00655} {arXiv:2302.00655 [hep-th]} \BibitemShut {NoStop}%
\bibitem [{\citenamefont {Pinol}\ \emph {et~al.}(2023)\citenamefont {Pinol}, \citenamefont {Renaux-Petel},\ and\ \citenamefont {Werth}}]{Pinol:2023oux}%
  \BibitemOpen
  \bibfield  {author} {\bibinfo {author} {\bibfnamefont {L.}~\bibnamefont {Pinol}}, \bibinfo {author} {\bibfnamefont {S.}~\bibnamefont {Renaux-Petel}}, \ and\ \bibinfo {author} {\bibfnamefont {D.}~\bibnamefont {Werth}},\ }\href@noop {} {\  (\bibinfo {year} {2023})},\ \Eprint {http://arxiv.org/abs/2312.06559} {arXiv:2312.06559 [astro-ph.CO]} \BibitemShut {NoStop}%
\bibitem [{\citenamefont {Wang}\ and\ \citenamefont {Mathews}(2002)}]{Wang:2000js}%
  \BibitemOpen
  \bibfield  {author} {\bibinfo {author} {\bibfnamefont {Y.}~\bibnamefont {Wang}}\ and\ \bibinfo {author} {\bibfnamefont {G.}~\bibnamefont {Mathews}},\ }\href {\doibase 10.1086/340492} {\bibfield  {journal} {\bibinfo  {journal} {Astrophys. J.}\ }\textbf {\bibinfo {volume} {573}},\ \bibinfo {pages} {1} (\bibinfo {year} {2002})},\ \Eprint {http://arxiv.org/abs/astro-ph/0011351} {arXiv:astro-ph/0011351} \BibitemShut {NoStop}%
\bibitem [{\citenamefont {{Adams}}\ \emph {et~al.}(2001)\citenamefont {{Adams}}, \citenamefont {{Cresswell}},\ and\ \citenamefont {{Easther}}}]{Adams:2001vc}%
  \BibitemOpen
  \bibfield  {author} {\bibinfo {author} {\bibfnamefont {J.}~\bibnamefont {{Adams}}}, \bibinfo {author} {\bibfnamefont {B.}~\bibnamefont {{Cresswell}}}, \ and\ \bibinfo {author} {\bibfnamefont {R.}~\bibnamefont {{Easther}}},\ }\href {\doibase 10.1103/PhysRevD.64.123514} {\bibfield  {journal} {\bibinfo  {journal} {Physics Review D}\ }\textbf {\bibinfo {volume} {64}},\ \bibinfo {pages} {123514} (\bibinfo {year} {2001})},\ \Eprint {http://arxiv.org/abs/astro-ph/0102236} {arXiv:astro-ph/0102236 [astro-ph]} \BibitemShut {NoStop}%
\bibitem [{\citenamefont {{Peiris}}\ \emph {et~al.}(2003)\citenamefont {{Peiris}}, \citenamefont {{Komatsu}}, \citenamefont {{Verde}}, \citenamefont {{Spergel}}, \citenamefont {{Bennett}}, \citenamefont {{Halpern}}, \citenamefont {{Hinshaw}}, \citenamefont {{Jarosik}}, \citenamefont {{Kogut}}, \citenamefont {{Limon}}, \citenamefont {{Meyer}}, \citenamefont {{Page}}, \citenamefont {{Tucker}}, \citenamefont {{Wollack}},\ and\ \citenamefont {{Wright}}}]{Peiris:2003ff}%
  \BibitemOpen
  \bibfield  {author} {\bibinfo {author} {\bibfnamefont {H.~V.}\ \bibnamefont {{Peiris}}}, \bibinfo {author} {\bibfnamefont {E.}~\bibnamefont {{Komatsu}}}, \bibinfo {author} {\bibfnamefont {L.}~\bibnamefont {{Verde}}}, \bibinfo {author} {\bibfnamefont {D.~N.}\ \bibnamefont {{Spergel}}}, \bibinfo {author} {\bibfnamefont {C.~L.}\ \bibnamefont {{Bennett}}}, \bibinfo {author} {\bibfnamefont {M.}~\bibnamefont {{Halpern}}}, \bibinfo {author} {\bibfnamefont {G.}~\bibnamefont {{Hinshaw}}}, \bibinfo {author} {\bibfnamefont {N.}~\bibnamefont {{Jarosik}}}, \bibinfo {author} {\bibfnamefont {A.}~\bibnamefont {{Kogut}}}, \bibinfo {author} {\bibfnamefont {M.}~\bibnamefont {{Limon}}}, \bibinfo {author} {\bibfnamefont {S.~S.}\ \bibnamefont {{Meyer}}}, \bibinfo {author} {\bibfnamefont {L.}~\bibnamefont {{Page}}}, \bibinfo {author} {\bibfnamefont {G.~S.}\ \bibnamefont {{Tucker}}}, \bibinfo {author} {\bibfnamefont {E.}~\bibnamefont {{Wollack}}}, \ and\ \bibinfo {author} {\bibfnamefont {E.~L.}\ \bibnamefont {{Wright}}},\ }\href
  {\doibase 10.1086/377228} {\bibfield  {journal} {\bibinfo  {journal} {apjs}\ }\textbf {\bibinfo {volume} {148}},\ \bibinfo {pages} {213} (\bibinfo {year} {2003})},\ \Eprint {http://arxiv.org/abs/astro-ph/0302225} {arXiv:astro-ph/0302225 [astro-ph]} \BibitemShut {NoStop}%
\bibitem [{\citenamefont {{Mukherjee}}\ and\ \citenamefont {{Wang}}(2003)}]{Mukherjee:2003ag}%
  \BibitemOpen
  \bibfield  {author} {\bibinfo {author} {\bibfnamefont {P.}~\bibnamefont {{Mukherjee}}}\ and\ \bibinfo {author} {\bibfnamefont {Y.}~\bibnamefont {{Wang}}},\ }\href {\doibase 10.1086/379161} {\bibfield  {journal} {\bibinfo  {journal} {apj}\ }\textbf {\bibinfo {volume} {599}},\ \bibinfo {pages} {1} (\bibinfo {year} {2003})},\ \Eprint {http://arxiv.org/abs/astro-ph/0303211} {arXiv:astro-ph/0303211 [astro-ph]} \BibitemShut {NoStop}%
\bibitem [{\citenamefont {{Covi}}\ \emph {et~al.}(2006)\citenamefont {{Covi}}, \citenamefont {{Hamann}}, \citenamefont {{Melchiorri}}, \citenamefont {{Slosar}},\ and\ \citenamefont {{Sorbera}}}]{Covi:2006ci}%
  \BibitemOpen
  \bibfield  {author} {\bibinfo {author} {\bibfnamefont {L.}~\bibnamefont {{Covi}}}, \bibinfo {author} {\bibfnamefont {J.}~\bibnamefont {{Hamann}}}, \bibinfo {author} {\bibfnamefont {A.}~\bibnamefont {{Melchiorri}}}, \bibinfo {author} {\bibfnamefont {A.}~\bibnamefont {{Slosar}}}, \ and\ \bibinfo {author} {\bibfnamefont {I.}~\bibnamefont {{Sorbera}}},\ }\href {\doibase 10.1103/PhysRevD.74.083509} {\bibfield  {journal} {\bibinfo  {journal} {Physics Review D}\ }\textbf {\bibinfo {volume} {74}},\ \bibinfo {eid} {083509} (\bibinfo {year} {2006})},\ \Eprint {http://arxiv.org/abs/astro-ph/0606452} {arXiv:astro-ph/0606452 [astro-ph]} \BibitemShut {NoStop}%
\bibitem [{\citenamefont {{Hamann}}\ \emph {et~al.}(2007)\citenamefont {{Hamann}}, \citenamefont {{Covi}}, \citenamefont {{Melchiorri}},\ and\ \citenamefont {{Slosar}}}]{Hamann:2007pa}%
  \BibitemOpen
  \bibfield  {author} {\bibinfo {author} {\bibfnamefont {J.}~\bibnamefont {{Hamann}}}, \bibinfo {author} {\bibfnamefont {L.}~\bibnamefont {{Covi}}}, \bibinfo {author} {\bibfnamefont {A.}~\bibnamefont {{Melchiorri}}}, \ and\ \bibinfo {author} {\bibfnamefont {A.}~\bibnamefont {{Slosar}}},\ }\href {\doibase 10.1103/PhysRevD.76.023503} {\bibfield  {journal} {\bibinfo  {journal} {Physics Review D}\ }\textbf {\bibinfo {volume} {76}},\ \bibinfo {eid} {023503} (\bibinfo {year} {2007})},\ \Eprint {http://arxiv.org/abs/astro-ph/0701380} {arXiv:astro-ph/0701380 [astro-ph]} \BibitemShut {NoStop}%
\bibitem [{\citenamefont {{Meerburg}}\ \emph {et~al.}(2012)\citenamefont {{Meerburg}}, \citenamefont {{Wijers}},\ and\ \citenamefont {{van der Schaar}}}]{Meerburg:2011gd}%
  \BibitemOpen
  \bibfield  {author} {\bibinfo {author} {\bibfnamefont {P.~D.}\ \bibnamefont {{Meerburg}}}, \bibinfo {author} {\bibfnamefont {R.~A.~M.~J.}\ \bibnamefont {{Wijers}}}, \ and\ \bibinfo {author} {\bibfnamefont {J.~P.}\ \bibnamefont {{van der Schaar}}},\ }\href {\doibase 10.1111/j.1365-2966.2011.20311.x} {\bibfield  {journal} {\bibinfo  {journal} {mnras}\ }\textbf {\bibinfo {volume} {421}},\ \bibinfo {pages} {369} (\bibinfo {year} {2012})},\ \Eprint {http://arxiv.org/abs/1109.5264} {arXiv:1109.5264 [astro-ph.CO]} \BibitemShut {NoStop}%
\bibitem [{\citenamefont {{Meerburg}}\ \emph {et~al.}(2014)\citenamefont {{Meerburg}}, \citenamefont {{Spergel}},\ and\ \citenamefont {{Wandelt}}}]{Meerburg:2013dla}%
  \BibitemOpen
  \bibfield  {author} {\bibinfo {author} {\bibfnamefont {P.~D.}\ \bibnamefont {{Meerburg}}}, \bibinfo {author} {\bibfnamefont {D.~N.}\ \bibnamefont {{Spergel}}}, \ and\ \bibinfo {author} {\bibfnamefont {B.~D.}\ \bibnamefont {{Wandelt}}},\ }\href {\doibase 10.1103/PhysRevD.89.063537} {\bibfield  {journal} {\bibinfo  {journal} {Physics Review D}\ }\textbf {\bibinfo {volume} {89}},\ \bibinfo {eid} {063537} (\bibinfo {year} {2014})},\ \Eprint {http://arxiv.org/abs/1308.3705} {arXiv:1308.3705 [astro-ph.CO]} \BibitemShut {NoStop}%
\bibitem [{\citenamefont {{Benetti}}(2013)}]{Benetti:2013cja}%
  \BibitemOpen
  \bibfield  {author} {\bibinfo {author} {\bibfnamefont {M.}~\bibnamefont {{Benetti}}},\ }\href {\doibase 10.1103/PhysRevD.88.087302} {\bibfield  {journal} {\bibinfo  {journal} {Physics Review D}\ }\textbf {\bibinfo {volume} {88}},\ \bibinfo {eid} {087302} (\bibinfo {year} {2013})},\ \Eprint {http://arxiv.org/abs/1308.6406} {arXiv:1308.6406 [astro-ph.CO]} \BibitemShut {NoStop}%
\bibitem [{\citenamefont {{Miranda}}\ and\ \citenamefont {{Hu}}(2014)}]{Miranda:2013wxa}%
  \BibitemOpen
  \bibfield  {author} {\bibinfo {author} {\bibfnamefont {V.}~\bibnamefont {{Miranda}}}\ and\ \bibinfo {author} {\bibfnamefont {W.}~\bibnamefont {{Hu}}},\ }\href {\doibase 10.1103/PhysRevD.89.083529} {\bibfield  {journal} {\bibinfo  {journal} {Physics Review D}\ }\textbf {\bibinfo {volume} {89}},\ \bibinfo {eid} {083529} (\bibinfo {year} {2014})},\ \Eprint {http://arxiv.org/abs/1312.0946} {arXiv:1312.0946 [astro-ph.CO]} \BibitemShut {NoStop}%
\bibitem [{\citenamefont {{Easther}}\ and\ \citenamefont {{Flauger}}(2014)}]{Easther:2013kla}%
  \BibitemOpen
  \bibfield  {author} {\bibinfo {author} {\bibfnamefont {R.}~\bibnamefont {{Easther}}}\ and\ \bibinfo {author} {\bibfnamefont {R.}~\bibnamefont {{Flauger}}},\ }\href {\doibase 10.1088/1475-7516/2014/02/037} {\bibfield  {journal} {\bibinfo  {journal} {jcap}\ }\textbf {\bibinfo {volume} {2014}},\ \bibinfo {eid} {037} (\bibinfo {year} {2014})},\ \Eprint {http://arxiv.org/abs/1308.3736} {arXiv:1308.3736 [astro-ph.CO]} \BibitemShut {NoStop}%
\bibitem [{\citenamefont {{Chen}}\ and\ \citenamefont {{Namjoo}}(2014)}]{Chen:2014joa}%
  \BibitemOpen
  \bibfield  {author} {\bibinfo {author} {\bibfnamefont {X.}~\bibnamefont {{Chen}}}\ and\ \bibinfo {author} {\bibfnamefont {M.~H.}\ \bibnamefont {{Namjoo}}},\ }\href {\doibase 10.1016/j.physletb.2014.11.002} {\bibfield  {journal} {\bibinfo  {journal} {Physics Letters B}\ }\textbf {\bibinfo {volume} {739}},\ \bibinfo {pages} {285} (\bibinfo {year} {2014})},\ \Eprint {http://arxiv.org/abs/1404.1536} {arXiv:1404.1536 [astro-ph.CO]} \BibitemShut {NoStop}%
\bibitem [{\citenamefont {{Hazra}}\ \emph {et~al.}(2014{\natexlab{a}})\citenamefont {{Hazra}}, \citenamefont {{Shafieloo}}, \citenamefont {{Smoot}},\ and\ \citenamefont {{Starobinsky}}}]{Hazra:2014goa}%
  \BibitemOpen
  \bibfield  {author} {\bibinfo {author} {\bibfnamefont {D.~K.}\ \bibnamefont {{Hazra}}}, \bibinfo {author} {\bibfnamefont {A.}~\bibnamefont {{Shafieloo}}}, \bibinfo {author} {\bibfnamefont {G.~F.}\ \bibnamefont {{Smoot}}}, \ and\ \bibinfo {author} {\bibfnamefont {A.~A.}\ \bibnamefont {{Starobinsky}}},\ }\href {\doibase 10.1088/1475-7516/2014/08/048} {\bibfield  {journal} {\bibinfo  {journal} {jcap}\ }\textbf {\bibinfo {volume} {2014}},\ \bibinfo {pages} {048} (\bibinfo {year} {2014}{\natexlab{a}})},\ \Eprint {http://arxiv.org/abs/1405.2012} {arXiv:1405.2012 [astro-ph.CO]} \BibitemShut {NoStop}%
\bibitem [{\citenamefont {{Hazra}}\ \emph {et~al.}(2014{\natexlab{b}})\citenamefont {{Hazra}}, \citenamefont {{Shafieloo}},\ and\ \citenamefont {{Souradeep}}}]{Hazra:2014jwa}%
  \BibitemOpen
  \bibfield  {author} {\bibinfo {author} {\bibfnamefont {D.~K.}\ \bibnamefont {{Hazra}}}, \bibinfo {author} {\bibfnamefont {A.}~\bibnamefont {{Shafieloo}}}, \ and\ \bibinfo {author} {\bibfnamefont {T.}~\bibnamefont {{Souradeep}}},\ }\href {\doibase 10.1088/1475-7516/2014/11/011} {\bibfield  {journal} {\bibinfo  {journal} {jcap}\ }\textbf {\bibinfo {volume} {2014}},\ \bibinfo {pages} {011} (\bibinfo {year} {2014}{\natexlab{b}})},\ \Eprint {http://arxiv.org/abs/1406.4827} {arXiv:1406.4827 [astro-ph.CO]} \BibitemShut {NoStop}%
\bibitem [{\citenamefont {{Hu}}\ and\ \citenamefont {{Torrado}}(2015)}]{Hu:2014hra}%
  \BibitemOpen
  \bibfield  {author} {\bibinfo {author} {\bibfnamefont {B.}~\bibnamefont {{Hu}}}\ and\ \bibinfo {author} {\bibfnamefont {J.}~\bibnamefont {{Torrado}}},\ }\href {\doibase 10.1103/PhysRevD.91.064039} {\bibfield  {journal} {\bibinfo  {journal} {Physics Review D}\ }\textbf {\bibinfo {volume} {91}},\ \bibinfo {eid} {064039} (\bibinfo {year} {2015})},\ \Eprint {http://arxiv.org/abs/1410.4804} {arXiv:1410.4804 [astro-ph.CO]} \BibitemShut {NoStop}%
\bibitem [{\citenamefont {Collaboration}(2016)}]{Ade:2015lrj}%
  \BibitemOpen
  \bibfield  {author} {\bibinfo {author} {\bibfnamefont {P.}~\bibnamefont {Collaboration}},\ }\href {\doibase 10.1051/0004-6361/201525898} {\bibfield  {journal} {\bibinfo  {journal} {aap}\ }\textbf {\bibinfo {volume} {594}},\ \bibinfo {eid} {A20} (\bibinfo {year} {2016})},\ \Eprint {http://arxiv.org/abs/1502.02114} {arXiv:1502.02114 [astro-ph.CO]} \BibitemShut {NoStop}%
\bibitem [{\citenamefont {{Gruppuso}}\ and\ \citenamefont {{Sagnotti}}(2015)}]{Gruppuso:2015zia}%
  \BibitemOpen
  \bibfield  {author} {\bibinfo {author} {\bibfnamefont {A.}~\bibnamefont {{Gruppuso}}}\ and\ \bibinfo {author} {\bibfnamefont {A.}~\bibnamefont {{Sagnotti}}},\ }\href {\doibase 10.1142/S0218271815440083} {\bibfield  {journal} {\bibinfo  {journal} {International Journal of Modern Physics D}\ }\textbf {\bibinfo {volume} {24}},\ \bibinfo {eid} {1544008-469} (\bibinfo {year} {2015})},\ \Eprint {http://arxiv.org/abs/1506.08093} {arXiv:1506.08093 [astro-ph.CO]} \BibitemShut {NoStop}%
\bibitem [{\citenamefont {{Gruppuso}}\ \emph {et~al.}(2016)\citenamefont {{Gruppuso}}, \citenamefont {{Kitazawa}}, \citenamefont {{Mandolesi}}, \citenamefont {{Natoli}},\ and\ \citenamefont {{Sagnotti}}}]{Gruppuso:2015xqa}%
  \BibitemOpen
  \bibfield  {author} {\bibinfo {author} {\bibfnamefont {A.}~\bibnamefont {{Gruppuso}}}, \bibinfo {author} {\bibfnamefont {N.}~\bibnamefont {{Kitazawa}}}, \bibinfo {author} {\bibfnamefont {N.}~\bibnamefont {{Mandolesi}}}, \bibinfo {author} {\bibfnamefont {P.}~\bibnamefont {{Natoli}}}, \ and\ \bibinfo {author} {\bibfnamefont {A.}~\bibnamefont {{Sagnotti}}},\ }\href {\doibase 10.1016/j.dark.2015.12.001} {\bibfield  {journal} {\bibinfo  {journal} {Physics of the Dark Universe}\ }\textbf {\bibinfo {volume} {11}},\ \bibinfo {pages} {68} (\bibinfo {year} {2016})},\ \Eprint {http://arxiv.org/abs/1508.00411} {arXiv:1508.00411 [astro-ph.CO]} \BibitemShut {NoStop}%
\bibitem [{\citenamefont {{Hazra}}\ \emph {et~al.}(2016)\citenamefont {{Hazra}}, \citenamefont {{Shafieloo}}, \citenamefont {{Smoot}},\ and\ \citenamefont {{Starobinsky}}}]{Hazra:2016fkm}%
  \BibitemOpen
  \bibfield  {author} {\bibinfo {author} {\bibfnamefont {D.~K.}\ \bibnamefont {{Hazra}}}, \bibinfo {author} {\bibfnamefont {A.}~\bibnamefont {{Shafieloo}}}, \bibinfo {author} {\bibfnamefont {G.~F.}\ \bibnamefont {{Smoot}}}, \ and\ \bibinfo {author} {\bibfnamefont {A.~A.}\ \bibnamefont {{Starobinsky}}},\ }\href {\doibase 10.1088/1475-7516/2016/09/009} {\bibfield  {journal} {\bibinfo  {journal} {JCAP}\ }\textbf {\bibinfo {volume} {2016}},\ \bibinfo {eid} {009} (\bibinfo {year} {2016})},\ \Eprint {http://arxiv.org/abs/1605.02106} {arXiv:1605.02106 [astro-ph.CO]} \BibitemShut {NoStop}%
\bibitem [{\citenamefont {{Ballardini}}(2019)}]{Ballardini:2018noo}%
  \BibitemOpen
  \bibfield  {author} {\bibinfo {author} {\bibfnamefont {M.}~\bibnamefont {{Ballardini}}},\ }\href {\doibase 10.1016/j.dark.2018.11.006} {\bibfield  {journal} {\bibinfo  {journal} {Physics of the Dark Universe}\ }\textbf {\bibinfo {volume} {23}},\ \bibinfo {eid} {100245} (\bibinfo {year} {2019})},\ \Eprint {http://arxiv.org/abs/1807.05521} {arXiv:1807.05521 [astro-ph.CO]} \BibitemShut {NoStop}%
\bibitem [{\citenamefont {{Zeng}}\ \emph {et~al.}(2019)\citenamefont {{Zeng}}, \citenamefont {{Kovetz}}, \citenamefont {{Chen}}, \citenamefont {{Gong}}, \citenamefont {{Mu{\~n}oz}},\ and\ \citenamefont {{Kamionkowski}}}]{Zeng:2018ufm}%
  \BibitemOpen
  \bibfield  {author} {\bibinfo {author} {\bibfnamefont {C.}~\bibnamefont {{Zeng}}}, \bibinfo {author} {\bibfnamefont {E.~D.}\ \bibnamefont {{Kovetz}}}, \bibinfo {author} {\bibfnamefont {X.}~\bibnamefont {{Chen}}}, \bibinfo {author} {\bibfnamefont {Y.}~\bibnamefont {{Gong}}}, \bibinfo {author} {\bibfnamefont {J.~B.}\ \bibnamefont {{Mu{\~n}oz}}}, \ and\ \bibinfo {author} {\bibfnamefont {M.}~\bibnamefont {{Kamionkowski}}},\ }\href {\doibase 10.1103/PhysRevD.99.043517} {\bibfield  {journal} {\bibinfo  {journal} {Physics Review D}\ }\textbf {\bibinfo {volume} {99}},\ \bibinfo {eid} {043517} (\bibinfo {year} {2019})},\ \Eprint {http://arxiv.org/abs/1812.05105} {arXiv:1812.05105 [astro-ph.CO]} \BibitemShut {NoStop}%
\bibitem [{\citenamefont {{Ca{\~n}as-Herrera}}\ \emph {et~al.}(2021)\citenamefont {{Ca{\~n}as-Herrera}}, \citenamefont {{Torrado}},\ and\ \citenamefont {{Ach{\'u}carro}}}]{Canas-Herrera:2020mme}%
  \BibitemOpen
  \bibfield  {author} {\bibinfo {author} {\bibfnamefont {G.}~\bibnamefont {{Ca{\~n}as-Herrera}}}, \bibinfo {author} {\bibfnamefont {J.}~\bibnamefont {{Torrado}}}, \ and\ \bibinfo {author} {\bibfnamefont {A.}~\bibnamefont {{Ach{\'u}carro}}},\ }\href {\doibase 10.1103/PhysRevD.103.123531} {\bibfield  {journal} {\bibinfo  {journal} {Physics Review D}\ }\textbf {\bibinfo {volume} {103}},\ \bibinfo {eid} {123531} (\bibinfo {year} {2021})},\ \Eprint {http://arxiv.org/abs/2012.04640} {arXiv:2012.04640 [astro-ph.CO]} \BibitemShut {NoStop}%
\bibitem [{\citenamefont {{Braglia}}\ \emph {et~al.}(2021)\citenamefont {{Braglia}}, \citenamefont {{Chen}},\ and\ \citenamefont {{Hazra}}}]{Braglia:2021ckn}%
  \BibitemOpen
  \bibfield  {author} {\bibinfo {author} {\bibfnamefont {M.}~\bibnamefont {{Braglia}}}, \bibinfo {author} {\bibfnamefont {X.}~\bibnamefont {{Chen}}}, \ and\ \bibinfo {author} {\bibfnamefont {D.~K.}\ \bibnamefont {{Hazra}}},\ }\href {\doibase 10.1088/1475-7516/2021/06/005} {\bibfield  {journal} {\bibinfo  {journal} {JCAP}\ }\textbf {\bibinfo {volume} {2021}},\ \bibinfo {eid} {005} (\bibinfo {year} {2021})},\ \Eprint {http://arxiv.org/abs/2103.03025} {arXiv:2103.03025 [astro-ph.CO]} \BibitemShut {NoStop}%
\bibitem [{\citenamefont {{Braglia}}\ \emph {et~al.}(2022{\natexlab{a}})\citenamefont {{Braglia}}, \citenamefont {{Chen}},\ and\ \citenamefont {{Hazra}}}]{Braglia:2021sun}%
  \BibitemOpen
  \bibfield  {author} {\bibinfo {author} {\bibfnamefont {M.}~\bibnamefont {{Braglia}}}, \bibinfo {author} {\bibfnamefont {X.}~\bibnamefont {{Chen}}}, \ and\ \bibinfo {author} {\bibfnamefont {D.~K.}\ \bibnamefont {{Hazra}}},\ }\href {\doibase 10.1140/epjc/s10052-022-10461-3} {\bibfield  {journal} {\bibinfo  {journal} {European Physical Journal C}\ }\textbf {\bibinfo {volume} {82}},\ \bibinfo {eid} {498} (\bibinfo {year} {2022}{\natexlab{a}})},\ \Eprint {http://arxiv.org/abs/2106.07546} {arXiv:2106.07546 [astro-ph.CO]} \BibitemShut {NoStop}%
\bibitem [{\citenamefont {{Naik}}\ \emph {et~al.}(2022)\citenamefont {{Naik}}, \citenamefont {{Furuuchi}},\ and\ \citenamefont {{Chingangbam}}}]{Naik:2022mxn}%
  \BibitemOpen
  \bibfield  {author} {\bibinfo {author} {\bibfnamefont {S.~S.}\ \bibnamefont {{Naik}}}, \bibinfo {author} {\bibfnamefont {K.}~\bibnamefont {{Furuuchi}}}, \ and\ \bibinfo {author} {\bibfnamefont {P.}~\bibnamefont {{Chingangbam}}},\ }\href {\doibase 10.1088/1475-7516/2022/07/016} {\bibfield  {journal} {\bibinfo  {journal} {JCAP}\ }\textbf {\bibinfo {volume} {2022}},\ \bibinfo {eid} {016} (\bibinfo {year} {2022})},\ \Eprint {http://arxiv.org/abs/2202.05862} {arXiv:2202.05862 [astro-ph.CO]} \BibitemShut {NoStop}%
\bibitem [{\citenamefont {{Hamann}}\ and\ \citenamefont {{Wons}}(2022)}]{Hamann:2021eyw}%
  \BibitemOpen
  \bibfield  {author} {\bibinfo {author} {\bibfnamefont {J.}~\bibnamefont {{Hamann}}}\ and\ \bibinfo {author} {\bibfnamefont {J.}~\bibnamefont {{Wons}}},\ }\href {\doibase 10.1088/1475-7516/2022/03/036} {\bibfield  {journal} {\bibinfo  {journal} {JCAP}\ }\textbf {\bibinfo {volume} {2022}},\ \bibinfo {eid} {036} (\bibinfo {year} {2022})},\ \Eprint {http://arxiv.org/abs/2112.08571} {arXiv:2112.08571 [astro-ph.CO]} \BibitemShut {NoStop}%
\bibitem [{\citenamefont {Gallego~Cadavid}\ \emph {et~al.}(2017)\citenamefont {Gallego~Cadavid}, \citenamefont {Romano},\ and\ \citenamefont {Gariazzo}}]{GallegoCadavid:2016wcz}%
  \BibitemOpen
  \bibfield  {author} {\bibinfo {author} {\bibfnamefont {A.}~\bibnamefont {Gallego~Cadavid}}, \bibinfo {author} {\bibfnamefont {A.~E.}\ \bibnamefont {Romano}}, \ and\ \bibinfo {author} {\bibfnamefont {S.}~\bibnamefont {Gariazzo}},\ }\href {\doibase 10.1140/epjc/s10052-017-4797-6} {\bibfield  {journal} {\bibinfo  {journal} {Eur. Phys. J. C}\ }\textbf {\bibinfo {volume} {77}},\ \bibinfo {pages} {242} (\bibinfo {year} {2017})},\ \Eprint {http://arxiv.org/abs/1612.03490} {arXiv:1612.03490 [astro-ph.CO]} \BibitemShut {NoStop}%
\bibitem [{\citenamefont {{Dom{\`e}nech}}\ and\ \citenamefont {{Kamionkowski}}(2019)}]{Domenech:2019cyh}%
  \BibitemOpen
  \bibfield  {author} {\bibinfo {author} {\bibfnamefont {G.}~\bibnamefont {{Dom{\`e}nech}}}\ and\ \bibinfo {author} {\bibfnamefont {M.}~\bibnamefont {{Kamionkowski}}},\ }\href {\doibase 10.1088/1475-7516/2019/11/040} {\bibfield  {journal} {\bibinfo  {journal} {JCAP}\ }\textbf {\bibinfo {volume} {2019}},\ \bibinfo {eid} {040} (\bibinfo {year} {2019})},\ \Eprint {http://arxiv.org/abs/1905.04323} {arXiv:1905.04323 [astro-ph.CO]} \BibitemShut {NoStop}%
\bibitem [{\citenamefont {{Dom{\`e}nech}}\ \emph {et~al.}(2020)\citenamefont {{Dom{\`e}nech}}, \citenamefont {{Chen}}, \citenamefont {{Kamionkowski}},\ and\ \citenamefont {{Loeb}}}]{Domenech:2020qay}%
  \BibitemOpen
  \bibfield  {author} {\bibinfo {author} {\bibfnamefont {G.}~\bibnamefont {{Dom{\`e}nech}}}, \bibinfo {author} {\bibfnamefont {X.}~\bibnamefont {{Chen}}}, \bibinfo {author} {\bibfnamefont {M.}~\bibnamefont {{Kamionkowski}}}, \ and\ \bibinfo {author} {\bibfnamefont {A.}~\bibnamefont {{Loeb}}},\ }\href {\doibase 10.1088/1475-7516/2020/10/005} {\bibfield  {journal} {\bibinfo  {journal} {JCAP}\ }\textbf {\bibinfo {volume} {2020}},\ \bibinfo {eid} {005} (\bibinfo {year} {2020})},\ \Eprint {http://arxiv.org/abs/2005.08998} {arXiv:2005.08998 [astro-ph.CO]} \BibitemShut {NoStop}%
\bibitem [{\citenamefont {{Braglia}}\ \emph {et~al.}(2022{\natexlab{b}})\citenamefont {{Braglia}}, \citenamefont {{Chen}},\ and\ \citenamefont {{Hazra}}}]{Braglia:2021rej}%
  \BibitemOpen
  \bibfield  {author} {\bibinfo {author} {\bibfnamefont {M.}~\bibnamefont {{Braglia}}}, \bibinfo {author} {\bibfnamefont {X.}~\bibnamefont {{Chen}}}, \ and\ \bibinfo {author} {\bibfnamefont {D.~K.}\ \bibnamefont {{Hazra}}},\ }\href {\doibase 10.1103/PhysRevD.105.103523} {\bibfield  {journal} {\bibinfo  {journal} {Physics Review D}\ }\textbf {\bibinfo {volume} {105}},\ \bibinfo {eid} {103523} (\bibinfo {year} {2022}{\natexlab{b}})},\ \Eprint {http://arxiv.org/abs/2108.10110} {arXiv:2108.10110 [astro-ph.CO]} \BibitemShut {NoStop}%
\bibitem [{\citenamefont {{Ballardini}}\ and\ \citenamefont {{Finelli}}(2022)}]{Ballardini:2022vzh}%
  \BibitemOpen
  \bibfield  {author} {\bibinfo {author} {\bibfnamefont {M.}~\bibnamefont {{Ballardini}}}\ and\ \bibinfo {author} {\bibfnamefont {F.}~\bibnamefont {{Finelli}}},\ }\href {\doibase 10.1088/1475-7516/2022/10/083} {\bibfield  {journal} {\bibinfo  {journal} {JCAP}\ }\textbf {\bibinfo {volume} {2022}},\ \bibinfo {eid} {083} (\bibinfo {year} {2022})},\ \Eprint {http://arxiv.org/abs/2207.14410} {arXiv:2207.14410 [astro-ph.CO]} \BibitemShut {NoStop}%
\bibitem [{\citenamefont {Hazra}\ \emph {et~al.}(2022)\citenamefont {Hazra}, \citenamefont {Antony},\ and\ \citenamefont {Shafieloo}}]{Hazra:2022rdl}%
  \BibitemOpen
  \bibfield  {author} {\bibinfo {author} {\bibfnamefont {D.~K.}\ \bibnamefont {Hazra}}, \bibinfo {author} {\bibfnamefont {A.}~\bibnamefont {Antony}}, \ and\ \bibinfo {author} {\bibfnamefont {A.}~\bibnamefont {Shafieloo}},\ }\href {\doibase 10.1088/1475-7516/2022/08/063} {\bibfield  {journal} {\bibinfo  {journal} {JCAP}\ }\textbf {\bibinfo {volume} {08}},\ \bibinfo {pages} {063} (\bibinfo {year} {2022})},\ \Eprint {http://arxiv.org/abs/2201.12000} {arXiv:2201.12000 [astro-ph.CO]} \BibitemShut {NoStop}%
\bibitem [{\citenamefont {Aiola}\ \emph {et~al.}(2020)\citenamefont {Aiola} \emph {et~al.}}]{ACT:2020gnv}%
  \BibitemOpen
  \bibfield  {author} {\bibinfo {author} {\bibfnamefont {S.}~\bibnamefont {Aiola}} \emph {et~al.} (\bibinfo {collaboration} {ACT}),\ }\href {\doibase 10.1088/1475-7516/2020/12/047} {\bibfield  {journal} {\bibinfo  {journal} {JCAP}\ }\textbf {\bibinfo {volume} {12}},\ \bibinfo {pages} {047} (\bibinfo {year} {2020})},\ \Eprint {http://arxiv.org/abs/2007.07288} {arXiv:2007.07288 [astro-ph.CO]} \BibitemShut {NoStop}%
\bibitem [{\citenamefont {Balkenhol}\ \emph {et~al.}(2023)\citenamefont {Balkenhol} \emph {et~al.}}]{SPT-3G:2022hvq}%
  \BibitemOpen
  \bibfield  {author} {\bibinfo {author} {\bibfnamefont {L.}~\bibnamefont {Balkenhol}} \emph {et~al.} (\bibinfo {collaboration} {SPT-3G}),\ }\href {\doibase 10.1103/PhysRevD.108.023510} {\bibfield  {journal} {\bibinfo  {journal} {Phys. Rev. D}\ }\textbf {\bibinfo {volume} {108}},\ \bibinfo {pages} {023510} (\bibinfo {year} {2023})},\ \Eprint {http://arxiv.org/abs/2212.05642} {arXiv:2212.05642 [astro-ph.CO]} \BibitemShut {NoStop}%
\bibitem [{\citenamefont {Ade}\ \emph {et~al.}(2019)\citenamefont {Ade} \emph {et~al.}}]{SimonsObservatory:2018koc}%
  \BibitemOpen
  \bibfield  {author} {\bibinfo {author} {\bibfnamefont {P.}~\bibnamefont {Ade}} \emph {et~al.} (\bibinfo {collaboration} {Simons Observatory}),\ }\href {\doibase 10.1088/1475-7516/2019/02/056} {\bibfield  {journal} {\bibinfo  {journal} {JCAP}\ }\textbf {\bibinfo {volume} {02}},\ \bibinfo {pages} {056} (\bibinfo {year} {2019})},\ \Eprint {http://arxiv.org/abs/1808.07445} {arXiv:1808.07445 [astro-ph.CO]} \BibitemShut {NoStop}%
\bibitem [{\citenamefont {Hazumi}\ \emph {et~al.}(2020)\citenamefont {Hazumi} \emph {et~al.}}]{LiteBIRD:2020khw}%
  \BibitemOpen
  \bibfield  {author} {\bibinfo {author} {\bibfnamefont {M.}~\bibnamefont {Hazumi}} \emph {et~al.} (\bibinfo {collaboration} {LiteBIRD}),\ }\href {\doibase 10.1117/12.2563050} {\bibfield  {journal} {\bibinfo  {journal} {Proc. SPIE Int. Soc. Opt. Eng.}\ }\textbf {\bibinfo {volume} {11443}},\ \bibinfo {pages} {114432F} (\bibinfo {year} {2020})},\ \Eprint {http://arxiv.org/abs/2101.12449} {arXiv:2101.12449 [astro-ph.IM]} \BibitemShut {NoStop}%
\bibitem [{\citenamefont {Abazajian}\ \emph {et~al.}(2019)\citenamefont {Abazajian} \emph {et~al.}}]{Abazajian:2019eic}%
  \BibitemOpen
  \bibfield  {author} {\bibinfo {author} {\bibfnamefont {K.}~\bibnamefont {Abazajian}} \emph {et~al.},\ }\href@noop {} {\  (\bibinfo {year} {2019})},\ \Eprint {http://arxiv.org/abs/1907.04473} {arXiv:1907.04473 [astro-ph.IM]} \BibitemShut {NoStop}%
\bibitem [{\citenamefont {Kogut}\ \emph {et~al.}(2024)\citenamefont {Kogut} \emph {et~al.}}]{Kogut:2024vbi}%
  \BibitemOpen
  \bibfield  {author} {\bibinfo {author} {\bibfnamefont {A.}~\bibnamefont {Kogut}} \emph {et~al.},\ }\href@noop {} {\  (\bibinfo {year} {2024})},\ \Eprint {http://arxiv.org/abs/2405.20403} {arXiv:2405.20403 [astro-ph.CO]} \BibitemShut {NoStop}%
\bibitem [{\citenamefont {Andr\'e}\ \emph {et~al.}(2014)\citenamefont {Andr\'e} \emph {et~al.}}]{PRISM:2013fvg}%
  \BibitemOpen
  \bibfield  {author} {\bibinfo {author} {\bibfnamefont {P.}~\bibnamefont {Andr\'e}} \emph {et~al.} (\bibinfo {collaboration} {PRISM}),\ }\href {\doibase 10.1088/1475-7516/2014/02/006} {\bibfield  {journal} {\bibinfo  {journal} {JCAP}\ }\textbf {\bibinfo {volume} {02}},\ \bibinfo {pages} {006} (\bibinfo {year} {2014})},\ \Eprint {http://arxiv.org/abs/1310.1554} {arXiv:1310.1554 [astro-ph.CO]} \BibitemShut {NoStop}%
\bibitem [{\citenamefont {Braglia}\ \emph {et~al.}(2023)\citenamefont {Braglia}, \citenamefont {Chen}, \citenamefont {Hazra},\ and\ \citenamefont {Pinol}}]{Braglia:2022ftm}%
  \BibitemOpen
  \bibfield  {author} {\bibinfo {author} {\bibfnamefont {M.}~\bibnamefont {Braglia}}, \bibinfo {author} {\bibfnamefont {X.}~\bibnamefont {Chen}}, \bibinfo {author} {\bibfnamefont {D.~K.}\ \bibnamefont {Hazra}}, \ and\ \bibinfo {author} {\bibfnamefont {L.}~\bibnamefont {Pinol}},\ }\href {\doibase 10.1088/1475-7516/2023/03/014} {\bibfield  {journal} {\bibinfo  {journal} {JCAP}\ }\textbf {\bibinfo {volume} {03}},\ \bibinfo {pages} {014} (\bibinfo {year} {2023})},\ \Eprint {http://arxiv.org/abs/2210.07028} {arXiv:2210.07028 [astro-ph.CO]} \BibitemShut {NoStop}%
\bibitem [{\citenamefont {Petretti}\ \emph {et~al.}(2024)\citenamefont {Petretti}, \citenamefont {Braglia}, \citenamefont {Chen}, \citenamefont {Hazra},\ and\ \citenamefont {Paban}}]{Petretti:2024mjy}%
  \BibitemOpen
  \bibfield  {author} {\bibinfo {author} {\bibfnamefont {C.}~\bibnamefont {Petretti}}, \bibinfo {author} {\bibfnamefont {M.}~\bibnamefont {Braglia}}, \bibinfo {author} {\bibfnamefont {X.}~\bibnamefont {Chen}}, \bibinfo {author} {\bibfnamefont {D.~K.}\ \bibnamefont {Hazra}}, \ and\ \bibinfo {author} {\bibfnamefont {S.}~\bibnamefont {Paban}},\ }\href@noop {} {\  (\bibinfo {year} {2024})},\ \Eprint {http://arxiv.org/abs/2411.03459} {arXiv:2411.03459 [astro-ph.CO]} \BibitemShut {NoStop}%
\bibitem [{\citenamefont {Riess}\ \emph {et~al.}(2022)\citenamefont {Riess} \emph {et~al.}}]{Riess:2021jrx}%
  \BibitemOpen
  \bibfield  {author} {\bibinfo {author} {\bibfnamefont {A.~G.}\ \bibnamefont {Riess}} \emph {et~al.},\ }\href {\doibase 10.3847/2041-8213/ac5c5b} {\bibfield  {journal} {\bibinfo  {journal} {Astrophys. J. Lett.}\ }\textbf {\bibinfo {volume} {934}},\ \bibinfo {pages} {L7} (\bibinfo {year} {2022})},\ \Eprint {http://arxiv.org/abs/2112.04510} {arXiv:2112.04510 [astro-ph.CO]} \BibitemShut {NoStop}%
\bibitem [{\citenamefont {Aghanim}\ \emph {et~al.}(2020)\citenamefont {Aghanim} \emph {et~al.}}]{Planck:2018vyg}%
  \BibitemOpen
  \bibfield  {author} {\bibinfo {author} {\bibfnamefont {N.}~\bibnamefont {Aghanim}} \emph {et~al.} (\bibinfo {collaboration} {Planck}),\ }\href {\doibase 10.1051/0004-6361/201833910} {\bibfield  {journal} {\bibinfo  {journal} {Astron. Astrophys.}\ }\textbf {\bibinfo {volume} {641}},\ \bibinfo {pages} {A6} (\bibinfo {year} {2020})},\ \bibinfo {note} {[Erratum: Astron.Astrophys. 652, C4 (2021)]},\ \Eprint {http://arxiv.org/abs/1807.06209} {arXiv:1807.06209 [astro-ph.CO]} \BibitemShut {NoStop}%
\bibitem [{\citenamefont {Asgari}\ \emph {et~al.}(2021)\citenamefont {Asgari} \emph {et~al.}}]{KiDS:2020suj}%
  \BibitemOpen
  \bibfield  {author} {\bibinfo {author} {\bibfnamefont {M.}~\bibnamefont {Asgari}} \emph {et~al.} (\bibinfo {collaboration} {KiDS}),\ }\href {\doibase 10.1051/0004-6361/202039070} {\bibfield  {journal} {\bibinfo  {journal} {Astron. Astrophys.}\ }\textbf {\bibinfo {volume} {645}},\ \bibinfo {pages} {A104} (\bibinfo {year} {2021})},\ \Eprint {http://arxiv.org/abs/2007.15633} {arXiv:2007.15633 [astro-ph.CO]} \BibitemShut {NoStop}%
\bibitem [{\citenamefont {Abbott}\ \emph {et~al.}(2022)\citenamefont {Abbott} \emph {et~al.}}]{DES:2021wwk}%
  \BibitemOpen
  \bibfield  {author} {\bibinfo {author} {\bibfnamefont {T.~M.~C.}\ \bibnamefont {Abbott}} \emph {et~al.} (\bibinfo {collaboration} {DES}),\ }\href {\doibase 10.1103/PhysRevD.105.023520} {\bibfield  {journal} {\bibinfo  {journal} {Phys. Rev. D}\ }\textbf {\bibinfo {volume} {105}},\ \bibinfo {pages} {023520} (\bibinfo {year} {2022})},\ \Eprint {http://arxiv.org/abs/2105.13549} {arXiv:2105.13549 [astro-ph.CO]} \BibitemShut {NoStop}%
\bibitem [{\citenamefont {Hikage}\ \emph {et~al.}(2019)\citenamefont {Hikage} \emph {et~al.}}]{HSC:2018mrq}%
  \BibitemOpen
  \bibfield  {author} {\bibinfo {author} {\bibfnamefont {C.}~\bibnamefont {Hikage}} \emph {et~al.} (\bibinfo {collaboration} {HSC}),\ }\href {\doibase 10.1093/pasj/psz010} {\bibfield  {journal} {\bibinfo  {journal} {Publ. Astron. Soc. Jap.}\ }\textbf {\bibinfo {volume} {71}},\ \bibinfo {pages} {43} (\bibinfo {year} {2019})},\ \Eprint {http://arxiv.org/abs/1809.09148} {arXiv:1809.09148 [astro-ph.CO]} \BibitemShut {NoStop}%
\bibitem [{\citenamefont {Chabanier}\ \emph {et~al.}(2019)\citenamefont {Chabanier} \emph {et~al.}}]{eBOSS:2018qyj}%
  \BibitemOpen
  \bibfield  {author} {\bibinfo {author} {\bibfnamefont {S.}~\bibnamefont {Chabanier}} \emph {et~al.} (\bibinfo {collaboration} {eBOSS}),\ }\href {\doibase 10.1088/1475-7516/2019/07/017} {\bibfield  {journal} {\bibinfo  {journal} {JCAP}\ }\textbf {\bibinfo {volume} {07}},\ \bibinfo {pages} {017} (\bibinfo {year} {2019})},\ \Eprint {http://arxiv.org/abs/1812.03554} {arXiv:1812.03554 [astro-ph.CO]} \BibitemShut {NoStop}%
\bibitem [{\citenamefont {Palanque-Delabrouille}\ \emph {et~al.}(2020)\citenamefont {Palanque-Delabrouille}, \citenamefont {Y\`eche}, \citenamefont {Sch\"oneberg}, \citenamefont {Lesgourgues}, \citenamefont {Walther}, \citenamefont {Chabanier},\ and\ \citenamefont {Armengaud}}]{Palanque-Delabrouille:2019iyz}%
  \BibitemOpen
  \bibfield  {author} {\bibinfo {author} {\bibfnamefont {N.}~\bibnamefont {Palanque-Delabrouille}}, \bibinfo {author} {\bibfnamefont {C.}~\bibnamefont {Y\`eche}}, \bibinfo {author} {\bibfnamefont {N.}~\bibnamefont {Sch\"oneberg}}, \bibinfo {author} {\bibfnamefont {J.}~\bibnamefont {Lesgourgues}}, \bibinfo {author} {\bibfnamefont {M.}~\bibnamefont {Walther}}, \bibinfo {author} {\bibfnamefont {S.}~\bibnamefont {Chabanier}}, \ and\ \bibinfo {author} {\bibfnamefont {E.}~\bibnamefont {Armengaud}},\ }\href {\doibase 10.1088/1475-7516/2020/04/038} {\bibfield  {journal} {\bibinfo  {journal} {JCAP}\ }\textbf {\bibinfo {volume} {04}},\ \bibinfo {pages} {038} (\bibinfo {year} {2020})},\ \Eprint {http://arxiv.org/abs/1911.09073} {arXiv:1911.09073 [astro-ph.CO]} \BibitemShut {NoStop}%
\bibitem [{\citenamefont {Goldstein}\ \emph {et~al.}(2023)\citenamefont {Goldstein}, \citenamefont {Hill}, \citenamefont {Ir\v{s}i\v{c}},\ and\ \citenamefont {Sherwin}}]{Goldstein:2023gnw}%
  \BibitemOpen
  \bibfield  {author} {\bibinfo {author} {\bibfnamefont {S.}~\bibnamefont {Goldstein}}, \bibinfo {author} {\bibfnamefont {J.~C.}\ \bibnamefont {Hill}}, \bibinfo {author} {\bibfnamefont {V.}~\bibnamefont {Ir\v{s}i\v{c}}}, \ and\ \bibinfo {author} {\bibfnamefont {B.~D.}\ \bibnamefont {Sherwin}},\ }\href {\doibase 10.1103/PhysRevLett.131.201001} {\bibfield  {journal} {\bibinfo  {journal} {Phys. Rev. Lett.}\ }\textbf {\bibinfo {volume} {131}},\ \bibinfo {pages} {201001} (\bibinfo {year} {2023})},\ \Eprint {http://arxiv.org/abs/2303.00746} {arXiv:2303.00746 [astro-ph.CO]} \BibitemShut {NoStop}%
\bibitem [{\citenamefont {Rogers}\ and\ \citenamefont {Poulin}(2023)}]{Rogers:2023upm}%
  \BibitemOpen
  \bibfield  {author} {\bibinfo {author} {\bibfnamefont {K.~K.}\ \bibnamefont {Rogers}}\ and\ \bibinfo {author} {\bibfnamefont {V.}~\bibnamefont {Poulin}},\ }\href@noop {} {\  (\bibinfo {year} {2023})},\ \Eprint {http://arxiv.org/abs/2311.16377} {arXiv:2311.16377 [astro-ph.CO]} \BibitemShut {NoStop}%
\bibitem [{\citenamefont {Freedman}(2021)}]{Freedman:2021ahq}%
  \BibitemOpen
  \bibfield  {author} {\bibinfo {author} {\bibfnamefont {W.~L.}\ \bibnamefont {Freedman}},\ }\href {\doibase 10.3847/1538-4357/ac0e95} {\bibfield  {journal} {\bibinfo  {journal} {Astrophys. J.}\ }\textbf {\bibinfo {volume} {919}},\ \bibinfo {pages} {16} (\bibinfo {year} {2021})},\ \Eprint {http://arxiv.org/abs/2106.15656} {arXiv:2106.15656 [astro-ph.CO]} \BibitemShut {NoStop}%
\bibitem [{\citenamefont {Amon}\ \emph {et~al.}(2023)\citenamefont {Amon} \emph {et~al.}}]{Amon:2022ycy}%
  \BibitemOpen
  \bibfield  {author} {\bibinfo {author} {\bibfnamefont {A.}~\bibnamefont {Amon}} \emph {et~al.},\ }\href {\doibase 10.1093/mnras/stac2938} {\bibfield  {journal} {\bibinfo  {journal} {Mon. Not. Roy. Astron. Soc.}\ }\textbf {\bibinfo {volume} {518}},\ \bibinfo {pages} {477} (\bibinfo {year} {2023})},\ \Eprint {http://arxiv.org/abs/2202.07440} {arXiv:2202.07440 [astro-ph.CO]} \BibitemShut {NoStop}%
\bibitem [{\citenamefont {Aric\`o}\ \emph {et~al.}(2023)\citenamefont {Aric\`o}, \citenamefont {Angulo}, \citenamefont {Zennaro}, \citenamefont {Contreras}, \citenamefont {Chen},\ and\ \citenamefont {Hern\'andez-Monteagudo}}]{Arico:2023ocu}%
  \BibitemOpen
  \bibfield  {author} {\bibinfo {author} {\bibfnamefont {G.}~\bibnamefont {Aric\`o}}, \bibinfo {author} {\bibfnamefont {R.~E.}\ \bibnamefont {Angulo}}, \bibinfo {author} {\bibfnamefont {M.}~\bibnamefont {Zennaro}}, \bibinfo {author} {\bibfnamefont {S.}~\bibnamefont {Contreras}}, \bibinfo {author} {\bibfnamefont {A.}~\bibnamefont {Chen}}, \ and\ \bibinfo {author} {\bibfnamefont {C.}~\bibnamefont {Hern\'andez-Monteagudo}},\ }\href {\doibase 10.1051/0004-6361/202346539} {\bibfield  {journal} {\bibinfo  {journal} {Astron. Astrophys.}\ }\textbf {\bibinfo {volume} {678}},\ \bibinfo {pages} {A109} (\bibinfo {year} {2023})},\ \Eprint {http://arxiv.org/abs/2303.05537} {arXiv:2303.05537 [astro-ph.CO]} \BibitemShut {NoStop}%
\bibitem [{\citenamefont {Fernandez}\ \emph {et~al.}(2024)\citenamefont {Fernandez}, \citenamefont {Bird},\ and\ \citenamefont {Ho}}]{Fernandez:2023grg}%
  \BibitemOpen
  \bibfield  {author} {\bibinfo {author} {\bibfnamefont {M.~A.}\ \bibnamefont {Fernandez}}, \bibinfo {author} {\bibfnamefont {S.}~\bibnamefont {Bird}}, \ and\ \bibinfo {author} {\bibfnamefont {M.-F.}\ \bibnamefont {Ho}},\ }\href {\doibase 10.1088/1475-7516/2024/07/029} {\bibfield  {journal} {\bibinfo  {journal} {JCAP}\ }\textbf {\bibinfo {volume} {07}},\ \bibinfo {pages} {029} (\bibinfo {year} {2024})},\ \Eprint {http://arxiv.org/abs/2309.03943} {arXiv:2309.03943 [astro-ph.CO]} \BibitemShut {NoStop}%
\bibitem [{\citenamefont {Walther}\ \emph {et~al.}(2024)\citenamefont {Walther} \emph {et~al.}}]{Walther:2024tcj}%
  \BibitemOpen
  \bibfield  {author} {\bibinfo {author} {\bibfnamefont {M.}~\bibnamefont {Walther}} \emph {et~al.},\ }\href@noop {} {\  (\bibinfo {year} {2024})},\ \Eprint {http://arxiv.org/abs/2412.05372} {arXiv:2412.05372 [astro-ph.CO]} \BibitemShut {NoStop}%
\bibitem [{\citenamefont {Freedman}\ \emph {et~al.}(2024)\citenamefont {Freedman}, \citenamefont {Madore}, \citenamefont {Jang}, \citenamefont {Hoyt}, \citenamefont {Lee},\ and\ \citenamefont {Owens}}]{Freedman:2024eph}%
  \BibitemOpen
  \bibfield  {author} {\bibinfo {author} {\bibfnamefont {W.~L.}\ \bibnamefont {Freedman}}, \bibinfo {author} {\bibfnamefont {B.~F.}\ \bibnamefont {Madore}}, \bibinfo {author} {\bibfnamefont {I.~S.}\ \bibnamefont {Jang}}, \bibinfo {author} {\bibfnamefont {T.~J.}\ \bibnamefont {Hoyt}}, \bibinfo {author} {\bibfnamefont {A.~J.}\ \bibnamefont {Lee}}, \ and\ \bibinfo {author} {\bibfnamefont {K.~A.}\ \bibnamefont {Owens}},\ }\href@noop {} {\  (\bibinfo {year} {2024})},\ \Eprint {http://arxiv.org/abs/2408.06153} {arXiv:2408.06153 [astro-ph.CO]} \BibitemShut {NoStop}%
\bibitem [{\citenamefont {Riess}\ \emph {et~al.}(2024)\citenamefont {Riess} \emph {et~al.}}]{Riess:2024vfa}%
  \BibitemOpen
  \bibfield  {author} {\bibinfo {author} {\bibfnamefont {A.~G.}\ \bibnamefont {Riess}} \emph {et~al.},\ }\href {\doibase 10.3847/1538-4357/ad8c21} {\bibfield  {journal} {\bibinfo  {journal} {Astrophys. J.}\ }\textbf {\bibinfo {volume} {977}},\ \bibinfo {pages} {120} (\bibinfo {year} {2024})},\ \Eprint {http://arxiv.org/abs/2408.11770} {arXiv:2408.11770 [astro-ph.CO]} \BibitemShut {NoStop}%
\bibitem [{\citenamefont {Abdalla}\ \emph {et~al.}(2022)\citenamefont {Abdalla} \emph {et~al.}}]{Abdalla:2022yfr}%
  \BibitemOpen
  \bibfield  {author} {\bibinfo {author} {\bibfnamefont {E.}~\bibnamefont {Abdalla}} \emph {et~al.},\ }\href {\doibase 10.1016/j.jheap.2022.04.002} {\bibfield  {journal} {\bibinfo  {journal} {JHEAp}\ }\textbf {\bibinfo {volume} {34}},\ \bibinfo {pages} {49} (\bibinfo {year} {2022})},\ \Eprint {http://arxiv.org/abs/2203.06142} {arXiv:2203.06142 [astro-ph.CO]} \BibitemShut {NoStop}%
\bibitem [{\citenamefont {{Liu}}\ and\ \citenamefont {{Huang}}(2020)}]{Liu:2019dxr}%
  \BibitemOpen
  \bibfield  {author} {\bibinfo {author} {\bibfnamefont {M.}~\bibnamefont {{Liu}}}\ and\ \bibinfo {author} {\bibfnamefont {Z.}~\bibnamefont {{Huang}}},\ }\href {\doibase 10.3847/1538-4357/ab982e} {\bibfield  {journal} {\bibinfo  {journal} {\apj}\ }\textbf {\bibinfo {volume} {897}},\ \bibinfo {eid} {166} (\bibinfo {year} {2020})},\ \Eprint {http://arxiv.org/abs/1910.05670} {arXiv:1910.05670 [astro-ph.CO]} \BibitemShut {NoStop}%
\bibitem [{\citenamefont {{Keeley}}\ \emph {et~al.}(2020)\citenamefont {{Keeley}}, \citenamefont {{Shafieloo}}, \citenamefont {{Hazra}},\ and\ \citenamefont {{Souradeep}}}]{Keeley:2020rmo}%
  \BibitemOpen
  \bibfield  {author} {\bibinfo {author} {\bibfnamefont {R.~E.}\ \bibnamefont {{Keeley}}}, \bibinfo {author} {\bibfnamefont {A.}~\bibnamefont {{Shafieloo}}}, \bibinfo {author} {\bibfnamefont {D.~K.}\ \bibnamefont {{Hazra}}}, \ and\ \bibinfo {author} {\bibfnamefont {T.}~\bibnamefont {{Souradeep}}},\ }\href {\doibase 10.1088/1475-7516/2020/09/055} {\bibfield  {journal} {\bibinfo  {journal} {JCAP}\ }\textbf {\bibinfo {volume} {2020}},\ \bibinfo {eid} {055} (\bibinfo {year} {2020})},\ \Eprint {http://arxiv.org/abs/2006.12710} {arXiv:2006.12710 [astro-ph.CO]} \BibitemShut {NoStop}%
\bibitem [{\citenamefont {{Antony}}\ \emph {et~al.}(2023)\citenamefont {{Antony}}, \citenamefont {{Finelli}}, \citenamefont {{Hazra}},\ and\ \citenamefont {{Shafieloo}}}]{Antony:2022ert}%
  \BibitemOpen
  \bibfield  {author} {\bibinfo {author} {\bibfnamefont {A.}~\bibnamefont {{Antony}}}, \bibinfo {author} {\bibfnamefont {F.}~\bibnamefont {{Finelli}}}, \bibinfo {author} {\bibfnamefont {D.~K.}\ \bibnamefont {{Hazra}}}, \ and\ \bibinfo {author} {\bibfnamefont {A.}~\bibnamefont {{Shafieloo}}},\ }\href {\doibase 10.1103/PhysRevLett.130.111001} {\bibfield  {journal} {\bibinfo  {journal} {\prl}\ }\textbf {\bibinfo {volume} {130}},\ \bibinfo {eid} {111001} (\bibinfo {year} {2023})},\ \Eprint {http://arxiv.org/abs/2202.14028} {arXiv:2202.14028 [astro-ph.CO]} \BibitemShut {NoStop}%
\bibitem [{\citenamefont {{Wang}}\ \emph {et~al.}(1999)\citenamefont {{Wang}}, \citenamefont {{Spergel}},\ and\ \citenamefont {{Strauss}}}]{Wang:1998gb}%
  \BibitemOpen
  \bibfield  {author} {\bibinfo {author} {\bibfnamefont {Y.}~\bibnamefont {{Wang}}}, \bibinfo {author} {\bibfnamefont {D.~N.}\ \bibnamefont {{Spergel}}}, \ and\ \bibinfo {author} {\bibfnamefont {M.~A.}\ \bibnamefont {{Strauss}}},\ }\href {\doibase 10.1086/306558} {\bibfield  {journal} {\bibinfo  {journal} {\apj}\ }\textbf {\bibinfo {volume} {510}},\ \bibinfo {pages} {20} (\bibinfo {year} {1999})},\ \Eprint {http://arxiv.org/abs/astro-ph/9802231} {arXiv:astro-ph/9802231 [astro-ph]} \BibitemShut {NoStop}%
\bibitem [{\citenamefont {{Zhan}}\ \emph {et~al.}(2006)\citenamefont {{Zhan}}, \citenamefont {{Knox}}, \citenamefont {{Tyson}},\ and\ \citenamefont {{Margoniner}}}]{Zhan:2005rz}%
  \BibitemOpen
  \bibfield  {author} {\bibinfo {author} {\bibfnamefont {H.}~\bibnamefont {{Zhan}}}, \bibinfo {author} {\bibfnamefont {L.}~\bibnamefont {{Knox}}}, \bibinfo {author} {\bibfnamefont {J.~A.}\ \bibnamefont {{Tyson}}}, \ and\ \bibinfo {author} {\bibfnamefont {V.}~\bibnamefont {{Margoniner}}},\ }\href {\doibase 10.1086/500077} {\bibfield  {journal} {\bibinfo  {journal} {\apj}\ }\textbf {\bibinfo {volume} {640}},\ \bibinfo {pages} {8} (\bibinfo {year} {2006})},\ \Eprint {http://arxiv.org/abs/astro-ph/0508119} {arXiv:astro-ph/0508119 [astro-ph]} \BibitemShut {NoStop}%
\bibitem [{\citenamefont {{Huang}}\ \emph {et~al.}(2012)\citenamefont {{Huang}}, \citenamefont {{Verde}},\ and\ \citenamefont {{Vernizzi}}}]{Huang:2012mr}%
  \BibitemOpen
  \bibfield  {author} {\bibinfo {author} {\bibfnamefont {Z.}~\bibnamefont {{Huang}}}, \bibinfo {author} {\bibfnamefont {L.}~\bibnamefont {{Verde}}}, \ and\ \bibinfo {author} {\bibfnamefont {F.}~\bibnamefont {{Vernizzi}}},\ }\href {\doibase 10.1088/1475-7516/2012/04/005} {\bibfield  {journal} {\bibinfo  {journal} {JCAP}\ }\textbf {\bibinfo {volume} {2012}},\ \bibinfo {eid} {005} (\bibinfo {year} {2012})},\ \Eprint {http://arxiv.org/abs/1201.5955} {arXiv:1201.5955 [astro-ph.CO]} \BibitemShut {NoStop}%
\bibitem [{\citenamefont {{Chen}}\ \emph {et~al.}(2016{\natexlab{a}})\citenamefont {{Chen}}, \citenamefont {{Dvorkin}}, \citenamefont {{Huang}}, \citenamefont {{Namjoo}},\ and\ \citenamefont {{Verde}}}]{Chen:2016vvw}%
  \BibitemOpen
  \bibfield  {author} {\bibinfo {author} {\bibfnamefont {X.}~\bibnamefont {{Chen}}}, \bibinfo {author} {\bibfnamefont {C.}~\bibnamefont {{Dvorkin}}}, \bibinfo {author} {\bibfnamefont {Z.}~\bibnamefont {{Huang}}}, \bibinfo {author} {\bibfnamefont {M.~H.}\ \bibnamefont {{Namjoo}}}, \ and\ \bibinfo {author} {\bibfnamefont {L.}~\bibnamefont {{Verde}}},\ }\href {\doibase 10.1088/1475-7516/2016/11/014} {\bibfield  {journal} {\bibinfo  {journal} {JCAP}\ }\textbf {\bibinfo {volume} {2016}},\ \bibinfo {eid} {014} (\bibinfo {year} {2016}{\natexlab{a}})},\ \Eprint {http://arxiv.org/abs/1605.09365} {arXiv:1605.09365 [astro-ph.CO]} \BibitemShut {NoStop}%
\bibitem [{\citenamefont {{Chen}}\ \emph {et~al.}(2016{\natexlab{b}})\citenamefont {{Chen}}, \citenamefont {{Meerburg}},\ and\ \citenamefont {{M{\"u}nchmeyer}}}]{Chen:2016zuu}%
  \BibitemOpen
  \bibfield  {author} {\bibinfo {author} {\bibfnamefont {X.}~\bibnamefont {{Chen}}}, \bibinfo {author} {\bibfnamefont {P.~D.}\ \bibnamefont {{Meerburg}}}, \ and\ \bibinfo {author} {\bibfnamefont {M.}~\bibnamefont {{M{\"u}nchmeyer}}},\ }\href {\doibase 10.1088/1475-7516/2016/09/023} {\bibfield  {journal} {\bibinfo  {journal} {JCAP}\ }\textbf {\bibinfo {volume} {2016}},\ \bibinfo {eid} {023} (\bibinfo {year} {2016}{\natexlab{b}})},\ \Eprint {http://arxiv.org/abs/1605.09364} {arXiv:1605.09364 [astro-ph.CO]} \BibitemShut {NoStop}%
\bibitem [{\citenamefont {{Ballardini}}\ \emph {et~al.}(2016)\citenamefont {{Ballardini}}, \citenamefont {{Finelli}}, \citenamefont {{Fedeli}},\ and\ \citenamefont {{Moscardini}}}]{Ballardini:2016hpi}%
  \BibitemOpen
  \bibfield  {author} {\bibinfo {author} {\bibfnamefont {M.}~\bibnamefont {{Ballardini}}}, \bibinfo {author} {\bibfnamefont {F.}~\bibnamefont {{Finelli}}}, \bibinfo {author} {\bibfnamefont {C.}~\bibnamefont {{Fedeli}}}, \ and\ \bibinfo {author} {\bibfnamefont {L.}~\bibnamefont {{Moscardini}}},\ }\href {\doibase 10.1088/1475-7516/2016/10/041} {\bibfield  {journal} {\bibinfo  {journal} {JCAP}\ }\textbf {\bibinfo {volume} {2016}},\ \bibinfo {eid} {041} (\bibinfo {year} {2016})},\ \Eprint {http://arxiv.org/abs/1606.03747} {arXiv:1606.03747 [astro-ph.CO]} \BibitemShut {NoStop}%
\bibitem [{\citenamefont {{Xu}}\ \emph {et~al.}(2016)\citenamefont {{Xu}}, \citenamefont {{Hamann}},\ and\ \citenamefont {{Chen}}}]{Xu:2016kwz}%
  \BibitemOpen
  \bibfield  {author} {\bibinfo {author} {\bibfnamefont {Y.}~\bibnamefont {{Xu}}}, \bibinfo {author} {\bibfnamefont {J.}~\bibnamefont {{Hamann}}}, \ and\ \bibinfo {author} {\bibfnamefont {X.}~\bibnamefont {{Chen}}},\ }\href {\doibase 10.1103/PhysRevD.94.123518} {\bibfield  {journal} {\bibinfo  {journal} {\prd}\ }\textbf {\bibinfo {volume} {94}},\ \bibinfo {eid} {123518} (\bibinfo {year} {2016})},\ \Eprint {http://arxiv.org/abs/1607.00817} {arXiv:1607.00817 [astro-ph.CO]} \BibitemShut {NoStop}%
\bibitem [{\citenamefont {{Ansari Fard}}\ and\ \citenamefont {{Baghram}}(2018)}]{Fard:2017oex}%
  \BibitemOpen
  \bibfield  {author} {\bibinfo {author} {\bibfnamefont {M.}~\bibnamefont {{Ansari Fard}}}\ and\ \bibinfo {author} {\bibfnamefont {S.}~\bibnamefont {{Baghram}}},\ }\href {\doibase 10.1088/1475-7516/2018/01/051} {\bibfield  {journal} {\bibinfo  {journal} {JCAP}\ }\textbf {\bibinfo {volume} {2018}},\ \bibinfo {eid} {051} (\bibinfo {year} {2018})},\ \Eprint {http://arxiv.org/abs/1709.05323} {arXiv:1709.05323 [astro-ph.CO]} \BibitemShut {NoStop}%
\bibitem [{\citenamefont {{Palma}}\ \emph {et~al.}(2018)\citenamefont {{Palma}}, \citenamefont {{Sapone}},\ and\ \citenamefont {{Sypsas}}}]{Palma:2017wxu}%
  \BibitemOpen
  \bibfield  {author} {\bibinfo {author} {\bibfnamefont {G.~A.}\ \bibnamefont {{Palma}}}, \bibinfo {author} {\bibfnamefont {D.}~\bibnamefont {{Sapone}}}, \ and\ \bibinfo {author} {\bibfnamefont {S.}~\bibnamefont {{Sypsas}}},\ }\href {\doibase 10.1088/1475-7516/2018/06/004} {\bibfield  {journal} {\bibinfo  {journal} {JCAP}\ }\textbf {\bibinfo {volume} {2018}},\ \bibinfo {eid} {004} (\bibinfo {year} {2018})},\ \Eprint {http://arxiv.org/abs/1710.02570} {arXiv:1710.02570 [astro-ph.CO]} \BibitemShut {NoStop}%
\bibitem [{\citenamefont {{Ballardini}}\ \emph {et~al.}(2018)\citenamefont {{Ballardini}}, \citenamefont {{Finelli}}, \citenamefont {{Maartens}},\ and\ \citenamefont {{Moscardini}}}]{Ballardini:2017qwq}%
  \BibitemOpen
  \bibfield  {author} {\bibinfo {author} {\bibfnamefont {M.}~\bibnamefont {{Ballardini}}}, \bibinfo {author} {\bibfnamefont {F.}~\bibnamefont {{Finelli}}}, \bibinfo {author} {\bibfnamefont {R.}~\bibnamefont {{Maartens}}}, \ and\ \bibinfo {author} {\bibfnamefont {L.}~\bibnamefont {{Moscardini}}},\ }\href {\doibase 10.1088/1475-7516/2018/04/044} {\bibfield  {journal} {\bibinfo  {journal} {JCAP}\ }\textbf {\bibinfo {volume} {2018}},\ \bibinfo {eid} {044} (\bibinfo {year} {2018})},\ \Eprint {http://arxiv.org/abs/1712.07425} {arXiv:1712.07425 [astro-ph.CO]} \BibitemShut {NoStop}%
\bibitem [{\citenamefont {{Ballardini}}\ \emph {et~al.}(2020)\citenamefont {{Ballardini}}, \citenamefont {{Murgia}}, \citenamefont {{Baldi}}, \citenamefont {{Finelli}},\ and\ \citenamefont {{Viel}}}]{Ballardini:2019tuc}%
  \BibitemOpen
  \bibfield  {author} {\bibinfo {author} {\bibfnamefont {M.}~\bibnamefont {{Ballardini}}}, \bibinfo {author} {\bibfnamefont {R.}~\bibnamefont {{Murgia}}}, \bibinfo {author} {\bibfnamefont {M.}~\bibnamefont {{Baldi}}}, \bibinfo {author} {\bibfnamefont {F.}~\bibnamefont {{Finelli}}}, \ and\ \bibinfo {author} {\bibfnamefont {M.}~\bibnamefont {{Viel}}},\ }\href {\doibase 10.1088/1475-7516/2020/04/030} {\bibfield  {journal} {\bibinfo  {journal} {JCAP}\ }\textbf {\bibinfo {volume} {2020}},\ \bibinfo {eid} {030} (\bibinfo {year} {2020})},\ \Eprint {http://arxiv.org/abs/1912.12499} {arXiv:1912.12499 [astro-ph.CO]} \BibitemShut {NoStop}%
\bibitem [{\citenamefont {{Debono}}\ \emph {et~al.}(2020)\citenamefont {{Debono}}, \citenamefont {{Hazra}}, \citenamefont {{Shafieloo}}, \citenamefont {{Smoot}},\ and\ \citenamefont {{Starobinsky}}}]{Debono:2020emh}%
  \BibitemOpen
  \bibfield  {author} {\bibinfo {author} {\bibfnamefont {I.}~\bibnamefont {{Debono}}}, \bibinfo {author} {\bibfnamefont {D.~K.}\ \bibnamefont {{Hazra}}}, \bibinfo {author} {\bibfnamefont {A.}~\bibnamefont {{Shafieloo}}}, \bibinfo {author} {\bibfnamefont {G.~F.}\ \bibnamefont {{Smoot}}}, \ and\ \bibinfo {author} {\bibfnamefont {A.~A.}\ \bibnamefont {{Starobinsky}}},\ }\href {\doibase 10.1093/mnras/staa1765} {\bibfield  {journal} {\bibinfo  {journal} {mnras}\ }\textbf {\bibinfo {volume} {496}},\ \bibinfo {pages} {3448} (\bibinfo {year} {2020})},\ \Eprint {http://arxiv.org/abs/2003.05262} {arXiv:2003.05262 [astro-ph.CO]} \BibitemShut {NoStop}%
\bibitem [{\citenamefont {{Li}}\ \emph {et~al.}(2022)\citenamefont {{Li}}, \citenamefont {{Zhu}},\ and\ \citenamefont {{Li}}}]{Li:2021jvz}%
  \BibitemOpen
  \bibfield  {author} {\bibinfo {author} {\bibfnamefont {Y.}~\bibnamefont {{Li}}}, \bibinfo {author} {\bibfnamefont {H.-M.}\ \bibnamefont {{Zhu}}}, \ and\ \bibinfo {author} {\bibfnamefont {B.}~\bibnamefont {{Li}}},\ }\href {\doibase 10.1093/mnras/stac1544} {\bibfield  {journal} {\bibinfo  {journal} {mnras}\ }\textbf {\bibinfo {volume} {514}},\ \bibinfo {pages} {4363} (\bibinfo {year} {2022})},\ \Eprint {http://arxiv.org/abs/2102.09007} {arXiv:2102.09007 [astro-ph.CO]} \BibitemShut {NoStop}%
\bibitem [{\citenamefont {Ballardini}\ \emph {et~al.}(2024)\citenamefont {Ballardini} \emph {et~al.}}]{Euclid:2023shr}%
  \BibitemOpen
  \bibfield  {author} {\bibinfo {author} {\bibfnamefont {M.}~\bibnamefont {Ballardini}} \emph {et~al.} (\bibinfo {collaboration} {Euclid}),\ }\href {\doibase 10.1051/0004-6361/202348162} {\bibfield  {journal} {\bibinfo  {journal} {Astron. Astrophys.}\ }\textbf {\bibinfo {volume} {683}},\ \bibinfo {pages} {A220} (\bibinfo {year} {2024})},\ \Eprint {http://arxiv.org/abs/2309.17287} {arXiv:2309.17287 [astro-ph.CO]} \BibitemShut {NoStop}%
\bibitem [{\citenamefont {Ballardini}\ and\ \citenamefont {Barbieri}(2024)}]{Ballardini:2024dto}%
  \BibitemOpen
  \bibfield  {author} {\bibinfo {author} {\bibfnamefont {M.}~\bibnamefont {Ballardini}}\ and\ \bibinfo {author} {\bibfnamefont {N.}~\bibnamefont {Barbieri}},\ }\href@noop {} {\  (\bibinfo {year} {2024})},\ \Eprint {http://arxiv.org/abs/2411.02261} {arXiv:2411.02261 [astro-ph.CO]} \BibitemShut {NoStop}%
\bibitem [{\citenamefont {Beutler}\ \emph {et~al.}(2019)\citenamefont {Beutler}, \citenamefont {Biagetti}, \citenamefont {Green}, \citenamefont {Slosar},\ and\ \citenamefont {Wallisch}}]{Beutler:2019ojk}%
  \BibitemOpen
  \bibfield  {author} {\bibinfo {author} {\bibfnamefont {F.}~\bibnamefont {Beutler}}, \bibinfo {author} {\bibfnamefont {M.}~\bibnamefont {Biagetti}}, \bibinfo {author} {\bibfnamefont {D.}~\bibnamefont {Green}}, \bibinfo {author} {\bibfnamefont {A.}~\bibnamefont {Slosar}}, \ and\ \bibinfo {author} {\bibfnamefont {B.}~\bibnamefont {Wallisch}},\ }\href {\doibase 10.1103/PhysRevResearch.1.033209} {\bibfield  {journal} {\bibinfo  {journal} {Phys. Rev. Res.}\ }\textbf {\bibinfo {volume} {1}},\ \bibinfo {pages} {033209} (\bibinfo {year} {2019})},\ \Eprint {http://arxiv.org/abs/1906.08758} {arXiv:1906.08758 [astro-ph.CO]} \BibitemShut {NoStop}%
\bibitem [{\citenamefont {Ballardini}\ \emph {et~al.}(2023)\citenamefont {Ballardini}, \citenamefont {Finelli}, \citenamefont {Marulli}, \citenamefont {Moscardini},\ and\ \citenamefont {Veropalumbo}}]{Ballardini:2022wzu}%
  \BibitemOpen
  \bibfield  {author} {\bibinfo {author} {\bibfnamefont {M.}~\bibnamefont {Ballardini}}, \bibinfo {author} {\bibfnamefont {F.}~\bibnamefont {Finelli}}, \bibinfo {author} {\bibfnamefont {F.}~\bibnamefont {Marulli}}, \bibinfo {author} {\bibfnamefont {L.}~\bibnamefont {Moscardini}}, \ and\ \bibinfo {author} {\bibfnamefont {A.}~\bibnamefont {Veropalumbo}},\ }\href {\doibase 10.1103/PhysRevD.107.043532} {\bibfield  {journal} {\bibinfo  {journal} {Phys. Rev. D}\ }\textbf {\bibinfo {volume} {107}},\ \bibinfo {pages} {043532} (\bibinfo {year} {2023})},\ \Eprint {http://arxiv.org/abs/2202.08819} {arXiv:2202.08819 [astro-ph.CO]} \BibitemShut {NoStop}%
\bibitem [{\citenamefont {Mergulh\~ao}\ \emph {et~al.}(2023)\citenamefont {Mergulh\~ao}, \citenamefont {Beutler},\ and\ \citenamefont {Peacock}}]{Mergulhao:2023ukp}%
  \BibitemOpen
  \bibfield  {author} {\bibinfo {author} {\bibfnamefont {T.}~\bibnamefont {Mergulh\~ao}}, \bibinfo {author} {\bibfnamefont {F.}~\bibnamefont {Beutler}}, \ and\ \bibinfo {author} {\bibfnamefont {J.~A.}\ \bibnamefont {Peacock}},\ }\href {\doibase 10.1088/1475-7516/2023/08/012} {\bibfield  {journal} {\bibinfo  {journal} {JCAP}\ }\textbf {\bibinfo {volume} {08}},\ \bibinfo {pages} {012} (\bibinfo {year} {2023})},\ \Eprint {http://arxiv.org/abs/2303.13946} {arXiv:2303.13946 [astro-ph.CO]} \BibitemShut {NoStop}%
\bibitem [{\citenamefont {Vlah}\ \emph {et~al.}(2016)\citenamefont {Vlah}, \citenamefont {Seljak}, \citenamefont {Chu},\ and\ \citenamefont {Feng}}]{Vlah:2015zda}%
  \BibitemOpen
  \bibfield  {author} {\bibinfo {author} {\bibfnamefont {Z.}~\bibnamefont {Vlah}}, \bibinfo {author} {\bibfnamefont {U.}~\bibnamefont {Seljak}}, \bibinfo {author} {\bibfnamefont {M.~Y.}\ \bibnamefont {Chu}}, \ and\ \bibinfo {author} {\bibfnamefont {Y.}~\bibnamefont {Feng}},\ }\href {\doibase 10.1088/1475-7516/2016/03/057} {\bibfield  {journal} {\bibinfo  {journal} {JCAP}\ }\textbf {\bibinfo {volume} {03}},\ \bibinfo {pages} {057} (\bibinfo {year} {2016})},\ \Eprint {http://arxiv.org/abs/1509.02120} {arXiv:1509.02120 [astro-ph.CO]} \BibitemShut {NoStop}%
\bibitem [{\citenamefont {Vasudevan}\ \emph {et~al.}(2019)\citenamefont {Vasudevan}, \citenamefont {Ivanov}, \citenamefont {Sibiryakov},\ and\ \citenamefont {Lesgourgues}}]{Vasudevan:2019ewf}%
  \BibitemOpen
  \bibfield  {author} {\bibinfo {author} {\bibfnamefont {A.}~\bibnamefont {Vasudevan}}, \bibinfo {author} {\bibfnamefont {M.~M.}\ \bibnamefont {Ivanov}}, \bibinfo {author} {\bibfnamefont {S.}~\bibnamefont {Sibiryakov}}, \ and\ \bibinfo {author} {\bibfnamefont {J.}~\bibnamefont {Lesgourgues}},\ }\href {\doibase 10.1088/1475-7516/2019/09/037} {\bibfield  {journal} {\bibinfo  {journal} {JCAP}\ }\textbf {\bibinfo {volume} {09}},\ \bibinfo {pages} {037} (\bibinfo {year} {2019})},\ \Eprint {http://arxiv.org/abs/1906.08697} {arXiv:1906.08697 [astro-ph.CO]} \BibitemShut {NoStop}%
\bibitem [{\citenamefont {Chen}\ \emph {et~al.}(2020)\citenamefont {Chen}, \citenamefont {Vlah},\ and\ \citenamefont {White}}]{Chen:2020ckc}%
  \BibitemOpen
  \bibfield  {author} {\bibinfo {author} {\bibfnamefont {S.-F.}\ \bibnamefont {Chen}}, \bibinfo {author} {\bibfnamefont {Z.}~\bibnamefont {Vlah}}, \ and\ \bibinfo {author} {\bibfnamefont {M.}~\bibnamefont {White}},\ }\href {\doibase 10.1088/1475-7516/2020/11/035} {\bibfield  {journal} {\bibinfo  {journal} {JCAP}\ }\textbf {\bibinfo {volume} {11}},\ \bibinfo {pages} {035} (\bibinfo {year} {2020})},\ \Eprint {http://arxiv.org/abs/2007.00704} {arXiv:2007.00704 [astro-ph.CO]} \BibitemShut {NoStop}%
\bibitem [{\citenamefont {Abolfathi}\ \emph {et~al.}(2018)\citenamefont {Abolfathi} \emph {et~al.}}]{Abolfathi_2018}%
  \BibitemOpen
  \bibfield  {author} {\bibinfo {author} {\bibnamefont {Abolfathi}} \emph {et~al.} (\bibinfo {collaboration} {SDSS}),\ }\href {\doibase 10.3847/1538-4365/aa9e8a} {\bibfield  {journal} {\bibinfo  {journal} {The Astrophysical Journal Supplement Series}\ }\textbf {\bibinfo {volume} {235}},\ \bibinfo {pages} {42} (\bibinfo {year} {2018})}\BibitemShut {NoStop}%
\bibitem [{\citenamefont {Troxel}\ \emph {et~al.}(2018)\citenamefont {Troxel} \emph {et~al.}}]{DES:2017qwj}%
  \BibitemOpen
  \bibfield  {author} {\bibinfo {author} {\bibfnamefont {M.~A.}\ \bibnamefont {Troxel}} \emph {et~al.} (\bibinfo {collaboration} {DES}),\ }\href {\doibase 10.1103/PhysRevD.98.043528} {\bibfield  {journal} {\bibinfo  {journal} {Phys. Rev. D}\ }\textbf {\bibinfo {volume} {98}},\ \bibinfo {pages} {043528} (\bibinfo {year} {2018})},\ \Eprint {http://arxiv.org/abs/1708.01538} {arXiv:1708.01538 [astro-ph.CO]} \BibitemShut {NoStop}%
\bibitem [{\citenamefont {Mu\~noz}\ \emph {et~al.}(2020)\citenamefont {Mu\~noz}, \citenamefont {Dvorkin},\ and\ \citenamefont {Cyr-Racine}}]{Munoz:2019hjh}%
  \BibitemOpen
  \bibfield  {author} {\bibinfo {author} {\bibfnamefont {J.~B.}\ \bibnamefont {Mu\~noz}}, \bibinfo {author} {\bibfnamefont {C.}~\bibnamefont {Dvorkin}}, \ and\ \bibinfo {author} {\bibfnamefont {F.-Y.}\ \bibnamefont {Cyr-Racine}},\ }\href {\doibase 10.1103/PhysRevD.101.063526} {\bibfield  {journal} {\bibinfo  {journal} {Phys. Rev. D}\ }\textbf {\bibinfo {volume} {101}},\ \bibinfo {pages} {063526} (\bibinfo {year} {2020})},\ \Eprint {http://arxiv.org/abs/1911.11144} {arXiv:1911.11144 [astro-ph.CO]} \BibitemShut {NoStop}%
\bibitem [{\citenamefont {Naik}\ \emph {et~al.}(2025)\citenamefont {Naik}, \citenamefont {Chingangbam}, \citenamefont {Singh}, \citenamefont {Mesinger},\ and\ \citenamefont {Furuuchi}}]{Naik:2025mba}%
  \BibitemOpen
  \bibfield  {author} {\bibinfo {author} {\bibfnamefont {S.~S.}\ \bibnamefont {Naik}}, \bibinfo {author} {\bibfnamefont {P.}~\bibnamefont {Chingangbam}}, \bibinfo {author} {\bibfnamefont {S.}~\bibnamefont {Singh}}, \bibinfo {author} {\bibfnamefont {A.}~\bibnamefont {Mesinger}}, \ and\ \bibinfo {author} {\bibfnamefont {K.}~\bibnamefont {Furuuchi}},\ }\href@noop {} {\  (\bibinfo {year} {2025})},\ \Eprint {http://arxiv.org/abs/2501.02538} {arXiv:2501.02538 [astro-ph.CO]} \BibitemShut {NoStop}%
\bibitem [{\citenamefont {Starobinsky}(1992)}]{Starobinsky:1992ts}%
  \BibitemOpen
  \bibfield  {author} {\bibinfo {author} {\bibfnamefont {A.~A.}\ \bibnamefont {Starobinsky}},\ }\href@noop {} {\bibfield  {journal} {\bibinfo  {journal} {JETP Lett.}\ }\textbf {\bibinfo {volume} {55}},\ \bibinfo {pages} {489} (\bibinfo {year} {1992})}\BibitemShut {NoStop}%
\bibitem [{\citenamefont {Hunt}\ and\ \citenamefont {Sarkar}(2004)}]{Hunt:2004vt}%
  \BibitemOpen
  \bibfield  {author} {\bibinfo {author} {\bibfnamefont {P.}~\bibnamefont {Hunt}}\ and\ \bibinfo {author} {\bibfnamefont {S.}~\bibnamefont {Sarkar}},\ }\href {\doibase 10.1103/PhysRevD.70.103518} {\bibfield  {journal} {\bibinfo  {journal} {Phys. Rev. D}\ }\textbf {\bibinfo {volume} {70}},\ \bibinfo {pages} {103518} (\bibinfo {year} {2004})},\ \Eprint {http://arxiv.org/abs/astro-ph/0408138} {arXiv:astro-ph/0408138} \BibitemShut {NoStop}%
\bibitem [{\citenamefont {Gong}(2005)}]{Gong:2005jr}%
  \BibitemOpen
  \bibfield  {author} {\bibinfo {author} {\bibfnamefont {J.-O.}\ \bibnamefont {Gong}},\ }\href {\doibase 10.1088/1475-7516/2005/07/015} {\bibfield  {journal} {\bibinfo  {journal} {JCAP}\ }\textbf {\bibinfo {volume} {07}},\ \bibinfo {pages} {015} (\bibinfo {year} {2005})},\ \Eprint {http://arxiv.org/abs/astro-ph/0504383} {arXiv:astro-ph/0504383} \BibitemShut {NoStop}%
\bibitem [{\citenamefont {Joy}\ \emph {et~al.}(2008)\citenamefont {Joy}, \citenamefont {Sahni},\ and\ \citenamefont {Starobinsky}}]{Joy:2007na}%
  \BibitemOpen
  \bibfield  {author} {\bibinfo {author} {\bibfnamefont {M.}~\bibnamefont {Joy}}, \bibinfo {author} {\bibfnamefont {V.}~\bibnamefont {Sahni}}, \ and\ \bibinfo {author} {\bibfnamefont {A.~A.}\ \bibnamefont {Starobinsky}},\ }\href {\doibase 10.1103/PhysRevD.77.023514} {\bibfield  {journal} {\bibinfo  {journal} {Phys. Rev. D}\ }\textbf {\bibinfo {volume} {77}},\ \bibinfo {pages} {023514} (\bibinfo {year} {2008})},\ \Eprint {http://arxiv.org/abs/0711.1585} {arXiv:0711.1585 [astro-ph]} \BibitemShut {NoStop}%
\bibitem [{\citenamefont {Dias}\ \emph {et~al.}(2016)\citenamefont {Dias}, \citenamefont {Frazer}, \citenamefont {Mulryne},\ and\ \citenamefont {Seery}}]{Dias:2016rjq}%
  \BibitemOpen
  \bibfield  {author} {\bibinfo {author} {\bibfnamefont {M.}~\bibnamefont {Dias}}, \bibinfo {author} {\bibfnamefont {J.}~\bibnamefont {Frazer}}, \bibinfo {author} {\bibfnamefont {D.~J.}\ \bibnamefont {Mulryne}}, \ and\ \bibinfo {author} {\bibfnamefont {D.}~\bibnamefont {Seery}},\ }\href {\doibase 10.1088/1475-7516/2016/12/033} {\bibfield  {journal} {\bibinfo  {journal} {JCAP}\ }\textbf {\bibinfo {volume} {12}},\ \bibinfo {pages} {033} (\bibinfo {year} {2016})},\ \Eprint {http://arxiv.org/abs/1609.00379} {arXiv:1609.00379 [astro-ph.CO]} \BibitemShut {NoStop}%
\bibitem [{\citenamefont {Mulryne}\ and\ \citenamefont {Ronayne}(2018)}]{Mulryne:2016mzv}%
  \BibitemOpen
  \bibfield  {author} {\bibinfo {author} {\bibfnamefont {D.~J.}\ \bibnamefont {Mulryne}}\ and\ \bibinfo {author} {\bibfnamefont {J.~W.}\ \bibnamefont {Ronayne}},\ }\href {\doibase 10.21105/joss.00494} {\bibfield  {journal} {\bibinfo  {journal} {J. Open Source Softw.}\ }\textbf {\bibinfo {volume} {3}},\ \bibinfo {pages} {494} (\bibinfo {year} {2018})},\ \Eprint {http://arxiv.org/abs/1609.00381} {arXiv:1609.00381 [astro-ph.CO]} \BibitemShut {NoStop}%
\bibitem [{\citenamefont {Ronayne}\ and\ \citenamefont {Mulryne}(2018)}]{Ronayne:2017qzn}%
  \BibitemOpen
  \bibfield  {author} {\bibinfo {author} {\bibfnamefont {J.~W.}\ \bibnamefont {Ronayne}}\ and\ \bibinfo {author} {\bibfnamefont {D.~J.}\ \bibnamefont {Mulryne}},\ }\href {\doibase 10.1088/1475-7516/2018/01/023} {\bibfield  {journal} {\bibinfo  {journal} {JCAP}\ }\textbf {\bibinfo {volume} {01}},\ \bibinfo {pages} {023} (\bibinfo {year} {2018})},\ \Eprint {http://arxiv.org/abs/1708.07130} {arXiv:1708.07130 [astro-ph.CO]} \BibitemShut {NoStop}%
\bibitem [{\citenamefont {Langlois}(1999)}]{Langlois:1999dw}%
  \BibitemOpen
  \bibfield  {author} {\bibinfo {author} {\bibfnamefont {D.}~\bibnamefont {Langlois}},\ }\href {\doibase 10.1103/PhysRevD.59.123512} {\bibfield  {journal} {\bibinfo  {journal} {Phys. Rev. D}\ }\textbf {\bibinfo {volume} {59}},\ \bibinfo {pages} {123512} (\bibinfo {year} {1999})},\ \Eprint {http://arxiv.org/abs/astro-ph/9906080} {arXiv:astro-ph/9906080} \BibitemShut {NoStop}%
\bibitem [{\citenamefont {Wands}\ \emph {et~al.}(2002)\citenamefont {Wands}, \citenamefont {Bartolo}, \citenamefont {Matarrese},\ and\ \citenamefont {Riotto}}]{Wands:2002bn}%
  \BibitemOpen
  \bibfield  {author} {\bibinfo {author} {\bibfnamefont {D.}~\bibnamefont {Wands}}, \bibinfo {author} {\bibfnamefont {N.}~\bibnamefont {Bartolo}}, \bibinfo {author} {\bibfnamefont {S.}~\bibnamefont {Matarrese}}, \ and\ \bibinfo {author} {\bibfnamefont {A.}~\bibnamefont {Riotto}},\ }\href {\doibase 10.1103/PhysRevD.66.043520} {\bibfield  {journal} {\bibinfo  {journal} {Phys. Rev. D}\ }\textbf {\bibinfo {volume} {66}},\ \bibinfo {pages} {043520} (\bibinfo {year} {2002})},\ \Eprint {http://arxiv.org/abs/astro-ph/0205253} {arXiv:astro-ph/0205253} \BibitemShut {NoStop}%
\bibitem [{\citenamefont {Lesgourgues}(2000)}]{Lesgourgues:1999uc}%
  \BibitemOpen
  \bibfield  {author} {\bibinfo {author} {\bibfnamefont {J.}~\bibnamefont {Lesgourgues}},\ }\href {\doibase 10.1016/S0550-3213(00)00301-1} {\bibfield  {journal} {\bibinfo  {journal} {Nucl. Phys. B}\ }\textbf {\bibinfo {volume} {582}},\ \bibinfo {pages} {593} (\bibinfo {year} {2000})},\ \Eprint {http://arxiv.org/abs/hep-ph/9911447} {arXiv:hep-ph/9911447} \BibitemShut {NoStop}%
\bibitem [{\citenamefont {Peterson}\ and\ \citenamefont {Tegmark}(2011)}]{Peterson:2010np}%
  \BibitemOpen
  \bibfield  {author} {\bibinfo {author} {\bibfnamefont {C.~M.}\ \bibnamefont {Peterson}}\ and\ \bibinfo {author} {\bibfnamefont {M.}~\bibnamefont {Tegmark}},\ }\href {\doibase 10.1103/PhysRevD.83.023522} {\bibfield  {journal} {\bibinfo  {journal} {Phys. Rev. D}\ }\textbf {\bibinfo {volume} {83}},\ \bibinfo {pages} {023522} (\bibinfo {year} {2011})},\ \Eprint {http://arxiv.org/abs/1005.4056} {arXiv:1005.4056 [astro-ph.CO]} \BibitemShut {NoStop}%
\bibitem [{\citenamefont {Achucarro}\ \emph {et~al.}(2011{\natexlab{a}})\citenamefont {Achucarro}, \citenamefont {Gong}, \citenamefont {Hardeman}, \citenamefont {Palma},\ and\ \citenamefont {Patil}}]{Achucarro:2010da}%
  \BibitemOpen
  \bibfield  {author} {\bibinfo {author} {\bibfnamefont {A.}~\bibnamefont {Achucarro}}, \bibinfo {author} {\bibfnamefont {J.-O.}\ \bibnamefont {Gong}}, \bibinfo {author} {\bibfnamefont {S.}~\bibnamefont {Hardeman}}, \bibinfo {author} {\bibfnamefont {G.~A.}\ \bibnamefont {Palma}}, \ and\ \bibinfo {author} {\bibfnamefont {S.~P.}\ \bibnamefont {Patil}},\ }\href {\doibase 10.1088/1475-7516/2011/01/030} {\bibfield  {journal} {\bibinfo  {journal} {JCAP}\ }\textbf {\bibinfo {volume} {01}},\ \bibinfo {pages} {030} (\bibinfo {year} {2011}{\natexlab{a}})},\ \Eprint {http://arxiv.org/abs/1010.3693} {arXiv:1010.3693 [hep-ph]} \BibitemShut {NoStop}%
\bibitem [{\citenamefont {Konieczka}\ \emph {et~al.}(2014)\citenamefont {Konieczka}, \citenamefont {Ribeiro},\ and\ \citenamefont {Turzynski}}]{Konieczka:2014zja}%
  \BibitemOpen
  \bibfield  {author} {\bibinfo {author} {\bibfnamefont {M.}~\bibnamefont {Konieczka}}, \bibinfo {author} {\bibfnamefont {R.~H.}\ \bibnamefont {Ribeiro}}, \ and\ \bibinfo {author} {\bibfnamefont {K.}~\bibnamefont {Turzynski}},\ }\href {\doibase 10.1088/1475-7516/2014/07/030} {\bibfield  {journal} {\bibinfo  {journal} {JCAP}\ }\textbf {\bibinfo {volume} {07}},\ \bibinfo {pages} {030} (\bibinfo {year} {2014})},\ \Eprint {http://arxiv.org/abs/1401.6163} {arXiv:1401.6163 [astro-ph.CO]} \BibitemShut {NoStop}%
\bibitem [{\citenamefont {Garcia-Saenz}\ \emph {et~al.}(2020)\citenamefont {Garcia-Saenz}, \citenamefont {Pinol},\ and\ \citenamefont {Renaux-Petel}}]{Garcia-Saenz:2019njm}%
  \BibitemOpen
  \bibfield  {author} {\bibinfo {author} {\bibfnamefont {S.}~\bibnamefont {Garcia-Saenz}}, \bibinfo {author} {\bibfnamefont {L.}~\bibnamefont {Pinol}}, \ and\ \bibinfo {author} {\bibfnamefont {S.}~\bibnamefont {Renaux-Petel}},\ }\href {\doibase 10.1007/JHEP01(2020)073} {\bibfield  {journal} {\bibinfo  {journal} {JHEP}\ }\textbf {\bibinfo {volume} {01}},\ \bibinfo {pages} {073} (\bibinfo {year} {2020})},\ \Eprint {http://arxiv.org/abs/1907.10403} {arXiv:1907.10403 [hep-th]} \BibitemShut {NoStop}%
\bibitem [{\citenamefont {Chen}\ and\ \citenamefont {Wang}(2010)}]{Chen:2009zp}%
  \BibitemOpen
  \bibfield  {author} {\bibinfo {author} {\bibfnamefont {X.}~\bibnamefont {Chen}}\ and\ \bibinfo {author} {\bibfnamefont {Y.}~\bibnamefont {Wang}},\ }\href {\doibase 10.1088/1475-7516/2010/04/027} {\bibfield  {journal} {\bibinfo  {journal} {JCAP}\ }\textbf {\bibinfo {volume} {04}},\ \bibinfo {pages} {027} (\bibinfo {year} {2010})},\ \Eprint {http://arxiv.org/abs/0911.3380} {arXiv:0911.3380 [hep-th]} \BibitemShut {NoStop}%
\bibitem [{\citenamefont {Dias}\ \emph {et~al.}(2015)\citenamefont {Dias}, \citenamefont {Frazer},\ and\ \citenamefont {Seery}}]{Dias:2015rca}%
  \BibitemOpen
  \bibfield  {author} {\bibinfo {author} {\bibfnamefont {M.}~\bibnamefont {Dias}}, \bibinfo {author} {\bibfnamefont {J.}~\bibnamefont {Frazer}}, \ and\ \bibinfo {author} {\bibfnamefont {D.}~\bibnamefont {Seery}},\ }\href {\doibase 10.1088/1475-7516/2015/12/030} {\bibfield  {journal} {\bibinfo  {journal} {JCAP}\ }\textbf {\bibinfo {volume} {12}},\ \bibinfo {pages} {030} (\bibinfo {year} {2015})},\ \Eprint {http://arxiv.org/abs/1502.03125} {arXiv:1502.03125 [astro-ph.CO]} \BibitemShut {NoStop}%
\bibitem [{\citenamefont {Gao}\ \emph {et~al.}(2012)\citenamefont {Gao}, \citenamefont {Langlois},\ and\ \citenamefont {Mizuno}}]{Gao:2012uq}%
  \BibitemOpen
  \bibfield  {author} {\bibinfo {author} {\bibfnamefont {X.}~\bibnamefont {Gao}}, \bibinfo {author} {\bibfnamefont {D.}~\bibnamefont {Langlois}}, \ and\ \bibinfo {author} {\bibfnamefont {S.}~\bibnamefont {Mizuno}},\ }\href {\doibase 10.1088/1475-7516/2012/10/040} {\bibfield  {journal} {\bibinfo  {journal} {JCAP}\ }\textbf {\bibinfo {volume} {10}},\ \bibinfo {pages} {040} (\bibinfo {year} {2012})},\ \Eprint {http://arxiv.org/abs/1205.5275} {arXiv:1205.5275 [hep-th]} \BibitemShut {NoStop}%
\bibitem [{\citenamefont {Gao}\ \emph {et~al.}(2013)\citenamefont {Gao}, \citenamefont {Langlois},\ and\ \citenamefont {Mizuno}}]{Gao:2013ota}%
  \BibitemOpen
  \bibfield  {author} {\bibinfo {author} {\bibfnamefont {X.}~\bibnamefont {Gao}}, \bibinfo {author} {\bibfnamefont {D.}~\bibnamefont {Langlois}}, \ and\ \bibinfo {author} {\bibfnamefont {S.}~\bibnamefont {Mizuno}},\ }\href {\doibase 10.1088/1475-7516/2013/10/023} {\bibfield  {journal} {\bibinfo  {journal} {JCAP}\ }\textbf {\bibinfo {volume} {10}},\ \bibinfo {pages} {023} (\bibinfo {year} {2013})},\ \Eprint {http://arxiv.org/abs/1306.5680} {arXiv:1306.5680 [hep-th]} \BibitemShut {NoStop}%
\bibitem [{\citenamefont {Cespedes}\ \emph {et~al.}(2012)\citenamefont {Cespedes}, \citenamefont {Atal},\ and\ \citenamefont {Palma}}]{Cespedes:2012hu}%
  \BibitemOpen
  \bibfield  {author} {\bibinfo {author} {\bibfnamefont {S.}~\bibnamefont {Cespedes}}, \bibinfo {author} {\bibfnamefont {V.}~\bibnamefont {Atal}}, \ and\ \bibinfo {author} {\bibfnamefont {G.~A.}\ \bibnamefont {Palma}},\ }\href {\doibase 10.1088/1475-7516/2012/05/008} {\bibfield  {journal} {\bibinfo  {journal} {JCAP}\ }\textbf {\bibinfo {volume} {05}},\ \bibinfo {pages} {008} (\bibinfo {year} {2012})},\ \Eprint {http://arxiv.org/abs/1201.4848} {arXiv:1201.4848 [hep-th]} \BibitemShut {NoStop}%
\bibitem [{\citenamefont {Achucarro}\ \emph {et~al.}(2012{\natexlab{a}})\citenamefont {Achucarro}, \citenamefont {Gong}, \citenamefont {Hardeman}, \citenamefont {Palma},\ and\ \citenamefont {Patil}}]{Achucarro:2012sm}%
  \BibitemOpen
  \bibfield  {author} {\bibinfo {author} {\bibfnamefont {A.}~\bibnamefont {Achucarro}}, \bibinfo {author} {\bibfnamefont {J.-O.}\ \bibnamefont {Gong}}, \bibinfo {author} {\bibfnamefont {S.}~\bibnamefont {Hardeman}}, \bibinfo {author} {\bibfnamefont {G.~A.}\ \bibnamefont {Palma}}, \ and\ \bibinfo {author} {\bibfnamefont {S.~P.}\ \bibnamefont {Patil}},\ }\href {\doibase 10.1007/JHEP05(2012)066} {\bibfield  {journal} {\bibinfo  {journal} {JHEP}\ }\textbf {\bibinfo {volume} {05}},\ \bibinfo {pages} {066} (\bibinfo {year} {2012}{\natexlab{a}})},\ \Eprint {http://arxiv.org/abs/1201.6342} {arXiv:1201.6342 [hep-th]} \BibitemShut {NoStop}%
\bibitem [{\citenamefont {Achucarro}\ \emph {et~al.}(2012{\natexlab{b}})\citenamefont {Achucarro}, \citenamefont {Atal}, \citenamefont {Cespedes}, \citenamefont {Gong}, \citenamefont {Palma},\ and\ \citenamefont {Patil}}]{Achucarro:2012yr}%
  \BibitemOpen
  \bibfield  {author} {\bibinfo {author} {\bibfnamefont {A.}~\bibnamefont {Achucarro}}, \bibinfo {author} {\bibfnamefont {V.}~\bibnamefont {Atal}}, \bibinfo {author} {\bibfnamefont {S.}~\bibnamefont {Cespedes}}, \bibinfo {author} {\bibfnamefont {J.-O.}\ \bibnamefont {Gong}}, \bibinfo {author} {\bibfnamefont {G.~A.}\ \bibnamefont {Palma}}, \ and\ \bibinfo {author} {\bibfnamefont {S.~P.}\ \bibnamefont {Patil}},\ }\href {\doibase 10.1103/PhysRevD.86.121301} {\bibfield  {journal} {\bibinfo  {journal} {Phys. Rev. D}\ }\textbf {\bibinfo {volume} {86}},\ \bibinfo {pages} {121301} (\bibinfo {year} {2012}{\natexlab{b}})},\ \Eprint {http://arxiv.org/abs/1205.0710} {arXiv:1205.0710 [hep-th]} \BibitemShut {NoStop}%
\bibitem [{\citenamefont {Achucarro}\ \emph {et~al.}(2011{\natexlab{b}})\citenamefont {Achucarro}, \citenamefont {Gong}, \citenamefont {Hardeman}, \citenamefont {Palma},\ and\ \citenamefont {Patil}}]{Achucarro:2010jv}%
  \BibitemOpen
  \bibfield  {author} {\bibinfo {author} {\bibfnamefont {A.}~\bibnamefont {Achucarro}}, \bibinfo {author} {\bibfnamefont {J.-O.}\ \bibnamefont {Gong}}, \bibinfo {author} {\bibfnamefont {S.}~\bibnamefont {Hardeman}}, \bibinfo {author} {\bibfnamefont {G.~A.}\ \bibnamefont {Palma}}, \ and\ \bibinfo {author} {\bibfnamefont {S.~P.}\ \bibnamefont {Patil}},\ }\href {\doibase 10.1103/PhysRevD.84.043502} {\bibfield  {journal} {\bibinfo  {journal} {Phys. Rev. D}\ }\textbf {\bibinfo {volume} {84}},\ \bibinfo {pages} {043502} (\bibinfo {year} {2011}{\natexlab{b}})},\ \Eprint {http://arxiv.org/abs/1005.3848} {arXiv:1005.3848 [hep-th]} \BibitemShut {NoStop}%
\bibitem [{\citenamefont {Palma}(2015)}]{Palma:2014hra}%
  \BibitemOpen
  \bibfield  {author} {\bibinfo {author} {\bibfnamefont {G.~A.}\ \bibnamefont {Palma}},\ }\href {\doibase 10.1088/1475-7516/2015/04/035} {\bibfield  {journal} {\bibinfo  {journal} {JCAP}\ }\textbf {\bibinfo {volume} {04}},\ \bibinfo {pages} {035} (\bibinfo {year} {2015})},\ \Eprint {http://arxiv.org/abs/1412.5615} {arXiv:1412.5615 [hep-th]} \BibitemShut {NoStop}%
\bibitem [{\citenamefont {Takahashi}\ \emph {et~al.}(2012)\citenamefont {Takahashi}, \citenamefont {Sato}, \citenamefont {Nishimichi}, \citenamefont {Taruya},\ and\ \citenamefont {Oguri}}]{Takahashi:2012em}%
  \BibitemOpen
  \bibfield  {author} {\bibinfo {author} {\bibfnamefont {R.}~\bibnamefont {Takahashi}}, \bibinfo {author} {\bibfnamefont {M.}~\bibnamefont {Sato}}, \bibinfo {author} {\bibfnamefont {T.}~\bibnamefont {Nishimichi}}, \bibinfo {author} {\bibfnamefont {A.}~\bibnamefont {Taruya}}, \ and\ \bibinfo {author} {\bibfnamefont {M.}~\bibnamefont {Oguri}},\ }\href {\doibase 10.1088/0004-637X/761/2/152} {\bibfield  {journal} {\bibinfo  {journal} {Astrophys. J.}\ }\textbf {\bibinfo {volume} {761}},\ \bibinfo {pages} {152} (\bibinfo {year} {2012})},\ \Eprint {http://arxiv.org/abs/1208.2701} {arXiv:1208.2701 [astro-ph.CO]} \BibitemShut {NoStop}%
\bibitem [{\citenamefont {Blas}\ \emph {et~al.}(2011)\citenamefont {Blas}, \citenamefont {Lesgourgues},\ and\ \citenamefont {Tram}}]{Blas:2011rf}%
  \BibitemOpen
  \bibfield  {author} {\bibinfo {author} {\bibfnamefont {D.}~\bibnamefont {Blas}}, \bibinfo {author} {\bibfnamefont {J.}~\bibnamefont {Lesgourgues}}, \ and\ \bibinfo {author} {\bibfnamefont {T.}~\bibnamefont {Tram}},\ }\href {\doibase 10.1088/1475-7516/2011/07/034} {\bibfield  {journal} {\bibinfo  {journal} {JCAP}\ }\textbf {\bibinfo {volume} {07}},\ \bibinfo {pages} {034} (\bibinfo {year} {2011})},\ \Eprint {http://arxiv.org/abs/1104.2933} {arXiv:1104.2933 [astro-ph.CO]} \BibitemShut {NoStop}%
\bibitem [{\citenamefont {Audren}\ \emph {et~al.}(2013)\citenamefont {Audren}, \citenamefont {Lesgourgues}, \citenamefont {Benabed},\ and\ \citenamefont {Prunet}}]{Audren:2012wb}%
  \BibitemOpen
  \bibfield  {author} {\bibinfo {author} {\bibfnamefont {B.}~\bibnamefont {Audren}}, \bibinfo {author} {\bibfnamefont {J.}~\bibnamefont {Lesgourgues}}, \bibinfo {author} {\bibfnamefont {K.}~\bibnamefont {Benabed}}, \ and\ \bibinfo {author} {\bibfnamefont {S.}~\bibnamefont {Prunet}},\ }\href {\doibase 10.1088/1475-7516/2013/02/001} {\bibfield  {journal} {\bibinfo  {journal} {JCAP}\ }\textbf {\bibinfo {volume} {1302}},\ \bibinfo {pages} {001} (\bibinfo {year} {2013})},\ \Eprint {http://arxiv.org/abs/1210.7183} {arXiv:1210.7183 [astro-ph.CO]} \BibitemShut {NoStop}%
\bibitem [{\citenamefont {Brinckmann}\ and\ \citenamefont {Lesgourgues}(2018)}]{Brinckmann:2018cvx}%
  \BibitemOpen
  \bibfield  {author} {\bibinfo {author} {\bibfnamefont {T.}~\bibnamefont {Brinckmann}}\ and\ \bibinfo {author} {\bibfnamefont {J.}~\bibnamefont {Lesgourgues}},\ }\href@noop {} {\  (\bibinfo {year} {2018})},\ \Eprint {http://arxiv.org/abs/1804.07261} {arXiv:1804.07261 [astro-ph.CO]} \BibitemShut {NoStop}%
\bibitem [{\citenamefont {Lewis}(2013)}]{Lewis:2013hha}%
  \BibitemOpen
  \bibfield  {author} {\bibinfo {author} {\bibfnamefont {A.}~\bibnamefont {Lewis}},\ }\href {\doibase 10.1103/PhysRevD.87.103529} {\bibfield  {journal} {\bibinfo  {journal} {\prd}\ }\textbf {\bibinfo {volume} {87}},\ \bibinfo {pages} {103529} (\bibinfo {year} {2013})},\ \Eprint {http://arxiv.org/abs/1304.4473} {arXiv:1304.4473 [astro-ph.CO]} \BibitemShut {NoStop}%
\bibitem [{\citenamefont {Gelman}\ and\ \citenamefont {Rubin}(1992)}]{Gelman:1992zz}%
  \BibitemOpen
  \bibfield  {author} {\bibinfo {author} {\bibfnamefont {A.}~\bibnamefont {Gelman}}\ and\ \bibinfo {author} {\bibfnamefont {D.~B.}\ \bibnamefont {Rubin}},\ }\href {\doibase 10.1214/ss/1177011136} {\bibfield  {journal} {\bibinfo  {journal} {Statist. Sci.}\ }\textbf {\bibinfo {volume} {7}},\ \bibinfo {pages} {457} (\bibinfo {year} {1992})}\BibitemShut {NoStop}%
\bibitem [{\citenamefont {Qu}\ \emph {et~al.}(2024)\citenamefont {Qu} \emph {et~al.}}]{ACT:2023dou}%
  \BibitemOpen
  \bibfield  {author} {\bibinfo {author} {\bibfnamefont {F.~J.}\ \bibnamefont {Qu}} \emph {et~al.} (\bibinfo {collaboration} {ACT}),\ }\href {\doibase 10.3847/1538-4357/acfe06} {\bibfield  {journal} {\bibinfo  {journal} {Astrophys. J.}\ }\textbf {\bibinfo {volume} {962}},\ \bibinfo {pages} {112} (\bibinfo {year} {2024})},\ \Eprint {http://arxiv.org/abs/2304.05202} {arXiv:2304.05202 [astro-ph.CO]} \BibitemShut {NoStop}%
\bibitem [{\citenamefont {Ge}\ \emph {et~al.}(2024)\citenamefont {Ge} \emph {et~al.}}]{SPT-3G:2024atg}%
  \BibitemOpen
  \bibfield  {author} {\bibinfo {author} {\bibfnamefont {F.}~\bibnamefont {Ge}} \emph {et~al.} (\bibinfo {collaboration} {SPT-3G}),\ }\href@noop {} {\  (\bibinfo {year} {2024})},\ \Eprint {http://arxiv.org/abs/2411.06000} {arXiv:2411.06000 [astro-ph.CO]} \BibitemShut {NoStop}%
\bibitem [{\citenamefont {Amon}\ and\ \citenamefont {Efstathiou}(2022)}]{Amon:2022azi}%
  \BibitemOpen
  \bibfield  {author} {\bibinfo {author} {\bibfnamefont {A.}~\bibnamefont {Amon}}\ and\ \bibinfo {author} {\bibfnamefont {G.}~\bibnamefont {Efstathiou}},\ }\href {\doibase 10.1093/mnras/stac2429} {\bibfield  {journal} {\bibinfo  {journal} {Mon. Not. Roy. Astron. Soc.}\ }\textbf {\bibinfo {volume} {516}},\ \bibinfo {pages} {5355} (\bibinfo {year} {2022})},\ \Eprint {http://arxiv.org/abs/2206.11794} {arXiv:2206.11794 [astro-ph.CO]} \BibitemShut {NoStop}%
\bibitem [{\citenamefont {Preston}\ \emph {et~al.}(2023)\citenamefont {Preston}, \citenamefont {Amon},\ and\ \citenamefont {Efstathiou}}]{Preston:2023uup}%
  \BibitemOpen
  \bibfield  {author} {\bibinfo {author} {\bibfnamefont {C.}~\bibnamefont {Preston}}, \bibinfo {author} {\bibfnamefont {A.}~\bibnamefont {Amon}}, \ and\ \bibinfo {author} {\bibfnamefont {G.}~\bibnamefont {Efstathiou}},\ }\href {\doibase 10.1093/mnras/stad2573} {\bibfield  {journal} {\bibinfo  {journal} {Mon. Not. Roy. Astron. Soc.}\ }\textbf {\bibinfo {volume} {525}},\ \bibinfo {pages} {5554} (\bibinfo {year} {2023})},\ \Eprint {http://arxiv.org/abs/2305.09827} {arXiv:2305.09827 [astro-ph.CO]} \BibitemShut {NoStop}%
\bibitem [{\citenamefont {Angulo}\ and\ \citenamefont {Hahn}(2021)}]{Angulo:2021kes}%
  \BibitemOpen
  \bibfield  {author} {\bibinfo {author} {\bibfnamefont {R.~E.}\ \bibnamefont {Angulo}}\ and\ \bibinfo {author} {\bibfnamefont {O.}~\bibnamefont {Hahn}},\ }\href {\doibase 10.1007/s41115-021-00013-z} {\  (\bibinfo {year} {2021}),\ 10.1007/s41115-021-00013-z},\ \Eprint {http://arxiv.org/abs/2112.05165} {arXiv:2112.05165 [astro-ph.CO]} \BibitemShut {NoStop}%
\bibitem [{\citenamefont {Springel}\ \emph {et~al.}(2021)\citenamefont {Springel}, \citenamefont {Pakmor}, \citenamefont {Zier},\ and\ \citenamefont {Reinecke}}]{Springel:2020plp}%
  \BibitemOpen
  \bibfield  {author} {\bibinfo {author} {\bibfnamefont {V.}~\bibnamefont {Springel}}, \bibinfo {author} {\bibfnamefont {R.}~\bibnamefont {Pakmor}}, \bibinfo {author} {\bibfnamefont {O.}~\bibnamefont {Zier}}, \ and\ \bibinfo {author} {\bibfnamefont {M.}~\bibnamefont {Reinecke}},\ }\href {\doibase 10.1093/mnras/stab1855} {\bibfield  {journal} {\bibinfo  {journal} {Mon. Not. Roy. Astron. Soc.}\ }\textbf {\bibinfo {volume} {506}},\ \bibinfo {pages} {2871} (\bibinfo {year} {2021})},\ \Eprint {http://arxiv.org/abs/2010.03567} {arXiv:2010.03567 [astro-ph.IM]} \BibitemShut {NoStop}%
\bibitem [{\citenamefont {Seo}\ \emph {et~al.}(2008)\citenamefont {Seo}, \citenamefont {Siegel}, \citenamefont {Eisenstein},\ and\ \citenamefont {White}}]{Seo:2008yx}%
  \BibitemOpen
  \bibfield  {author} {\bibinfo {author} {\bibfnamefont {H.-J.}\ \bibnamefont {Seo}}, \bibinfo {author} {\bibfnamefont {E.~R.}\ \bibnamefont {Siegel}}, \bibinfo {author} {\bibfnamefont {D.~J.}\ \bibnamefont {Eisenstein}}, \ and\ \bibinfo {author} {\bibfnamefont {M.}~\bibnamefont {White}},\ }\href {\doibase 10.1086/589921} {\bibfield  {journal} {\bibinfo  {journal} {Astrophys. J.}\ }\textbf {\bibinfo {volume} {686}},\ \bibinfo {pages} {13} (\bibinfo {year} {2008})},\ \Eprint {http://arxiv.org/abs/0805.0117} {arXiv:0805.0117 [astro-ph]} \BibitemShut {NoStop}%
\bibitem [{\citenamefont {Buckley}\ \emph {et~al.}(2014)\citenamefont {Buckley}, \citenamefont {Zavala}, \citenamefont {Cyr-Racine}, \citenamefont {Sigurdson},\ and\ \citenamefont {Vogelsberger}}]{Buckley:2014hja}%
  \BibitemOpen
  \bibfield  {author} {\bibinfo {author} {\bibfnamefont {M.~R.}\ \bibnamefont {Buckley}}, \bibinfo {author} {\bibfnamefont {J.}~\bibnamefont {Zavala}}, \bibinfo {author} {\bibfnamefont {F.-Y.}\ \bibnamefont {Cyr-Racine}}, \bibinfo {author} {\bibfnamefont {K.}~\bibnamefont {Sigurdson}}, \ and\ \bibinfo {author} {\bibfnamefont {M.}~\bibnamefont {Vogelsberger}},\ }\href {\doibase 10.1103/PhysRevD.90.043524} {\bibfield  {journal} {\bibinfo  {journal} {Phys. Rev. D}\ }\textbf {\bibinfo {volume} {90}},\ \bibinfo {pages} {043524} (\bibinfo {year} {2014})},\ \Eprint {http://arxiv.org/abs/1405.2075} {arXiv:1405.2075 [astro-ph.CO]} \BibitemShut {NoStop}%
\bibitem [{\citenamefont {Schaeffer}\ and\ \citenamefont {Schneider}(2021)}]{Schaeffer:2021qwm}%
  \BibitemOpen
  \bibfield  {author} {\bibinfo {author} {\bibfnamefont {T.}~\bibnamefont {Schaeffer}}\ and\ \bibinfo {author} {\bibfnamefont {A.}~\bibnamefont {Schneider}},\ }\href {\doibase 10.1093/mnras/stab1116} {\bibfield  {journal} {\bibinfo  {journal} {Mon. Not. Roy. Astron. Soc.}\ }\textbf {\bibinfo {volume} {504}},\ \bibinfo {pages} {3773} (\bibinfo {year} {2021})},\ \Eprint {http://arxiv.org/abs/2101.12229} {arXiv:2101.12229 [astro-ph.CO]} \BibitemShut {NoStop}%
\bibitem [{\citenamefont {Stahl}\ \emph {et~al.}(2024)\citenamefont {Stahl}, \citenamefont {Famaey}, \citenamefont {Ibata}, \citenamefont {Hahn}, \citenamefont {Martinet},\ and\ \citenamefont {Montandon}}]{Stahl:2024stz}%
  \BibitemOpen
  \bibfield  {author} {\bibinfo {author} {\bibfnamefont {C.}~\bibnamefont {Stahl}}, \bibinfo {author} {\bibfnamefont {B.}~\bibnamefont {Famaey}}, \bibinfo {author} {\bibfnamefont {R.}~\bibnamefont {Ibata}}, \bibinfo {author} {\bibfnamefont {O.}~\bibnamefont {Hahn}}, \bibinfo {author} {\bibfnamefont {N.}~\bibnamefont {Martinet}}, \ and\ \bibinfo {author} {\bibfnamefont {T.}~\bibnamefont {Montandon}},\ }\href {\doibase 10.1103/PhysRevD.110.063501} {\bibfield  {journal} {\bibinfo  {journal} {Phys. Rev. D}\ }\textbf {\bibinfo {volume} {110}},\ \bibinfo {pages} {063501} (\bibinfo {year} {2024})},\ \Eprint {http://arxiv.org/abs/2404.03244} {arXiv:2404.03244 [astro-ph.CO]} \BibitemShut {NoStop}%
\bibitem [{\citenamefont {Fiorino}\ \emph {et~al.}(2024)\citenamefont {Fiorino}, \citenamefont {Contarini}, \citenamefont {Marulli}, \citenamefont {Sanchez}, \citenamefont {Baldi}, \citenamefont {Fiorilli},\ and\ \citenamefont {Moscardini}}]{Fiorino:2024ncx}%
  \BibitemOpen
  \bibfield  {author} {\bibinfo {author} {\bibfnamefont {L.}~\bibnamefont {Fiorino}}, \bibinfo {author} {\bibfnamefont {S.}~\bibnamefont {Contarini}}, \bibinfo {author} {\bibfnamefont {F.}~\bibnamefont {Marulli}}, \bibinfo {author} {\bibfnamefont {A.~G.}\ \bibnamefont {Sanchez}}, \bibinfo {author} {\bibfnamefont {M.}~\bibnamefont {Baldi}}, \bibinfo {author} {\bibfnamefont {A.}~\bibnamefont {Fiorilli}}, \ and\ \bibinfo {author} {\bibfnamefont {L.}~\bibnamefont {Moscardini}},\ }\href@noop {} {\  (\bibinfo {year} {2024})},\ \Eprint {http://arxiv.org/abs/2410.21457} {arXiv:2410.21457 [astro-ph.CO]} \BibitemShut {NoStop}%
\bibitem [{\citenamefont {Murgia}\ \emph {et~al.}(2017)\citenamefont {Murgia}, \citenamefont {Merle}, \citenamefont {Viel}, \citenamefont {Totzauer},\ and\ \citenamefont {Schneider}}]{Murgia:2017lwo}%
  \BibitemOpen
  \bibfield  {author} {\bibinfo {author} {\bibfnamefont {R.}~\bibnamefont {Murgia}}, \bibinfo {author} {\bibfnamefont {A.}~\bibnamefont {Merle}}, \bibinfo {author} {\bibfnamefont {M.}~\bibnamefont {Viel}}, \bibinfo {author} {\bibfnamefont {M.}~\bibnamefont {Totzauer}}, \ and\ \bibinfo {author} {\bibfnamefont {A.}~\bibnamefont {Schneider}},\ }\href {\doibase 10.1088/1475-7516/2017/11/046} {\bibfield  {journal} {\bibinfo  {journal} {JCAP}\ }\textbf {\bibinfo {volume} {11}},\ \bibinfo {pages} {046} (\bibinfo {year} {2017})},\ \Eprint {http://arxiv.org/abs/1704.07838} {arXiv:1704.07838 [astro-ph.CO]} \BibitemShut {NoStop}%
\bibitem [{\citenamefont {St\"ucker}\ \emph {et~al.}(2021)\citenamefont {St\"ucker}, \citenamefont {Angulo}, \citenamefont {Hahn},\ and\ \citenamefont {White}}]{Stucker:2021vyx}%
  \BibitemOpen
  \bibfield  {author} {\bibinfo {author} {\bibfnamefont {J.}~\bibnamefont {St\"ucker}}, \bibinfo {author} {\bibfnamefont {R.~E.}\ \bibnamefont {Angulo}}, \bibinfo {author} {\bibfnamefont {O.}~\bibnamefont {Hahn}}, \ and\ \bibinfo {author} {\bibfnamefont {S.~D.~M.}\ \bibnamefont {White}},\ }\href {\doibase 10.1093/mnras/stab3078} {\bibfield  {journal} {\bibinfo  {journal} {Mon. Not. Roy. Astron. Soc.}\ }\textbf {\bibinfo {volume} {509}},\ \bibinfo {pages} {1703} (\bibinfo {year} {2021})},\ \Eprint {http://arxiv.org/abs/2109.09760} {arXiv:2109.09760 [astro-ph.CO]} \BibitemShut {NoStop}%
\bibitem [{\citenamefont {Cyr-Racine}\ \emph {et~al.}(2016)\citenamefont {Cyr-Racine}, \citenamefont {Sigurdson}, \citenamefont {Zavala}, \citenamefont {Bringmann}, \citenamefont {Vogelsberger},\ and\ \citenamefont {Pfrommer}}]{Cyr-Racine:2015ihg}%
  \BibitemOpen
  \bibfield  {author} {\bibinfo {author} {\bibfnamefont {F.-Y.}\ \bibnamefont {Cyr-Racine}}, \bibinfo {author} {\bibfnamefont {K.}~\bibnamefont {Sigurdson}}, \bibinfo {author} {\bibfnamefont {J.}~\bibnamefont {Zavala}}, \bibinfo {author} {\bibfnamefont {T.}~\bibnamefont {Bringmann}}, \bibinfo {author} {\bibfnamefont {M.}~\bibnamefont {Vogelsberger}}, \ and\ \bibinfo {author} {\bibfnamefont {C.}~\bibnamefont {Pfrommer}},\ }\href {\doibase 10.1103/PhysRevD.93.123527} {\bibfield  {journal} {\bibinfo  {journal} {Phys. Rev. D}\ }\textbf {\bibinfo {volume} {93}},\ \bibinfo {pages} {123527} (\bibinfo {year} {2016})},\ \Eprint {http://arxiv.org/abs/1512.05344} {arXiv:1512.05344 [astro-ph.CO]} \BibitemShut {NoStop}%
\bibitem [{\citenamefont {Enzi}\ \emph {et~al.}(2021)\citenamefont {Enzi} \emph {et~al.}}]{Enzi:2020ieg}%
  \BibitemOpen
  \bibfield  {author} {\bibinfo {author} {\bibfnamefont {W.}~\bibnamefont {Enzi}} \emph {et~al.},\ }\href {\doibase 10.1093/mnras/stab1960} {\bibfield  {journal} {\bibinfo  {journal} {Mon. Not. Roy. Astron. Soc.}\ }\textbf {\bibinfo {volume} {506}},\ \bibinfo {pages} {5848} (\bibinfo {year} {2021})},\ \Eprint {http://arxiv.org/abs/2010.13802} {arXiv:2010.13802 [astro-ph.CO]} \BibitemShut {NoStop}%
\bibitem [{\citenamefont {Das}\ \emph {et~al.}(2022)\citenamefont {Das}, \citenamefont {Maharana}, \citenamefont {Poulin},\ and\ \citenamefont {Sharma}}]{Das:2021pof}%
  \BibitemOpen
  \bibfield  {author} {\bibinfo {author} {\bibfnamefont {S.}~\bibnamefont {Das}}, \bibinfo {author} {\bibfnamefont {A.}~\bibnamefont {Maharana}}, \bibinfo {author} {\bibfnamefont {V.}~\bibnamefont {Poulin}}, \ and\ \bibinfo {author} {\bibfnamefont {R.~K.}\ \bibnamefont {Sharma}},\ }\href {\doibase 10.1103/PhysRevD.105.103503} {\bibfield  {journal} {\bibinfo  {journal} {Phys. Rev. D}\ }\textbf {\bibinfo {volume} {105}},\ \bibinfo {pages} {103503} (\bibinfo {year} {2022})},\ \Eprint {http://arxiv.org/abs/2104.03329} {arXiv:2104.03329 [astro-ph.CO]} \BibitemShut {NoStop}%
\bibitem [{\citenamefont {Stahl}\ \emph {et~al.}(2025)\citenamefont {Stahl}, \citenamefont {Famaey}, \citenamefont {Ibata}, \citenamefont {Kraljic},\ and\ \citenamefont {Castillo}}]{Stahl:2025nta}%
  \BibitemOpen
  \bibfield  {author} {\bibinfo {author} {\bibfnamefont {C.}~\bibnamefont {Stahl}}, \bibinfo {author} {\bibfnamefont {B.}~\bibnamefont {Famaey}}, \bibinfo {author} {\bibfnamefont {R.}~\bibnamefont {Ibata}}, \bibinfo {author} {\bibfnamefont {K.}~\bibnamefont {Kraljic}}, \ and\ \bibinfo {author} {\bibfnamefont {F.}~\bibnamefont {Castillo}},\ }\href@noop {} {\  (\bibinfo {year} {2025})},\ \Eprint {http://arxiv.org/abs/2501.03672} {arXiv:2501.03672 [astro-ph.CO]} \BibitemShut {NoStop}%
\bibitem [{\citenamefont {Bohr}\ \emph {et~al.}(2020)\citenamefont {Bohr}, \citenamefont {Zavala}, \citenamefont {Cyr-Racine}, \citenamefont {Vogelsberger}, \citenamefont {Bringmann},\ and\ \citenamefont {Pfrommer}}]{Bohr:2020yoe}%
  \BibitemOpen
  \bibfield  {author} {\bibinfo {author} {\bibfnamefont {S.}~\bibnamefont {Bohr}}, \bibinfo {author} {\bibfnamefont {J.}~\bibnamefont {Zavala}}, \bibinfo {author} {\bibfnamefont {F.-Y.}\ \bibnamefont {Cyr-Racine}}, \bibinfo {author} {\bibfnamefont {M.}~\bibnamefont {Vogelsberger}}, \bibinfo {author} {\bibfnamefont {T.}~\bibnamefont {Bringmann}}, \ and\ \bibinfo {author} {\bibfnamefont {C.}~\bibnamefont {Pfrommer}},\ }\href {\doibase 10.1093/mnras/staa2579} {\bibfield  {journal} {\bibinfo  {journal} {Mon. Not. Roy. Astron. Soc.}\ }\textbf {\bibinfo {volume} {498}},\ \bibinfo {pages} {3403} (\bibinfo {year} {2020})},\ \Eprint {http://arxiv.org/abs/2006.01842} {arXiv:2006.01842 [astro-ph.CO]} \BibitemShut {NoStop}%
\bibitem [{\citenamefont {Bohr}\ \emph {et~al.}(2021)\citenamefont {Bohr}, \citenamefont {Zavala}, \citenamefont {Cyr-Racine},\ and\ \citenamefont {Vogelsberger}}]{Bohr:2021bdm}%
  \BibitemOpen
  \bibfield  {author} {\bibinfo {author} {\bibfnamefont {S.}~\bibnamefont {Bohr}}, \bibinfo {author} {\bibfnamefont {J.}~\bibnamefont {Zavala}}, \bibinfo {author} {\bibfnamefont {F.-Y.}\ \bibnamefont {Cyr-Racine}}, \ and\ \bibinfo {author} {\bibfnamefont {M.}~\bibnamefont {Vogelsberger}},\ }\href {\doibase 10.1093/mnras/stab1758} {\bibfield  {journal} {\bibinfo  {journal} {Mon. Not. Roy. Astron. Soc.}\ }\textbf {\bibinfo {volume} {506}},\ \bibinfo {pages} {128} (\bibinfo {year} {2021})},\ \Eprint {http://arxiv.org/abs/2101.08790} {arXiv:2101.08790 [astro-ph.CO]} \BibitemShut {NoStop}%
\bibitem [{\citenamefont {Koyama}\ \emph {et~al.}(2024)\citenamefont {Koyama} \emph {et~al.}}]{Euclid:2024xfd}%
  \BibitemOpen
  \bibfield  {author} {\bibinfo {author} {\bibfnamefont {K.}~\bibnamefont {Koyama}} \emph {et~al.} (\bibinfo {collaboration} {Euclid}),\ }\href@noop {} {\  (\bibinfo {year} {2024})},\ \Eprint {http://arxiv.org/abs/2409.03524} {arXiv:2409.03524 [astro-ph.CO]} \BibitemShut {NoStop}%
\bibitem [{\citenamefont {Smith}\ \emph {et~al.}(2024)\citenamefont {Smith}, \citenamefont {Mylova}, \citenamefont {Brax}, \citenamefont {van~de Bruck}, \citenamefont {Burgess},\ and\ \citenamefont {Davis}}]{Smith:2024ibv}%
  \BibitemOpen
  \bibfield  {author} {\bibinfo {author} {\bibfnamefont {A.}~\bibnamefont {Smith}}, \bibinfo {author} {\bibfnamefont {M.}~\bibnamefont {Mylova}}, \bibinfo {author} {\bibfnamefont {P.}~\bibnamefont {Brax}}, \bibinfo {author} {\bibfnamefont {C.}~\bibnamefont {van~de Bruck}}, \bibinfo {author} {\bibfnamefont {C.~P.}\ \bibnamefont {Burgess}}, \ and\ \bibinfo {author} {\bibfnamefont {A.-C.}\ \bibnamefont {Davis}},\ }\href@noop {} {\  (\bibinfo {year} {2024})},\ \Eprint {http://arxiv.org/abs/2410.11099} {arXiv:2410.11099 [hep-th]} \BibitemShut {NoStop}%
\bibitem [{\citenamefont {Cannone}\ \emph {et~al.}(2014)\citenamefont {Cannone}, \citenamefont {Bartolo},\ and\ \citenamefont {Matarrese}}]{Cannone:2014qna}%
  \BibitemOpen
  \bibfield  {author} {\bibinfo {author} {\bibfnamefont {D.}~\bibnamefont {Cannone}}, \bibinfo {author} {\bibfnamefont {N.}~\bibnamefont {Bartolo}}, \ and\ \bibinfo {author} {\bibfnamefont {S.}~\bibnamefont {Matarrese}},\ }\href {\doibase 10.1103/PhysRevD.89.127301} {\bibfield  {journal} {\bibinfo  {journal} {Phys. Rev. D}\ }\textbf {\bibinfo {volume} {89}},\ \bibinfo {pages} {127301} (\bibinfo {year} {2014})},\ \Eprint {http://arxiv.org/abs/1402.2258} {arXiv:1402.2258 [astro-ph.CO]} \BibitemShut {NoStop}%
\bibitem [{\citenamefont {Rosenberg}\ \emph {et~al.}(2022)\citenamefont {Rosenberg}, \citenamefont {Gratton},\ and\ \citenamefont {Efstathiou}}]{Rosenberg:2022sdy}%
  \BibitemOpen
  \bibfield  {author} {\bibinfo {author} {\bibfnamefont {E.}~\bibnamefont {Rosenberg}}, \bibinfo {author} {\bibfnamefont {S.}~\bibnamefont {Gratton}}, \ and\ \bibinfo {author} {\bibfnamefont {G.}~\bibnamefont {Efstathiou}},\ }\href {\doibase 10.1093/mnras/stac2744} {\bibfield  {journal} {\bibinfo  {journal} {Mon. Not. Roy. Astron. Soc.}\ }\textbf {\bibinfo {volume} {517}},\ \bibinfo {pages} {4620} (\bibinfo {year} {2022})},\ \Eprint {http://arxiv.org/abs/2205.10869} {arXiv:2205.10869 [astro-ph.CO]} \BibitemShut {NoStop}%
\bibitem [{\citenamefont {Tristram}\ \emph {et~al.}(2024)\citenamefont {Tristram} \emph {et~al.}}]{Tristram:2023haj}%
  \BibitemOpen
  \bibfield  {author} {\bibinfo {author} {\bibfnamefont {M.}~\bibnamefont {Tristram}} \emph {et~al.},\ }\href {\doibase 10.1051/0004-6361/202348015} {\bibfield  {journal} {\bibinfo  {journal} {Astron. Astrophys.}\ }\textbf {\bibinfo {volume} {682}},\ \bibinfo {pages} {A37} (\bibinfo {year} {2024})},\ \Eprint {http://arxiv.org/abs/2309.10034} {arXiv:2309.10034 [astro-ph.CO]} \BibitemShut {NoStop}%
\bibitem [{\citenamefont {Pajer}\ \emph {et~al.}(2013)\citenamefont {Pajer}, \citenamefont {Schmidt},\ and\ \citenamefont {Zaldarriaga}}]{Pajer:2013ana}%
  \BibitemOpen
  \bibfield  {author} {\bibinfo {author} {\bibfnamefont {E.}~\bibnamefont {Pajer}}, \bibinfo {author} {\bibfnamefont {F.}~\bibnamefont {Schmidt}}, \ and\ \bibinfo {author} {\bibfnamefont {M.}~\bibnamefont {Zaldarriaga}},\ }\href {\doibase 10.1103/PhysRevD.88.083502} {\bibfield  {journal} {\bibinfo  {journal} {Phys. Rev. D}\ }\textbf {\bibinfo {volume} {88}},\ \bibinfo {pages} {083502} (\bibinfo {year} {2013})},\ \Eprint {http://arxiv.org/abs/1305.0824} {arXiv:1305.0824 [astro-ph.CO]} \BibitemShut {NoStop}%
\bibitem [{\citenamefont {{Turk}}\ \emph {et~al.}(2011)\citenamefont {{Turk}}, \citenamefont {{Smith}}, \citenamefont {{Oishi}}, \citenamefont {{Skory}}, \citenamefont {{Skillman}}, \citenamefont {{Abel}},\ and\ \citenamefont {{Norman}}}]{Turk:2010ah}%
  \BibitemOpen
  \bibfield  {author} {\bibinfo {author} {\bibfnamefont {M.~J.}\ \bibnamefont {{Turk}}}, \bibinfo {author} {\bibfnamefont {B.~D.}\ \bibnamefont {{Smith}}}, \bibinfo {author} {\bibfnamefont {J.~S.}\ \bibnamefont {{Oishi}}}, \bibinfo {author} {\bibfnamefont {S.}~\bibnamefont {{Skory}}}, \bibinfo {author} {\bibfnamefont {S.~W.}\ \bibnamefont {{Skillman}}}, \bibinfo {author} {\bibfnamefont {T.}~\bibnamefont {{Abel}}}, \ and\ \bibinfo {author} {\bibfnamefont {M.~L.}\ \bibnamefont {{Norman}}},\ }\href {\doibase 10.1088/0067-0049/192/1/9} {\bibfield  {journal} {\bibinfo  {journal} {apjs}\ }\textbf {\bibinfo {volume} {192}},\ \bibinfo {eid} {9} (\bibinfo {year} {2011})},\ \Eprint {http://arxiv.org/abs/1011.3514} {arXiv:1011.3514 [astro-ph.IM]} \BibitemShut {NoStop}%
\bibitem [{\citenamefont {{Villaescusa-Navarro}}(2018)}]{Pylians}%
  \BibitemOpen
  \bibfield  {author} {\bibinfo {author} {\bibfnamefont {F.}~\bibnamefont {{Villaescusa-Navarro}}},\ }\href@noop {} {\enquote {\bibinfo {title} {{Pylians: Python libraries for the analysis of numerical simulations}},}\ }\bibinfo {howpublished} {Astrophysics Source Code Library, record ascl:1811.008} (\bibinfo {year} {2018}),\ \Eprint {http://arxiv.org/abs/1811.008} {ascl:1811.008} \BibitemShut {NoStop}%
\bibitem [{\citenamefont {Perez}\ and\ \citenamefont {Granger}(2007)}]{Perez:2007emg}%
  \BibitemOpen
  \bibfield  {author} {\bibinfo {author} {\bibfnamefont {F.}~\bibnamefont {Perez}}\ and\ \bibinfo {author} {\bibfnamefont {B.~E.}\ \bibnamefont {Granger}},\ }\href {\doibase 10.1109/MCSE.2007.53} {\bibfield  {journal} {\bibinfo  {journal} {Comput. Sci. Eng.}\ }\textbf {\bibinfo {volume} {9}},\ \bibinfo {pages} {21} (\bibinfo {year} {2007})}\BibitemShut {NoStop}%
\bibitem [{\citenamefont {Hunter}(2007)}]{Hunter:2007ouj}%
  \BibitemOpen
  \bibfield  {author} {\bibinfo {author} {\bibfnamefont {J.~D.}\ \bibnamefont {Hunter}},\ }\href {\doibase 10.1109/MCSE.2007.55} {\bibfield  {journal} {\bibinfo  {journal} {Comput. Sci. Eng.}\ }\textbf {\bibinfo {volume} {9}},\ \bibinfo {pages} {90} (\bibinfo {year} {2007})}\BibitemShut {NoStop}%
\bibitem [{\citenamefont {van~der Walt}\ \emph {et~al.}(2011)\citenamefont {van~der Walt}, \citenamefont {Colbert},\ and\ \citenamefont {Varoquaux}}]{vanderWalt:2011bqk}%
  \BibitemOpen
  \bibfield  {author} {\bibinfo {author} {\bibfnamefont {S.}~\bibnamefont {van~der Walt}}, \bibinfo {author} {\bibfnamefont {S.~C.}\ \bibnamefont {Colbert}}, \ and\ \bibinfo {author} {\bibfnamefont {G.}~\bibnamefont {Varoquaux}},\ }\href {\doibase 10.1109/MCSE.2011.37} {\bibfield  {journal} {\bibinfo  {journal} {Comput. Sci. Eng.}\ }\textbf {\bibinfo {volume} {13}},\ \bibinfo {pages} {22} (\bibinfo {year} {2011})},\ \Eprint {http://arxiv.org/abs/1102.1523} {arXiv:1102.1523 [cs.MS]} \BibitemShut {NoStop}%
\bibitem [{\citenamefont {Berthoud}\ \emph {et~al.}(2020)\citenamefont {Berthoud}, \citenamefont {Bzeznik}, \citenamefont {Gibelin}, \citenamefont {Laurens}, \citenamefont {Bonamy}, \citenamefont {Morel},\ and\ \citenamefont {Schwindenhammer}}]{berthoud}%
  \BibitemOpen
  \bibfield  {author} {\bibinfo {author} {\bibfnamefont {F.}~\bibnamefont {Berthoud}}, \bibinfo {author} {\bibfnamefont {B.}~\bibnamefont {Bzeznik}}, \bibinfo {author} {\bibfnamefont {N.}~\bibnamefont {Gibelin}}, \bibinfo {author} {\bibfnamefont {M.}~\bibnamefont {Laurens}}, \bibinfo {author} {\bibfnamefont {C.}~\bibnamefont {Bonamy}}, \bibinfo {author} {\bibfnamefont {M.}~\bibnamefont {Morel}}, \ and\ \bibinfo {author} {\bibfnamefont {X.}~\bibnamefont {Schwindenhammer}},\ }\href {https://hal.archives-ouvertes.fr/hal-02549565} {\emph {\bibinfo {title} {{Estimation de l'empreinte carbone d'une heure.coeur de calcul}}}},\ \bibinfo {type} {Research Report}\ (\bibinfo  {institution} {{UGA - Universit{\'e} Grenoble Alpes ; CNRS ; INP Grenoble ; INRIA}},\ \bibinfo {year} {2020})\BibitemShut {NoStop}%
\bibitem [{\citenamefont {Weinberg}(2005)}]{Weinberg:2005vy}%
  \BibitemOpen
  \bibfield  {author} {\bibinfo {author} {\bibfnamefont {S.}~\bibnamefont {Weinberg}},\ }\href {\doibase 10.1103/PhysRevD.72.043514} {\bibfield  {journal} {\bibinfo  {journal} {Phys. Rev. D}\ }\textbf {\bibinfo {volume} {72}},\ \bibinfo {pages} {043514} (\bibinfo {year} {2005})},\ \Eprint {http://arxiv.org/abs/hep-th/0506236} {arXiv:hep-th/0506236} \BibitemShut {NoStop}%
\bibitem [{\citenamefont {Chen}\ \emph {et~al.}(2017)\citenamefont {Chen}, \citenamefont {Wang},\ and\ \citenamefont {Xianyu}}]{Chen:2017ryl}%
  \BibitemOpen
  \bibfield  {author} {\bibinfo {author} {\bibfnamefont {X.}~\bibnamefont {Chen}}, \bibinfo {author} {\bibfnamefont {Y.}~\bibnamefont {Wang}}, \ and\ \bibinfo {author} {\bibfnamefont {Z.-Z.}\ \bibnamefont {Xianyu}},\ }\href {\doibase 10.1088/1475-7516/2017/12/006} {\bibfield  {journal} {\bibinfo  {journal} {JCAP}\ }\textbf {\bibinfo {volume} {12}},\ \bibinfo {pages} {006} (\bibinfo {year} {2017})},\ \Eprint {http://arxiv.org/abs/1703.10166} {arXiv:1703.10166 [hep-th]} \BibitemShut {NoStop}%
\bibitem [{\citenamefont {Feroz}\ \emph {et~al.}(2009)\citenamefont {Feroz}, \citenamefont {Hobson},\ and\ \citenamefont {Bridges}}]{Feroz:2008xx}%
  \BibitemOpen
  \bibfield  {author} {\bibinfo {author} {\bibfnamefont {F.}~\bibnamefont {Feroz}}, \bibinfo {author} {\bibfnamefont {M.~P.}\ \bibnamefont {Hobson}}, \ and\ \bibinfo {author} {\bibfnamefont {M.}~\bibnamefont {Bridges}},\ }\href {\doibase 10.1111/j.1365-2966.2009.14548.x} {\bibfield  {journal} {\bibinfo  {journal} {Mon. Not. Roy. Astron. Soc.}\ }\textbf {\bibinfo {volume} {398}},\ \bibinfo {pages} {1601} (\bibinfo {year} {2009})},\ \Eprint {http://arxiv.org/abs/0809.3437} {arXiv:0809.3437 [astro-ph]} \BibitemShut {NoStop}%
\bibitem [{\citenamefont {Karwal}\ \emph {et~al.}(2024)\citenamefont {Karwal}, \citenamefont {Patel}, \citenamefont {Bartlett}, \citenamefont {Poulin}, \citenamefont {Smith},\ and\ \citenamefont {Pfeffer}}]{Karwal:2024qpt}%
  \BibitemOpen
  \bibfield  {author} {\bibinfo {author} {\bibfnamefont {T.}~\bibnamefont {Karwal}}, \bibinfo {author} {\bibfnamefont {Y.}~\bibnamefont {Patel}}, \bibinfo {author} {\bibfnamefont {A.}~\bibnamefont {Bartlett}}, \bibinfo {author} {\bibfnamefont {V.}~\bibnamefont {Poulin}}, \bibinfo {author} {\bibfnamefont {T.~L.}\ \bibnamefont {Smith}}, \ and\ \bibinfo {author} {\bibfnamefont {D.~N.}\ \bibnamefont {Pfeffer}},\ }\href@noop {} {\  (\bibinfo {year} {2024})},\ \Eprint {http://arxiv.org/abs/2401.14225} {arXiv:2401.14225 [astro-ph.CO]} \BibitemShut {NoStop}%
\end{thebibliography}%
\let\addcontentsline\oldaddcontentsline

\end{document}